Московский Физико-Технический Институт (Государственный Университет)

Гематологический Научный Центр РАМН

КАТРУХА

Евгений Александрович

# ДИНАМИЧЕСКИЕ НЕУСТОЙЧИВОСТИ В КИНЕТИКЕ РОСТА И ДЕПОЛИМЕРИЗАЦИИ ТУБУЛИНОВЫХ МИКРОТРУБОЧЕК

специальность 03.00.02 (биофизика)

ДИССЕРТАЦИЯ

на соискание ученой степени

кандидата физико-математических наук

Научный руководитель:

д.ф.-м.н, Г.Т. Гурия

Москва – 2007

# СОДЕРЖАНИЕ

# Список основных используемых сокращений

МТ      – микротрубочка (она же тубулиновое волокно)
ГТФ     – гуанозин-трифосфат
ГДФ     – гуанозин-дифосфат
АТФ     – аденозин-трифосфат
MAP     – microtubule associated proteins, влияющие на динамику микротрубочки белки
GAP     – GTPase activating proteins, белки, активирующие ГТФазы
GEF     – guanine nucleotide exchange factors, факторы регуляции обмена между ГТФ и ГДФ
DLA     – diffusion limited aggregation, лимитированная диффузией агрегация
ELAD    – energetically limited aggregation disaggregation, энергетически лимитированная агрегация-дисагрегация
SOC     – self-organized criticality, самоорганизующаяся критичность
РДП     – реакционно-диффузионно-преципитационная
Tu-ГТФ  – молекула тубулина, связанная с ГТФ
Tu-ГДФ  – молекула тубулина, связанная с ГДФ
$MT_n$  – микротрубочка, состоящая из n молекул тубулина
$N_A$   – число Авогадро
МФТИ    – Московский физико-технически институт
ФМБФ    – Факультет молекулярной и биологической физики
ГНЦ     – Гематологический научный центр
РАМН    – Российская академия медицинских наук
РФФИ    – Российский фонд фундаментальных исследований

# ВВЕДЕНИЕ

Тубулиновые микротрубочки (наряду с актиновыми и промежуточными филаментами) являются основными компонентами цитоскелета эукариотических клеток. Поведение микротрубочек на определенных стадиях клеточного цикла и в условиях *in vitro* характеризуется крупномасштабными нестационарными пульсациями их длины. Нестационарные апериодические режимы, типа изображенного на рис.В.1, имеют вид, характерный для релаксационных хаотических колебаний [Анищенко, 2000; Кузнецов, 2001; Данилов, 2001; Трубецков, 2004]. Особенностью этих колебаний являются отчетливо выраженные моменты переключения микротрубочки со стадии роста на деполимеризацию и наоборот. В биологической литературе именно эти моменты в динамике микротрубочек принято называть «катастрофами» и «спасениями»[1].

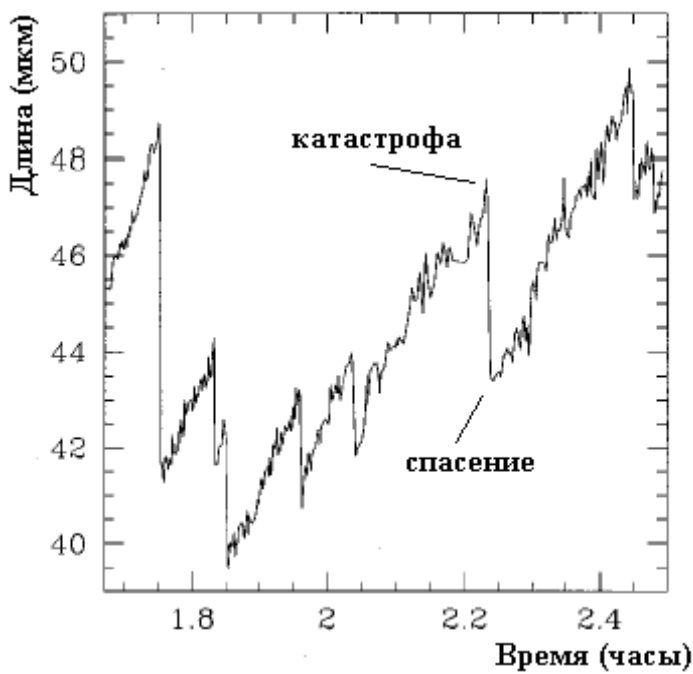

**Рис.В.1.** Динамическая нестабильность микротрубочки [Fygenson et al., 1994]. По оси абсцисс отложено время, по оси координат – размер индивидуальной микротрубочки.

В современной литературе под термином «динамическая нестабильность» микротрубочек понимается круг явлений, связанных с «катастрофами» и «спасениями» [Alberts et al., 2002]. Во избежание недоразумений в дальнейшем под термином «катастрофа» будет подразумеваться динамическая неустойчивость роста микротрубочек, а под термином «спасение» – неустойчивость процесса деполимеризации.

---

[1] В англоязычной литературе catastrophe и rescue event.

По существующим представлениям динамические нестабильности в кинетике роста и деполимеризации микротрубочек лежат в основе крупномасштабных трансформаций цитоскелета, свойственных процессам пролиферации, дифференцировки и миграции клеток в норме и патологии [Jordan and Wilson 2004; Watanabe T. et al., 2005; Honore et al., 2005].

В частности, крупномасштабные пространственно-временные трансформации цитоскелета наблюдаются на стадии деления клеток. В этой связи большое значение приобретает вопрос о том, в какой мере процессы деления атипичных и/или трансформированных клеток обуславливаются условиями возникновения и особенностями развития динамических неустойчивостей микротрубочек [Scholey et al., 2003; Канцерогенез, 2004; Honore et al., 2005].

С момента открытия самого явления динамической нестабильности микротрубочек в 1984 году, накоплен обширный экспериментальный материал и предложено несколько теоретических моделей явления. Проведенный в работе анализ имеющихся данных и теоретических подходов показал, что динамические неустойчивости, свойственные отдельным микротрубочкам и их ансамблям, перспективно изучать в свете современной теории критических явлений и неравновесных структур [Haken, 1977; Николис и Пригожин, 1979; Эбелинг, 1979].

Целью настоящей работы являлось нахождение необходимых и достаточных условий параметрической дестабилизации микротрубочек, выяснение кинетических механизмов регуляции динамических нестабильностей при росте и деполимеризации микротрубочек.

В работе решались следующие задачи:

1. Методами стохастического анализа выяснить условия развития структурных неустойчивостей, сопровождающихся деполимеризацией или фрагментацией микротрубочек.

2. Построить кинетическую модель процессов сорбции молекул тубулина на плюс-концы микротрубочек из раствора (десорбции в раствор).

3. Выяснить условия потери устойчивости стационарных распределений микротрубочек по длинам. Построить диаграммы состояния, позволяющие определять характер динамики микротрубочек при различных значениях параметров системы.

4. Исследовать механизмы формирования концентрационных колебаний и волн тубулина, ассоциированных с крупномасштабными трансформациями цитоскелета.

5. Проанализировать сочетанное действие биохимических агентов различных типов на смену динамических режимов, свойственных тубулиновым волокнам.

Развит последовательный физико-математический подход к описанию катастроф в динамике микротрубочек. Наблюдаемые экспериментально катастрофы трактуются, как результат развития процессов нелинейного взаимодействия структурных дефектов при их кластеризации в микротрубочках. При этом структурные катастрофы и эффекты фрагментации микротрубочек, подобно тому, как это имеет место в физике фазовых переходов первого рода, описываются в рамках феноменологических кинетических уравнений. Использование численных методов и компьютерных алгоритмов так называемой теории клеточных автоматов дало возможность довести результаты расчетов до стадий, допускающих прямое сопоставление с данными экспериментов.

Впервые удалось с единых позиций объяснить целый ряд нетривиальных эффектов, характерных для динамики микротрубочек: степенную зависимость частоты катастроф от скорости роста микротрубочек, дробно-линейную зависимость времени задержки деполимеризации микротрубочек от величины их длины и концентрации тубулина в растворе и т.д.

Впервые исследована обусловленность динамических нестабильностей роста и деполимеризации микротрубочек неустойчивостями кинетических процессов сорбции молекул тубулина из раствора (и десорбции с плюс-концов микротрубочек в раствор). При этом содержащая микротрубочки реконструированная система трактовалась, как существенно <u>двуфазная</u>, состоящая из реакционно-диффузионной части и собственно микротрубочек (тубулиновых волокон), выступающих в качестве конденсированной фазы. Последняя характеризуется наличием черт дальнего порядка в пространственном упорядочении составляющих микротрубочку элементов (тубулиновых димеров), свойственных твердым телам.

Впервые не постулированы, а исследованы условия потери устойчивости этой двуфазной системы. Показано, что в рассмотренной системе при приближении к критическим условиям имеет место увеличение радиуса пространственно-временных корреляций. Концентрационные флуктуации в растворной части системы трансформируются в автоволны конечной амплитуды, управляющие крупномасштабными трепетаниями микротрубочек.

Впервые построена параметрическая диаграмма состояния, на которой отображены границы областей устойчивости, отвечающие стационарным и нестационарным режимам поведения системы. На основе диаграммы состояния дана классификация воздействий цитостатическими агентами на тубулиновые микротрубочки. Выделено четыре основных класса воздействий. Построена таблица, отражающая наличие эффектов взаимного усиления/ослабления действия цитостатических агентов при совместном применении агентов различных классов.

Полученные в работе результаты имеют важное значение для поиска эффективных режимов управления крупномасштабной трансформацией цитоскелета на ключевых стадиях клеточного цикла. Построенная диаграмма состояний микротрубочек позволяет уже на стадии проектирования экспериментов *in vitro* целенаправленно проводить подбор необходимых параметров (концентрации тубулина, ГТФ, ГДФ), соответствующих искомому динамическому режиму системы микротрубочки-раствор. Установленный характер сочетаемости одновременного воздействия биохимических агентов и цитостатических препаратов различных классов открыл принципиально новую возможность для поиска путей снижения побочного токсического действия. Открылась возможность для целенаправленного поиска и разработки комплексных противоопухолевых препаратов, обладающих совместным (синергетическим) эффектом действия входящих в их состав компонентов.

# Глава 1. Обзор литературы

## § 1.1 Общие представления о структуре микротрубочек

Как известно, микротрубочки, наряду с актиновыми и промежуточными филаментами, являются важной частью цитоскелета эукариотических клеток [Alberts et al., 2002]. В 70-80-х годах XX столетия было обнаружено, что динамика микротрубочек на определенных стадиях клеточного цикла в условиях *in vivo* характеризуется нестабильностями [Inoue and Ritter, 1975; Inoue, 1981; Sammak and Borisy, 1988]. Отдельные микротрубочки в состоянии резко укорачиваться, после чего повторно возобновлять свой рост. Явление стремительного укорачивания микротрубочек также удалось обнаружить и в системах *in vitro*, приготовленных по известным методикам и состоящих из строго определенного набора очищенных биохимических компонентов [Mitchison and Kirschner, 1984; Walker et al., 1988].

Оказалось, что существует прямая связь между динамическими нестабильностями микротрубочек и энергетическим метаболизмом клетки [Olmsted and Borisy, 1975; Carlier and Pantaloni, 1981; Melki et al., 1988]. В настоящее время существует большое количество данных, указывающих на то, какие из эндогенных факторов регуляции энергообмена в клетке оказывают наибольшее воздействие на трансформацию клеточного цитоскелета, в том числе, на трансформацию его микротубулиновой части [Janmey, 1998; Valiron et al., 2001; Anesti and Scorrano, 2006; Grigoriev et al., 2006].

Существует несколько разных точек зрения на природу биологической регуляции динамической нестабильности [Holy and Leibler, 1994; Wilson et al., 1999; Caudron et al., 2005].

В силу наличия большого числа регуляторных эндогенных факторов в системах *in vivo*, для изучения природы динамических нестабильностей представляется более удобным обратиться сначала к анализу реконструируемых систем *in vitro*. Активное изучение последних в связи с проблемами микротубулиновой динамики началось с цикла работ в начале 70-х годов XX века [Weisenberg, 1972; Borisy and Olmsted, 1972; Shelanski et al., 1973] и продолжается до настоящего времени [Caudron et al., 2000; Valiron et al., 2001; Howard and Hyman, 2003]. При анализе динамики микротрубочек в реконструируемых системах *in vitro*, удается выделять фазы, когда микротрубочки являются стабильными, т.е. их длина не изменяется во времени. При этом в микротубулиновых сетях наблюдается стационарное распределение микротрубочек по длинам.

Наряду с этим бывают ситуации, в которых микротрубочки демонстрируют отчетливо выраженные крупномасштабные трепетания своей длины [Mitchison and Kirschner, 1984; Walker et al., 1988; Fygenson et al, 1994]. При помощи методов оптического рассеяния света и ренгеновского рассеяния удается наблюдать отчетливо выраженные изменения длины, присущие целым ансамблям микротрубочек [Carlier et al., 1987; Mandelkow et al., 1988]. В то время, как трепетания и изменения длины отдельных микротрубочек наиболее часто наблюдаются при помощи средств видеомикроскопии: методов флуоресценции, темного поля и дифференциально-интерференционного контраста (см. рис.1.1) [Mitchison and Kirschner, 1984; Walker et al., 1988; Walker et al., 1991].

Иными словами, в настоящее время развиты хорошо воспроизводимые методы контроля и регистрации статических и динамических характеристик тубулиновых волокон в системах *in vitro*.

При анализе реконструированных систем *in vitro*, представляется важным отметить, что по своей природе они являются системами конечномерными, т.к. речь идет о приготовленных искусственно системах, в которые входят несколько основных компонентов. Прежде всего, это молекулы тубулина, связанные с ГТФ или ГДФ. В ряде случаев, одновременно проводится дополнительное насыщение систем энергетически богатыми факторами (ГТФ, АТФ) [Olmsted and Borisy, 1975; Melki et al., 1988]. Кроме того, в реконструируемые системы добавляются центросомы или же аксонемы в качестве нуклеационных затравок. При этом контролируется общий pH и температура.

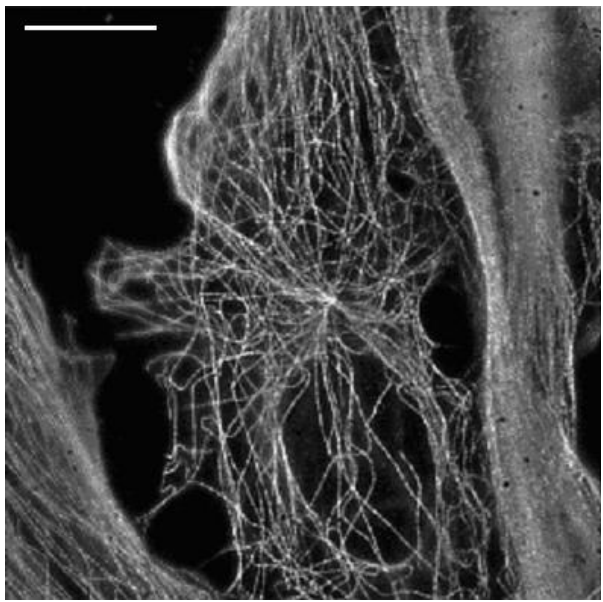

**Рис.1.1.** Микротрубочки, окрашенные флуоресцентным белком в клетках Vero (эпителий почки зеленой африканской мартышки) [Коваленко, 2005]. Размер светлого прямоугольника в левом верхнем углу соответствует 10 мкм.

Роль молекул ГТФ и ГДФ в реакциях полимеризации тубулина была выяснена в цикле работ в начале 80-х годов прошлого столетия [Carlier and Pantaloni, 1981; Carlier, 1982]. Было установлено, что на растущий конец микротрубочки из раствора сорбируются молекулы тубулина, связанные с ГТФ. Затем имеет место гидролиз ГТФ до ГДФ. В результате основное тело микротрубочки состоит из молекул тубулина, связанных с молекулами ГДФ[1] [Carlier and Pantaloni, 1981]. Эти же молекулы возвращаются в раствор в ходе реакций десорбции (деадсорбции) из микротрубочки при ее деполимеризации [Carlier, 1982].

С точки зрения современных методов формальной кинетики, реконструированные системы *in vitro* представляют собой объекты, которые уместно отнести к числу традиционно изучаемых в рамках теории динамических систем [Аносов и Арнольд, 1985; Кузнецов, 2001]. Это утверждение справедливо ровно в той мере, в которой удается поддерживать параметры рассматриваемых систем на неизменном уровне. К числу такого рода параметров может относиться и уровень содержания отдельных ингредиентов системы. В частности, например, если в реконструированную систему добавляется биохимическая ферментативная система[2], позволяющая поддерживать уровень ГТФ на определенном практически неизменном уровне, протекание основных реакций полимеризации и деполимеризации молекул тубулина, будет происходить с постоянными константами скоростей [Melki at al., 1988]. Важно отметить, что при отсутствии механизмов поддержания энергообеспеченности, реконструированная система, предоставленная самой себе, с течением времени, в силу исчерпания энергетических ресурсов, придет к своему термодинамическому равновесному состоянию, в котором все микротрубочки исчезнут [Carlier et al., 1987; Melki at al., 1988].

Тем самым ясно, что тубулиновые микротрубочки являются существенно неравновесным объектом. Они имеют место только в системах, далеких от своего термодинамического равновесия. В связи с этим явление динамической нестабильности микротрубочек в данной работе будет рассматриваться с позиций теории неравновесных систем. При этом крупномасштабные трепетания микротрубочек будут трактоваться, как разновидность неравновесных режимов поведения, обладающих чертами пространственно-временной упорядоченности.

Прежде, чем подробно остановиться на теоретических методах описания динамических нестабильностей, представляется уместным изложить основные структурные особенности строения микротрубочек.

Микротрубочки представляют собой полые цилиндрические супрамолекулярные структуры с внешним диаметром 24 нм, внутренним диаметром 13 нм и длиной до 100 мкм, состоящие из молекул белка тубулина (см. рис.1.2) [Nogales et al., 1998; Li et al., 2002].

---

[1] Всюду в дальнейшем молекулы тубулина, связанные с молекулами ГТФ и ГДФ, будут обозначаться, как тубулин-ГТФ и тубулин-ГДФ.
[2] На практике используется киназная система, способная восстанавливать ГТФ из ГДФ.

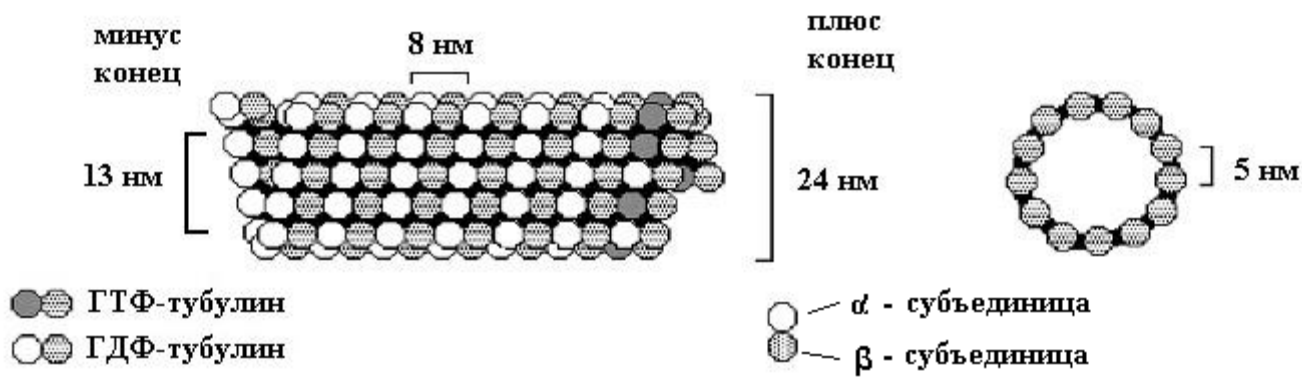

**Рис.1.2**. Структура и характерные размеры микротрубочки и составляющих ее элементов.

Молекула тубулина состоит из двух субъединиц: α- и β-форм, поэтому тубулин также называют димером. По данным электронной микроскопии микротрубочки представляют собой 3-х заходные правозакрученные спиральные структуры [Nogales et al., 1998]. В силу химической анизотропии, свойственной отдельным молекулам тубулина и стерическим ориентационным эффектам при их полимеризации, у каждой микротрубочки принято различать "плюс" и "минус" концы [Alberts et al., 2002]. Минус-конец обычно закрепляется вблизи ядра клетки, на центросоме, а плюс-конец у внешней мембраны. Структурные аспекты строения микротрубочек (зафиксированных в тех или иных экспериментальных условиях) изучены достаточно подробно как методами электронной микроскопии, так и ренгеноструктурного анализа [Nogales et al., 1998; Nogales, 2001; Li et al., 2002].

Динамические черты поведения отдельных микротрубочек и их ансамблей активно изучаются с экспериментальной [Caudron et al., 2001; Pearson et al., 2006; Grigoriev et al., 2006] и теоретической точек зрения [Flyvbjerg et al., 1996; Houchmandzadeh and Vallade, 1996; Hammele and Zimmermann, 2003]. В связи с тем, что микротрубочки играют важную роль в клеточном цитоскелете и особенно активно они перестраиваются на стадии митоза, понятно, что функциональное назначение микротрубочек прежде всего связано с их динамическими возможностями, в частности, с их способностью к стремительному изменению длины [Holy and Leibler, 1994; Karsenti and Vernos, 2001]. Этот круг вопросов продолжает интенсивно изучаться [Potapova et al., 2006; Grigoriev et al., 2006; Howard and Hyman, 2007].

## § 1.2 Динамические особенности поведения микротрубочек и их математические модели

### § 1.2.1 Модели, основанные на предположении о существовании ГТФ-крышки

Ранние попытки теоретического описания динамической нестабильности, представлены в цикле работ Хилла и соавторов [Hill and Carlier, 1983; Chen and Hill, 1983]. В них выдвинуто предположение, что природа динамической нестабильности обуславливается гидролизом связанной с тубулином молекулы ГТФ, сопровождающим полимеризацию (гипотеза ГТФ-крышки). Полагается, что гидролиз связанной с тубулином молекулы ГТФ в стенке микротрубочки происходит самопроизвольно. Скорость полимеризации микротрубочек зависит от концентрации тубулина-ГТФ в растворе. Авторы постулируют, что полимеризуется только тубулин-ГТФ из раствора (но не тубулин-ГДФ) и он может присоединяться только к тем молекулам тубулина на конце микротрубочки, которые ассоциированы с молекулой ГТФ (но не ГДФ), т.е. еще не успели гидролизоваться после присоединения к концу микротрубочки.

В качестве одной из основных характеристик динамики микротрубочек, авторы используют величину стационарного потока присоединяющихся к микротрубочке молекул тубулина. В рамках модели авторы выделяют два предельных случая. Согласно первому, если скорость присоединения превышает скорость гидролиза, микротрубочка растет с постоянной скоростью, зависящей от концентрации тубулина-ГТФ в растворе. При этом микротрубочка состоит из двух частей: основной части, состоящей из тубулина-ГДФ, и конечного участка, расположенного на конце микротрубочки, состоящего из молекул тубулина-ГТФ. Этот участок принято называть ГТФ-крышкой[3]. Во втором выделенном авторами предельном случае скорость гидролиза превышает скорость полимеризации. При этом вся микротрубочка полагается состоящей из тубулина-ГДФ и в силу невозможности присоединения молекул тубулина-ГТФ из раствора, имеют место лишь реакции отсоединения молекул тубулина-ГДФ, что и проявляется в виде деполимеризации микротрубочки.

Основными параметрами модели являются: концентрация молекул тубулина-ГТФ в растворе, константы скоростей реакций присоединения и отсоединения молекул тубулина-ГТФ и тубулина-ГДФ и константа скорости реакции гидролиза. Используя методы имитационного моделирования, известные, как методы Монте-Карло, авторам удалось показать наличие порогового эффекта в кинетике роста микротрубочек. При концентрациях тубулина-ГТФ в растворе выше некоторой критической $Tu_{cr}$, наблюдается стационарный рост микротрубочеки. Если же концентрация тубулина в растворе ниже пороговой, то происходит деполимеризация. Скорость реакции деполимеризации при этом нелинейно зависит от концентрации тубулина-ГТФ в растворе [Chen and Hill, 1983].

Данный результат был поставлен под сомнение в работе Митчисона и Киршнера [Mitchison and Kirschner, 1984], в которой с помощью видеомикроскопии наблюдались отдельные микротрубочки *in vitro*. Полное отсутствие полимеризации микротрубочек наблюдалось при концентрации тубулина $Tu_0$, значительно меньшей по сравнению с $Tu_{cr}$, рассчитанной ранее с помощью модели [Chen and Hill, 1983]. При этом в работе [Mitchison and Kirschner, 1984] было показано, что при концентрациях тубулина в растворе $Tu$ таких, что $Tu_0 < Tu < Tu_{cr}$, скорость роста микротрубочек не является стационарной величиной, а подвергается значительным флуктуациям. За стадиями монотонного роста следуют стадии стремительной, полной деполимеризации. Подобное поведение получило название динамической нестабильности, а переход из состояния роста к укорачиванию получил название катастрофы. Явление динамической нестабильности получило подтверждение в последующем ряде работ независимых исследователей (см.рис.1.3) [Walker et al., 1988; Fygenson et al., 1994].

---

[3] В англоязычной литературе устоялся термин «GTP cap».

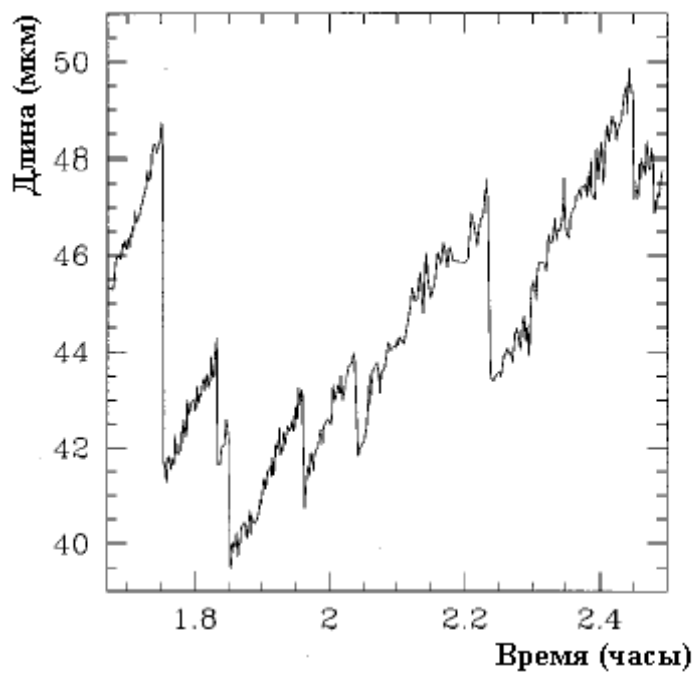

**Рис.1.3.** Динамическая нестабильность микротрубочки [Fygenson et al., 1994]. По оси абсцисс отложено время, по оси координат – размер индивидуальной микротрубочки.

Авторы исходной гипотезы о существовании ГТФ-крышки модифицировали свою раннюю модель [Hill and Chen, 1984; Hill, 1984; Chen and Hill, 1985]. Согласно новым выдвинутым предположениям, длина ГТФ-крышки не постоянна, а может претерпевать значительные флуктуации во времени. Флуктуации, в частности, могут приводить к полному исчезновению фрагмента из тубулина-ГТФ на конце микротрубочки, что по мнению авторов должно приводить к переключению динамики микротрубочек из фазы роста в фазу деполимеризации. Проведенный авторами численный анализ показал, что в новой интерпретации теоретическая модель лучше описывает экспериментальные данные [Chen and Hill, 1985].

В дальнейшем, однако, появились ряд экспериментальные работы, в которых гипотеза ГТФ-крышки была вновь поставлена под сомнение [O'Brien et al., 1987; Schilstra et al., 1987]. Эксперименты по определению длины гипотетической ГТФ-крышки показали [O'Brien et al., 1987; Schilstra et al., 1987], что она либо не существует вовсе, либо ее размер крайне мал (порядка 10-20 молекул тубулина), что значительно меньше оценок, приводимых в предыдущих теоретических работах (1000 молекул тубулина) [Hill and Chen, 1984; Chen and Hill, 1985].

Для преодоления этого противоречия в работе Бейли и соавторов [Bayley et al., 1989] была предложена модель так называемой «латеральной» ГТФ-крышки (lateral GTP-cap). В ней предполагается, что гидролиз молекулы тубулина-ГТФ, находящейся на краю микротрубочки, осуществляется лишь вследствие того, что очередная молекула тубулина-ГТФ уже из раствора, присоединяется к микротрубочке. Таким образом, вся микротрубочка полагается состоящей из молекул тубулина-ГДФ, за исключением молекул тубулина-ГТФ, находящихся на самом крае микротрубочки.

В случае, если большая часть краевых молекул тубулина-ГТФ десорбируется, по мнению авторов, запускается процесс стремительной деполимеризации всей микротрубочки. Иными словами, предполагается, что ГТФ-крышка состоит из одного слоя молекул тубулина-ГТФ. Численное моделирование методом Монте-Карло позволило авторам в рамках упомянутых предположений произвести расчет пороговой концентрации тубулина, а также найти зависимость средних времен нахождения ансамбля микротрубочек в фазе роста или деполимеризации от концентрации тубулина в растворе [Bayley et al., 1990].

Хорошее согласие полученных результатов с упомянутыми выше экспериментальными данными является достоинством данной работы. Однако, подход, в целом основанный на идее «латеральной ГТФ-крышки» оказался недостаточно эффективным при попытках использовать его для объяснения результатов экспериментов по вымыванию тубулина-ГТФ из раствора, в котором находятся микротруочки [Walker et al., 1991; Voter et al.,1991].

## § 1.2.2 Кинетическая модель Флайберга, Холи и Лейблера

Четыре года спустя, в 1994 году, вышеперечисленные работы были подвергнуты критике в публикациях Флайберга, Холи и Лейблера [Flyvbjerg et al., 1994; Flyvbjerg et al., 1996]. Авторами было отмечено, что теоретические подходы, основанные на гипотезе ГТФ-крышки, не позволяют в полной мере описывать большой круг экспериментальных данных. В частности, эксперименты по вымыванию димеров тубулина-ГТФ из раствора и эксперименты, в ходе которых регистрировалась зависимость частоты катастроф от концентрации тубулина [Walker et al., 1988; Walker et al., 1991; Voter et al.,1991; Drechsel et al., 1992]. Одновременно Флайберг и соавторы отмечают, что на основании использованных ранее экспериментальных данных в принципе не представляется возможным оценить величины констант скоростей реакций сорбции, десорбции и гидролиза для отдельных тубулиновых димеров. При моделировании динамики микротрубочек численным методом Монте-Карло, значения констант скоростей реакций фигурируют в качестве свободных параметров [Chen and Hill, 1985; Bayley et al., 1990]. Поэтому Флайберг и соавторы предлагают построение нового подхода, содержащего как можно меньшее количество свободных параметров.

Основными параметрами в работах [Flyvbjerg et al., 1994; Flyvbjerg et al., 1996] являются средняя скорость роста микротрубочки и средняя скорость гидролиза тубулина-ГТФ в микротрубочке. По Флайбергу, если рост микротрубочки опережает гидролиз присоединяющихся молекул тубулина-ГТФ, то формируется ГТФ-крышка. Полагается, что внутри ГТФ-крышки может происходить самопроизвольный гидролиз молекул тубулина-ГТФ. Решая кинетическое уравнение изменения во времени длины ГТФ-крышки, авторы получили характерный вид зависимости частоты катастроф от концентрации тубулина. Кроме того, авторам удалось теоретически найти зависимость времени ожидания деполимеризации микротрубочек после вымывания из раствора молекул тубулина-ГТФ. Обе зависимости оказались в удовлетворительном согласии с экспериментальными данными. Наиболее ценным с теоретической точки зрения явилось и то, что обе зависимости были впервые объяснены в рамках одного подхода.

### § 1.2.3 Модели структурных крышек

В середине 90-х годов, путем сверхбыстрого замораживания растущих и деполимеризующихся трубочек, с последующим изучением их под электронным микроскопом удалось получить фотографии концов микротрубочек с высоким разрешением (рис.1.4) [Chretien et al., 1995; Janosi et al., 1998]. Оказалось, что протофиламенты на конце полимеризующихся микротрубочек в основной своей массе вытянуты и слабо отклоняются от оси симметрии микротрубочек. Тогда как на конце у стремительно разрушающихся микротрубочек они изогнуты в форме завитков, напоминающих лепестки ромашки (при этом конец микротрубочки напоминает раструб музыкальной трубы, см. рис.1.4А).

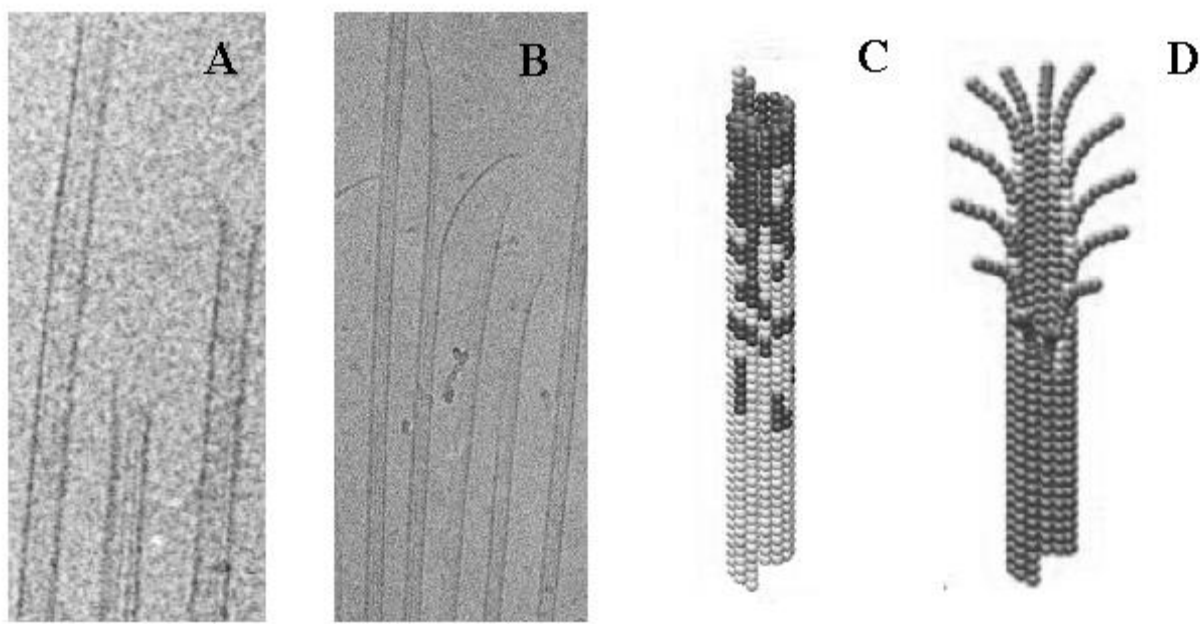

**Рис.1.4.** Фотографии формы концов деполимеризующихся (**A**) и растущих (**B**) микротрубочек, полученные методами электронной микроскопии [Chretien et al., 1995]. Пространственный масштаб для каждой фотографии А и Б может быть определен исходя из толщины волкон, приблизительно равной 24 нм. (**C**), (**D**) – трехмерная визуализация структуры концов микротрубочек, согласно работе [Van Buren et al., 2005]

Объяснение этих различий было предложено в работах Джанози и соавторов [Janosi et al., 1998; Janosi et al., 2002]. Авторы этих работ ввели новое понятие: понятие «структурной крышки». Как и в ряде предыдущих работ полагается, что на конце микротрубочки существует участок, состоящий из молекул тубулина-ГТФ, тогда как основная часть микротрубочек состоит из молекул тубулина-ГДФ.

При этом предполагается, что микротрубочка является эластичным полимером цилиндрической формы, по всей длине которого могут существовать механические напряжения, обусловленные гетерогенной структурой составляющих его элементов. Внутренние напряжения возникают за счет того, что по мнению авторов, молекула тубулина-ГДФ в обычной конформации имеет изогнутую форму, тогда как молекула тубулина-ГТФ – прямую. Следовательно, при переходе из одной формы в другую (гидролизе ГТФ) уже в стенках микротрубочки, за счет энергии гидролиза ГТФ, молекулы димера изгибаются, что приводит к росту внутреннего механического напряжения в микротрубочке.

Таким образом, часть микротрубочки, состоящая из тубулина-ГДФ, содержит накопленную энергию, высвобождению которой мешает нахождение на конце микротрубочки стабилизирующего сегмента из тубулина-ГТФ, для которого прямая форма микротрубочки является равновесной. Как только он исчезает, микротрубочка переходит в метастабильное состояние и прекращает рост. Далее могут следовать два сценария. Согласно первому, на край микротрубочки могут сорбироваться тубулиновые димеры из раствора, при этом рост микротрубочки возобновляется. Если же крышка не восстанавливается, тепловой флуктуацией запускается процесс деполимеризации, в данном случае энергетически выгодный, поскольку способствует сбросу накопленного механического напряжения микротрубочки.

Развитый авторами подход апеллирует к очевидности того факта, что на краю полимеризующейся микротрубочки имеется зона, содержащая не искривленные тубулиновые протофиламенты. Именно с ней авторы предлагают ассоциировать понятие «структурной ГТФ-крышки». Размер этой зоны значительно превышает монослой, что, по мнению авторов, свидетельствует о том, что реакции гидролиза ГТФ занимают конечное время и они сопряжены с механической структурной реорганизацией в зоне предполагаемой ГТФ-крышки.

В основе модели «структурной ГТФ-крышки» лежат статистические данные. В опубликованных авторами работах они используют методы, широко применяемые в теории сопротивления материалов для вычисления степени статистической напряженности структур разных типов.

Стоит отметить, что у моделей структурных крышек, на наш взгляд, есть ряд внутренних недостатков. В частности, авторы не проводят кинетический анализ предлагаемой модели, т.е. фактически кинетика процессов роста и деполимеризации микротрубочек никак не связана с динамической неустойчивостью. Также остается непонятным, каким образом небольшая часть тубулина-ГТФ на конце микротрубочки способна сдерживать делокализованные напряжения, вызываемые значительно большим количеством молекул тубулина-ГДФ. Это особенно трудно себе представить, если принять во внимание, что латеральное взаимодействие между молекулами тубулина-ГТФ в микротрубочке невелико [VanBuren et al., 2002; Sept et al., 2003].

В соответствии с гипотезой о роли структурной крышки, определяющие динамику тубулиновых волокон процессы должны происходить у конца микротрубочки. Эти данные, однако, не согласуются с результатами экспериментов, которые показывают, что большая часть MAP-белков сорбируется на поверхность по всей длине микротрубочек, при этом влияя на скорость роста и частоту катастроф [Pryer et al., 1992; Kowalski and Williams, 1993; Brandt and Lee, 1994].

На сегодняшний день, ввиду обилия противоречивых экспериментальных данных, не существует единого мнения о существовании и роли ГТФ-крышки в динамической нестабильности, хотя представляется очевидным, что процесс гидролиза непонятным пока образом все же связан с процессом полимеризации и может влиять на потерю устойчивости роста микротрубочек [Erickson and O'Brien, 1992; Caudron et al., 2000].

## § 1.2.4 Модель, основанная на представлении о роли вакансий в динамике микротрубочки

В 1996 году появилась статья М.Семенова [Semenov, 1996], в которой автор впервые предположил, что в упорядоченной структуре микротрубочки могут появляться дефекты (см. рис.1.5). Происхождение дефектов может быть обусловлено разными причинами: присоединением к микротрубочке поврежденных молекул, сложным механизмом образования продольных и поперечных связей, процессами, сопряженными с гидролизом ГТФ и процессами релаксации механических напряжений, а также случающимися в процессе роста микротрубочки продолжительными паузами [Walker et al., 1988]. Доказательствами реального существования дефектов служат следующие экспериментальные данные.

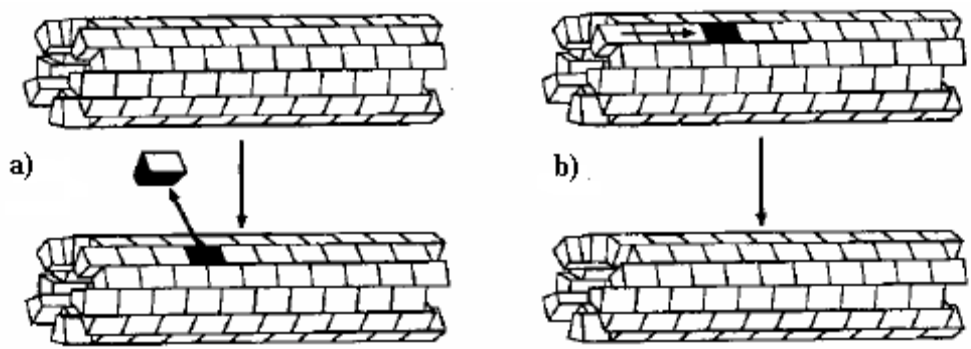

**Рис.1.5.** Иллюстрация гипотезы Семенова об образовании **a)** и перемещении **b)** дефектов в структуре микротрубочки. Рисунок взят из оригинальной работы [Semenov, 1996].

Известно, что укорачивание микротрубочки имеет место при уменьшении концентрации тубулина в растворе. Это может быть вызвано как деполимеризацией на концах, так и выходом димеров из стенок. Косвенное доказательство факта обмена димерами между стенками и раствором дано в эксперименте, в котором микротрубочки вовсе не имели свободных концов. При понижении концентрации наблюдалось укорачивание тубулиновых волокон [Koshland et al., 1988].

Согласно данным электронной микроскопии, появление дефектов крайне вероятно в областях, где изменяется спиралевидная структура микротрубочек (переходы между структурами, состоящими из 13-ти, 14-ти или 15-ти протофиламентов) [Chretien et al., 1992]. Добавление химических агентов (таких, как таксол и паклитаксел) способно увеличивать количество областей, в которых наблюдаются подобного рода структурные дефекты [Arnal and Wade, 1995; Diaz et al., 1998]. Исследования способного вызывать «разлом» тубулиновых волокон белка катанина (katanin) показали, что местом его предпочтительного связывания являются те места стенки микротрубочек, в которых наблюдается накопление дефектов [Davis et al., 2002]. Непосредственная визуализация точечных дефектов решетки микротрубочек и растущего конца была получена с помощью сканирующей силовой микроскопии в работе [Schaap et al., 2004]. Подводя общий итог, можно отметить, что за последние десять лет существование дефектов в упорядоченной структуре микротрубочек из теоретической гипотезы перешло в раздел подтвержденных фактов.

Согласно гипотезе Семенова, дефекты также могут перемещаться вдоль волокна, изменяя свое местоположение [Semenov, 1996]. Эффективное блуждание дефекта проявляется в том, что один из соседних продольных тубулиновых димеров сдвинется на место дырки

В работе Семенова, дается оценка силы, которую может развивать микротрубочка, сокращающаяся за счет «испарения» отдельных тубулиновых молекул. Автор исходит из положения, что растяжение микротрубочки должно сопровождаться увеличением числа вакансий в них, а сжатие, напротив, вести к уменьшению числа вакансий.

В ходе своего анализа Семенов не использовал кинетических моделей движения дефектов, ограничиваясь термодинамическими соображениями. Используя, по сути, квазиравновесный подход, Семенов не претендует на описание крупномасштабных динамических нестабильностей, свойственных микротрубочкам.

Гипотеза Семенова о существовании и движении дефектов, дополненная соображениями о возможных механизмах коллективного взаимодействия, легла в основу нового теоретического подхода описания динамической нестабильности микротрубочек, развиваемого в данной работе (см. главу 2).

### § 1.2.5 Кинетические модели

Помимо приведенного выше круга моделей, явно основанных на гипотезах о существовании своеобразных ГТФ-крышек, для описания наблюдающихся трепетаний микротрубочек и их коллективных колебаний было предложено несколько кинетических моделей феноменологического характера. В данных моделях не анализируются характеристики, присущие отдельным микротрубочкам (внутренняя структура, частоты катастроф, спасений, характер гидролиза). Основной акцент делается на рассмотрении коллективных характеристик ансамблей микротрубочек (распределению микротрубочек по длинам, временной динамике общей массы полимеризованного тубулина, влиянию внешних параметров).

Ранние теоретические работы в этой области принадлежат Хиллу, Чену и Карльер [Carlier et al., 1987; Chen and Hill, 1987; Melki et al., 1988]. В этом цикле работ была построена и исследована модель, объясняющая коллективные колебательные процессы, наблюдающиеся путем регистрации оптической плотности в растворе, содержащем полимеризующиеся микротрубочки [Carlier et al., 1987]. Авторы используют следующие основные переменные для описания системы микротрубочек *in vitro*: концентрации тубулина-ГТФ и тубулина-ГДФ в растворе, общее количество полимеризованного тубулина и долю микротрубочек, находящихся в растущем и (деполимеризующемся) состоянии [Chen and Hill, 1987]. В работе [Carlier et al., 1987] впервые предложена базовая реакционная схема, согласно которой, в растворе происходят следующие процессы:

А) рост тубулиновых волокон происходит за счет сорбции молекул тубулина-ГТФ из раствора. Сразу после присоединения молекулы тубулина-ГТФ к микротрубочке происходит гидролиз связанной молекулы ГТФ.

Б) с определенными частотами микротрубочки способны переключаться из состояния роста в состояние деполимеризации и обратно.

В) в результате деполимеризации из микротрубочек в раствор выходят молекулы тубулина-ГДФ.

Г) тубулин-ГДФ способен восстанавливаться в растворе до тубулина-ГТФ.

На основании предложенной схемы авторами была построена система кинетических уравнений:

$$dc_p/dt = KJ_1 c^n, \quad dp_1/dt = \tilde{k}(1-p_1) - kp_1 \qquad (1.1)$$

$$dc/dt = -J_1 p_1 c_p + \eta c_D \qquad (1.2)$$

$$dc_D/dt = -J_2(1-p_1)c_p - \eta c_D \qquad (1.3)$$

$$c_p^* = \bar{N} c_p = c_t - c - c_D \qquad (1.4)$$

где $c_p$ – концентрация микротрубочек, $K$ – константа, характеризующая скорость реакции образования нуклеационной затравки, $p_1$ – доля концов микротрубочек, растущих со скоростью $J_1$, $J_2$ – скорость деполимеризации, $c$ и $c_D$ – концентрации соответственно тубулина-ГТФ и тубулина-ГДФ в растворе, $k$ – константа скорости перехода микротрубочки из растущего в деполимеризующееся состояние, $\tilde{k}$ – константа скорости обратного перехода, $\eta$ – константа скорости реакции восстановления тубулина-ГТФ из тубулина-ГДФ в растворе, $\bar{N}$ - среднее число молекул тубулина в микротрубочке, $c_t$ – концентрация тубулина-ГТФ в начальный момент времени. Среди решений системы (1.1)-(1.4) авторам удается выделить решения, отвечающие временным колебаниям. Указанные решения качественно воспроизводят основные черты поведения ансамбля микротрубочек, наблюдающиеся в системах *in vitro* методами оптического рассеяния (см. рис.1.6).

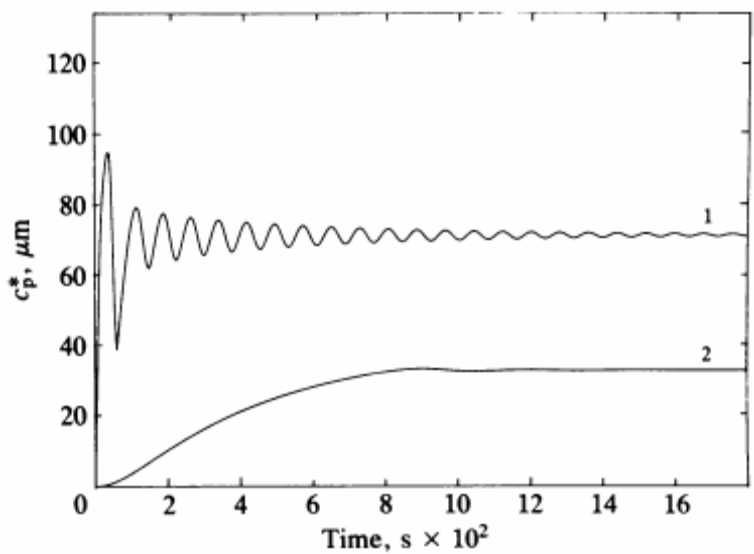

**Рис.1.6.** Теоретическая зависимость количества тубулина, находящегося в полимеризованной фазе от времени, согласно работе [Carlier et al., 1987]. 1) – колебательное решение; 2) – решение, релаксирующее к стационарному.

Не сложно видеть, что в развитом подходе все переменные являются независящими от пространственных координат. Это подразумевает, что в рассматриваемой системе реализуется полное перемешивание. В принципе, такое возможно в системах, в которых скорость диффузионных процессов сильно превышает скорость реакционных [Франк-Каменецкий, 1947]. Условия выполнимости этого предположения в системах *in vitro*, содержащих микротрубочки подробно обсуждаются ниже, в разделе 1.2.7.

### § 1.2.6 Аксиоматические модели Лейблера и соавторов

В работах Станисласа Лейблера и соавторов [Verde et al., 1992; Dogterom and Leibler, 1993; Dogterom et al., 1995] аксиоматически постулируется, что микротрубочки могут находиться только в двух состояниях – растущем и деполимеризующемся. Для обоих состояний вводятся функции плотности вероятности обнаружения плюс-конца в данном месте – $z$, в момент времени – $t$. Частота переключений между двумя упомянутыми выше состояниями полагается изменяющейся стохастически [Dogterom and Leibler, 1993].

Скорость процесса полимеризации полагается зависящей от концентрации тубулина, связанного с ГТФ в растворе, который в свою очередь, полагается способным диффундировать по пространству. Присутствием молекул тубулина-ГДФ в растворе авторы пренебрегают. На основании этих соображений была составлена система кинетических уравнений, описывающих пространственную динамику плюс-концов растущих и деполимеризующихся микротрубочек:

$$\partial_t p_+ = -f_{+-} p_+ + f_{-+} p_- - v_+ \partial_z (c p_+) \qquad (1.5)$$

$$\partial_t p_- = +f_{+-} p_+ - f_{-+} p_- + v_- \partial_z p_- \qquad (1.6)$$

$$\partial_t c = -v_+ p_+ + D \partial_{zz} c \qquad (1.7)$$

где $p_+(z,t)$ и $p_-(z,t)$ обозначают плотности вероятности нахождения плюс-конца микротрубочки в растущем и деполимеризующемся состоянии соответственно, $f_{+-}$ – частота переключения из растущего состояния в деполимеризующееся, $f_{-+}$ – соответственно, из деполимеризующегося в растущее, $c$ – концентрация тубулина-ГТФ в растворе, $v_+$ и $v_+$ – константы скоростей реакций роста и деполимеризации соответственно, $D$ – коэффициент диффузии тубулина-ГТФ в растворе. Решения системы (1.5)-(1.7) находились аналитически и методом Монте-Карло [Dogterom and Leibler, 1993].

В результате среди решений удалось выделить два класса, соответствующих ограниченному (bounded) и неограниченному (unbounded) росту микротрубочки (см.рис.1.7). Первый характеризуется стохастическими пульсациями длины микротрубочки в области центра нуклеации (центросомы), а второй – пульсациями роста по отношению к бегущей с постоянной скоростью в бесконечность волне плотности плюс-концов микротрубочек (показано пунктиром на рис.1.7). Выяснилось, что смена типа поведения микротрубочки происходит пороговым образом, при выполнении условия $v_- f_{+-} = v_+ f_{-+}$ на параметры системы, характеризующие средние скорости роста и деполимеризации [Dogterom and Leibler, 1993].

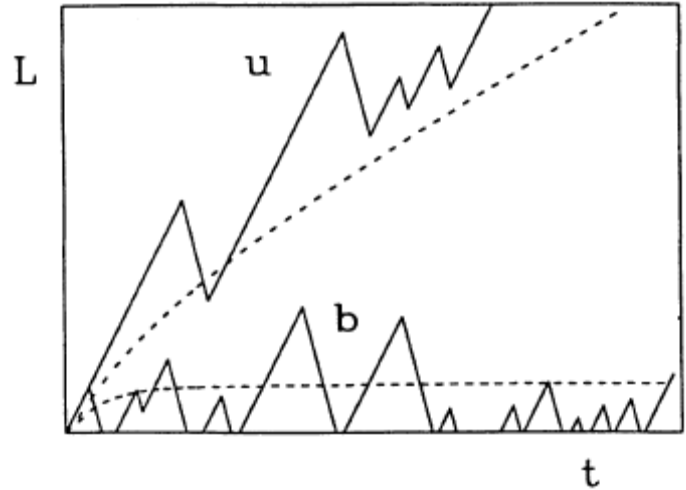

**Рис.1.7.** Изменение длины микротрубочки (L) во времени (t), согласно работе [Dogterom and Leibler, 1993]. Буквами обозначены два характерных решения: (u) неограниченного и (b) ограниченного роста.

В 1995 году появилась публикация этих же авторов [Dogterom et al., 1995], в которой основное внимание уделялось эффекту диффузионного лимитирования роста микротрубочек. В рамках данной работы полагалось, что рост микротрубочек инициируется на центросоме и происходит в радиальном направлении. Авторы исследовали следующую систему уравнений:

$$\partial_t p_+ = -f_{+-} p_+ + f_{-+} p_- - \partial_r (v_+ p_+) \qquad (1.8)$$

$$\partial_t p_- = +f_{+-} p_+ - f_{-+} p_- + v_- \partial_r p_- \qquad (1.9)$$

$$\partial_t c_T = -v_+ s_0 \left(\frac{R^2}{r^2}\right) p_+ + k c_D + D \nabla^2 c_T \qquad (1.10)$$

$$\partial_t c_D = +v_- s_0 \left(\frac{R^2}{r^2}\right) p_- - k c_D + D \nabla^2 c_D \qquad (1.11)$$

где $p_+(r,t)$ и $p_-(r,t)$ обозначают плотность вероятности нахождения плюс-конца микротрубочки в растущем и деполимеризующемся состоянии соответственно, $f_{+-}$ – частота переключения из растущего состояния деполимеризующееся, $f_{-+}$ – соответственно, из деполимеризующегося в растущее, $c_T(r,t)$ и $c_D(r,t)$ – концентрации молекул тубулина-ГТФ и тубулина-ГДФ в растворе в момент времени $t$ на расстоянии $r$ от нуклеационного центра, $v_+$ и $v_-$ – параметры, характеризующие скорости реакций роста и деполимеризации соответственно, $k$ – константа скорости реакции восстановления тубулина-ГТФ из тубулина-ГДФ в растворе, $s_0$ – плотность сайтов нуклеации на центросоме, $R$ – радиус центросомы, $D$ – коэффициент диффузии молекул тубулина-ГТФ и тубулина-ГДФ в растворе. Полагалось также, что выполняются условия:

- $v_+ = u_+ c_T$, где $u_+$ константа скорости сорбции молекул тубулин-ГТФ из раствора;
- $f_{-+} = w c_T$, где $w$ – константа;
- $f_{+-} = A \exp(-\sigma c_T)$, где $A$ и $\sigma$ являются константами.

Последнее выражение означает, что частота катастроф (перехода от роста к деполимеризации), по мнению авторов, должна уменьшаться при увеличении концентрации тубулина-ГТФ в растворе.

Решая уравнения (1.8)-(1.11) аналитически и численно (методом Монте-Карло), авторы находили зависимость от времени количества нуклеационных сайтов центросомы, занятых растущими микротрубочками. Оказалось, что решение для случая, в котором распределение молекул тубулина в растворе полагается однородным, значительно отличается от решения, когда коэффициент диффузии молекул тубулина-ГТФ и тубулина-ГДФ в растворе полагался конечным.

Варьируя количество активных нуклеационных центров на центросоме $s_0$, авторы установили наличие двух предельных режимов роста микротрубочек. Первый режим соответствует малым $s_0$ ($0 < s_0 < 0{,}5$), при этом количество растущих на центросоме микротрубочек, определяется исключительно значением $s_0$. Во втором режиме, при $s_0 > 0{,}5$ количество растущих на центросоме микротрубочек много меньше числа нуклеационных сайтов и практически не зависит от $s_0$. Выяснилось, что количество микротрубочек в этом случае определяется диффузей молекул тубулина в растворе. Другими словами, авторы работы [Dogterom et al., 1995] установили, что при определенных условиях рост микротрубочек может лимитироваться диффузионными процессами в растворе.

Авторы получили оценочное значение для характерной области возмущения однородного распределения концентраций тубулина в растворе, вызываемого ростом микротрубочки, оно составляет 10 мкм.

### § 1.2.7 Диффузионные ограничения на рост микротрубочек

Данная оценка, а также сама возможность влияния процессов диффузии и концентрационных градиентов тубулина в растворе на динамику микротрубочек была подвергнута критике в публикации Дэвида Оддэ [Odde, 1997]. В свое работе автор предпринял попытку оценить, насколько велико влияние диффузионных факторов на динамику роста и деполимеризации микротрубочек. В предположении, что микротрубочка растет с постоянной скоростью, т.е. стационарно, автор исследовал стационарный профиль концентрации тубулина-ГТФ по пространству. Основное уравнение, решаемое Оддэ, выглядит следующим образом:

$$0 = D\left(\frac{1}{r}\frac{\partial}{\partial r}\left(r\frac{\partial c}{\partial r}\right) + \frac{\partial^2 c}{\partial z^2}\right) - V\frac{\partial c}{\partial z} \quad (1.12)$$

где $c(r,z)$ – концентрация тубулина-ГТФ, $D$ – коэффициент диффузии тубулина-ГТФ, $V$ – скорость роста микротрубочки. Граничное условие, отражающее процесс сорбции молекул тубулина-ГТФ на конец растущей микротрубочки имеет вид:

$$4\pi s^2 D\left(\frac{\partial c}{\partial s}\right)_z = KV \quad \text{при} \quad s = s_0 \quad (1.13)$$

где $s = \sqrt{r^2 + z^2}$, $s_0$ – радиус растущего конца микротрубочки, а $K$ – эффективная константа скорости сорбции тубулина-ГТФ из раствора.

Автором показано, что в условиях, когда рост микротрубочек происходит квазистационарным образом, диффузионные факторы действительно обуславливают образование пространственного градиента молекул тубулина-ГТФ в растворе. Однако они, по мнению автора, не сказываются на росте микротрубочек до тех пор, пока величина скорости роста не превосходит значения 65 мкм/мин. Поскольку эта величина значительно выше скоростей, реально наблюдающихся в системах *in vitro* и *in vivo*, автор сделал вывод, что в физиологических условиях и реконструированных системах никогда не бывает столь больших градиентов, какие предсказываются в работе Лейблера и соавторов [Dogterom et al., 1995]. Сверх того, автор делает вывод о невозможности диффузионной лимитированности процессов роста и деполимеризации микротрубочек.

Работу Оддэ многократно цитируют в последующих публикациях, не подвергая сомнению его выводы [Houchmandzadeh and Vallade, 1996; Sept et al., 1999; Hammele and Zimmermann, 2003].

С нашей точки зрения, сделанные Оддэ выводы относятся исключительно к той модели, которую он проанализировал, а именно: к модели стационарного роста микротрубочки. Им полагалось, что пространственные распределения концентрации тубулина устойчивы, а скорость роста микротрубочки постоянна и не зависит от концентрации тубулина в растворе. Выяснение степени влияния градиентов концентрации тубулина на динамические нестабильности роста микротрубочек в нестационарных условиях требует отдельного рассмотрения. По крайней мере представляется очевидным, что о роли градиентов в нестационарной динамике микротрубочек нельзя сделать каких-либо выводов на основании развитых в работе [Odde, 1997] представлений.

На наш взгляд, авторы последовавших за работой Оддэ публикаций по моделированию кинетики роста и деполимеризации микротрубочек [Houchmandzadeh and Vallade, 1996; Sept et al., 1999; Hammele and Zimmermann, 2003] ошибочно полагают, что работа Одде дает основание во всех ситуациях исключать диффузионный фактор из рассмотрения. Представляется, что проблема степени диффузионной лимитированности процессов роста и деполимеризации микротрубочек остается открытой, требующей более углубленного изучения.

## § 1.2.8 Диффузионная модель динамики микротрубочек И. Воробьева и соавторов

В цикле работ И. Воробьева и соавторов [Vorobjev et al., 1997; Vorobjev et al., 1999; Воробьев и др., 2000; Komarova et al., 2002] было исследовано поведение микротрубочек *in vivo*, в культивируемых клетках при помощи флуоресцентной видеомикроскопии. Авторы показали, что общий круговорот тубулина в клетке (тубулин->микротрубочки->тубулин) обуславливается как трепетаниями плюс-концов, так и процессами деполимеризации тубулина с минус-конца [Vorobjev et al., 1999]. Минус-концы могут образовываться как путем фрагментации микротрубочек, растущих от центросомы, так и за счет реакций спонтанной нуклеации микротрубочек в цитоплазме.

Авторами была предложена так называемая диффузионная модель динамики плюс и минус концов микротрубочек, согласно которой поведение концов микротрубочек рассматривалось по аналогии с задачей о случайном одномерном блуждании (одномерная диффузия). Полагая, что микротрубочки расположены радиально, распределение их концов описывалось уравнением [Воробьев и др., 2000]:

$$\frac{\partial c(x,t)}{\partial t} = \frac{B(c)}{2}\frac{\partial^2 c(x,t)}{\partial x^2} - A(c)\frac{\partial c(x,t)}{\partial x} \qquad (1.14)$$

где $c(x,t)$ соответствует концентрации концов микротрубочек вдоль радиуса клетки, $B$ – коэффициенту диффузии концов микротрубочек, $A$ – коэффициенту сноса, отражающему рост микротрубочек.

На основании экспериментальных наблюдений динамики концов микротрубочек в разных клеточных культурах на стадии интерфазы, авторами проведен расчет величин коэффициентов сноса и диффузии концов микротрубочек при анализе стационарных распределений микротрубочек [Vorobjev et al., 1997; Vorobjev et al., 1999; Воробьев и др., 2000]. Авторами показано, что стационарное решение уравнения (1.14), экспоненциально возрастающее при приближении к границе клетки, согласуется с распределениями, наблюдаемыми экспериментально *in vivo* на стадии интерфазы [Komarova et al., 2002]. Полученные авторами значения коэффициентов *A* и *B* плюс-концов микротрубочек для различных клеточных культур представлены в таблице 1.1.

**Таблица 1.1.** Величины коэффициентов сноса *A* и диффузии *B* плюс-концов при стационарных распределениях микротрубочек по длинам в различных клеточных культурах, согласно работе [Воробьев и др., 2000].

| Культура<br>Коэффициент | $PtK_1$ (почка кенгуровой крысы) | МЭФ (мышиные эмбриональные фибробласты) | Vero (почка зеленой африканской мартышки) |
|---|---|---|---|
| *A* [мкм/мин] | 0.03 ± 0.70 | 0.76 ± 0.67 | 2.9 ± 0.5 |
| *B* [мкм$^2$/мин] | 3.4 ± 0.2 | 15.0 ± 1.4 | 17.8 ± 2.6 |

Подробный численный анализ данной модели и сопоставление данного подхода с работами, опирающимися на гипотезу о существовании «ГТФ-крышки» [Hill and Chen, 1984], были проведены в работе [Maly, 2002]. Автором было показано, что распределения микротрубочек по длинам, расчитанные согласно «диффузионной» модели удовлетворительно совпадают с вычислениями в рамках подхода, основанного на гипотезе «ГТФ-крышки» в случаях, когда ансамбли микротрубочек в клетках рассматриваются на продолжительных временах наблюдения.

Как видно, развитый в работах И. Воробьева и соавторов подход основывается на представлении о стохастической природе процессов укорочения и удлинения микротрубочек *in vivo*. При этом длина отдельной микротрубочки в клетке полагается случайной величиной, обладающей собственной функцией распределения. В отличие от развитых Лейблером и соавторами кинетических подходов (см. § 1.2.6) , в которых, следуя общим идеям Лэкса [Ван Кампен, 1990], описание строится с упором на прямой учет вероятностей перехода микротрубочек из растущего состояния в состояние деполимеризации и обратно, в работах И. Воробьева и соавторов развивается подход, восходящий идейно к подходам к стохастическим процессам, развитым Фоккером и Планком [Fokker, 1914; Plank, 1917]. Этот подход представляется предпочтительным, поскольку авторам не приходится делать сложных допущений о элементарных случайных процессах, лежащих в основании процессов удлинения и укорочения микротрубочек. Им достаточно использовать первые два момента статистической функции распределения микротрубочек по длинам, которые доступны прямому экспериментальному измерению в реальных системах (первый момент соответствует коэффициенту сноса $A$, второй момент соответствует коэффициенту диффузии концов микротрубочек $B$).

Следует отметить, что до работ И. Воробьева и соавторов так называемый «сносовый» член включался в математические модели роста микротрубочек [Chen and Hill, 1987; Dogterom et al., 1995; Flyvbjerg et al., 1996]. В то время как диффузионный, отражающий стохастические особенности процессов роста и деполимеризации, в явном виде не учитывался.

С нашей точки зрения, диффузионная модель в полной мере продемонстрировала свою эффективность применительно к проблемам интерпретации свойств стационарных распределений микротрубочек. Хотя, как несложно видеть, само уравнение (1.14) допускает и нестационарные решения. Их нахождение и последующий анализ откроют дорогу для интерпретации нестационарных явлений, свойственных микротрубочкам в системах in vivo.

## § 1.3 Нелинейные явления в динамике микротрубочек

С момента первой экспериментальной регистрации микротрубочек в 1963 году [Slautterback, 1963] и до настоящего времени они продолжают оставаться объектом интенсивного экспериментального изучения [Howard and Hyman, 2003; Pearson et al., 2006; Grigoriev et al., 2006]. Особое внимание исследователи уделяют необычным явлениям и чертам поведения микротрубочек. Среди которых центральное место занимают явления динамической нестабильности. Представляет интерес и проблема разнообразия микротрубочек по длинам в системах *in vivo* и *in vitro*.

Исследования, проведенные оптическими методами, показали, что распределения микротрубочек по длинам в реконструированных системах могут быть как стационарными, так и нестационарными [Carlier et al., 1987; Fygenson et al., 1994; Odde et al., 1995; Caudron et al., 2000]. Выяснилось, что в системах, содержащих тубулин, переходы от стационарных режимов к нестационарным в целом ряде случаев носят отчетливо выраженный _пороговый характер_. Пороговые эффекты были

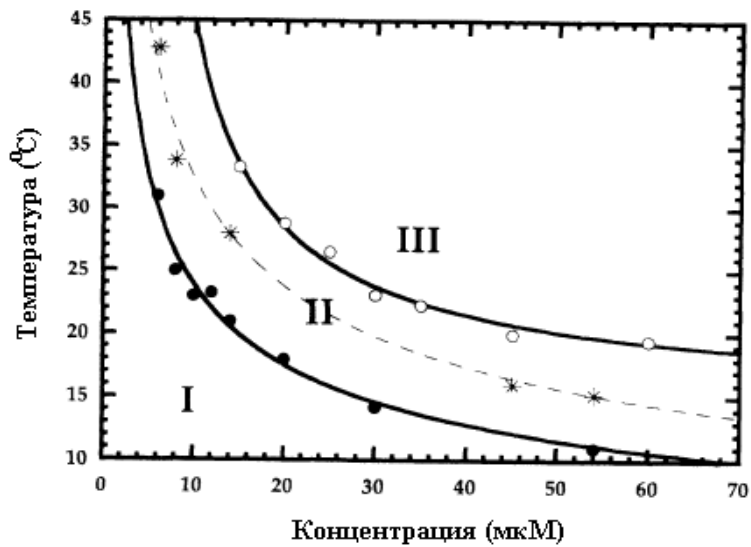

**Рис.1.8**. "Фазовая" диаграмма для системы микротрубочек _in vitro_, согласно работе [Fygenson et al., 1994]. Область **I** соответствует отсутствию микротрубочек, область **II** – их росту с нуклеационных затравок, область **III** – присутствию спонтанной нуклеации. Пунктирная линия разбивает область **II** на две части: в нижней наблюдается ограниченный (bounded) рост, тогда как в верхней – неограниченный (unbounded) (см. §1.2.6 и рис.1.7).

обнаружены по концентрациям всех основных ингридиентов: тубулина [Mitchison and Kirschner, 1984; Carlier et al., 1984], ГТФ [O'Brien et al., 1987; Melki et al., 1988], ионов $Mg^{2+}$[Carlier et al., 1987; O'Brien et al., 1990], а также при варьировании температуры [Fygenson et al., 1994] (см. рис.1.8).

Наличие пороговых эффектов свидетельствует о нелинейном характере механизмов регуляции роста и деполимеризации микротрубочек. Незначительные изменения параметров системы сопровождаются качественной трансформацией ее поведения. В качестве примера, рассмотрим результаты, полученные в работе Фигенсон и соавторов [Fygenson et al., 1994]. В наиболее наглядной форме они представлены на рис. 1.8, представляющем собой, как пишут авторы, «фазовую» диаграмму[4].

---

[4] Строго говоря, термодинамическое понятие фазы подразумевает, что данное состояние является неизменным по времени и однородным по пространству [Gibbs, 1928], поэтому «фазы», представленные на диаграмме [Fygenson, 1994], по нашему мнению, корректней называть неравновесными состояниями.

Стационарные распределения микротрубочек по длинам в системах *in vitro* подробно изучались в работах Карлиер и соавторов [Carlier et al., 1987] и Фигенсон и соавторов [Fygenson et al., 1994]. Характерные распределения представлены на рис. 1.9a и 1.9b.

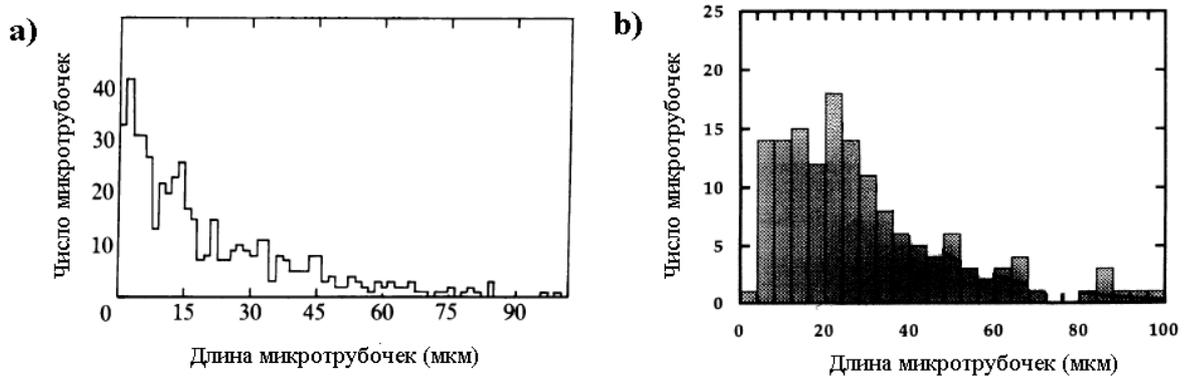

**Рис. 1.9.** Стационарные распределения микротрубочек по длинам в системах *in vitro*, согласно работам **a)** [Carlier et al., 1987] , **b)** [Fygenson et al., 1994].

Примечательно, что в системах *in vivo* спектр наблюдаемых распределений микротрубочек по длинам более широк (см. рис. 1.10a и 1.10b) [Alieva and Vorobjev, 2000; Komarova et al., 2002].

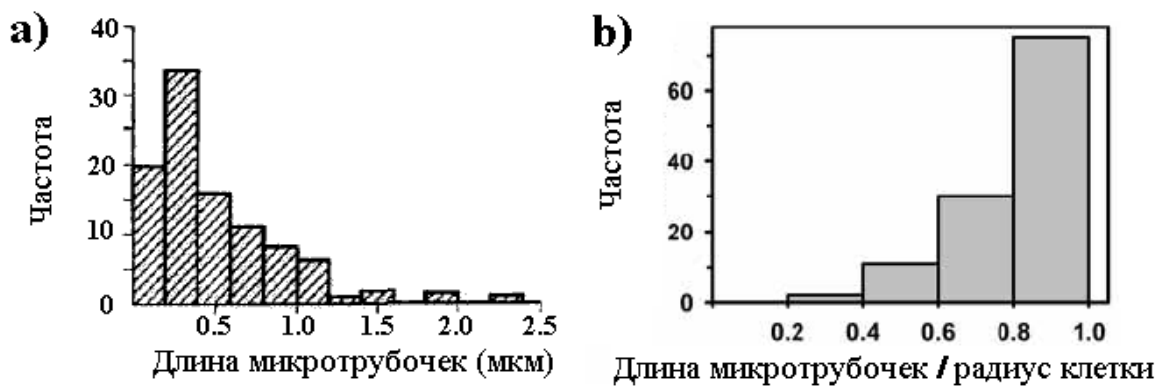

**Рис. 1.10.** Распределение микротрубочек по длинам в клетках, согласно работам **a)** [Alieva and Vorobjev, 2000] и **b)** [Komarova et al., 2002].

Поскольку микротрубочки наблюдаются не во всех фазах клеточного деления, исследователи уверены, что в системах *in vivo* существуют эндогенные факторы и механизмы регуляции [Karsenti and Vernos, 2001; Scholey et al., 2003]. К числу таких эндогенных факторов относятся широко известные регуляторные белки tau [Drechsel et al., 1992], MAP2 [Itoh and Hotani, 1994], MAP4 [Nguyuen et al., 1999], статмин [Niethammer et al., 2004], циклин зависимые киназы [Verde et al., 1992], ГТФазы семейств Rho и Ran [Wilde and Zheng, 1999; Grigoriev et al., 2006]. Некоторые из этих агентов обеспечивают триггерно-пороговое запускание процессов формирования микротрубочек [Karsenti and Vernos, 2001; Potapova et al., 2006].

Помимо эндогенных факторов, известен целый ряд химических препаратов и веществ, способных кардинально менять динамику микротрубочек в системах in vivo и in vitro [Wilson et al., 1999; Honore et al., 2005]. В частности, к числу такого рода веществ относятся широко применяемые в медицинской практике цитостатики. Действие последних сопровождается либо мгновенным «замораживанием» микротрубулинового цитоскелета в клетке [Vasques et al., 1997], либо полной деполимеризацией микротрубочек [Jordan, 2002]. Способность к триггерному, пороговому переключению динамических режимов поведения микротрубочек в статические лежит в основе действия цитостатиков.

Активно изучаются и обратные процессы, в ходе которых стационарные режимы поведения ансамблей микротрубочек трансформируются в нестационарные, например, колебательные [Carlier et al., 1987; Mandelkow et al., 1988; Melki et al., 1988]. Переходы от стационарных режимов к нестационарным могут происходить не только вследствие параметрических неустойчивостей, но и вследствие неустойчивостей динамических. Последние наиболее часто реализуются при запороговой дестабилизации метастабильных стационарных состояний [Tran et al., 1997].

Нестационарные режимы поведения микротрубочек могут быть периодическими [Carlier et al., 1987; Melki et al., 1988] и апериодическими

[Mitchison and Kirschner, 1984; Walker et al., 1988], синхронными по пространству [Carlier et al., 1987] и асинхронными [Walker et al., 1988] (см. рис. 1.11a,b и 1.12). Как несложно видеть из рис. 1.11a, колебания длины микротрубочек сопряжены с колебаниями уровня молекул ГТФ и ГДФ в системе. Исчерпание молекул ГТФ ведет к затуханию колебаний. Если же в систему добавляется киназные комплексы регенерации ГТФ из ГДФ, то частота колебаний и их продолжительность увеличиваются (см. рис. 1.11b).

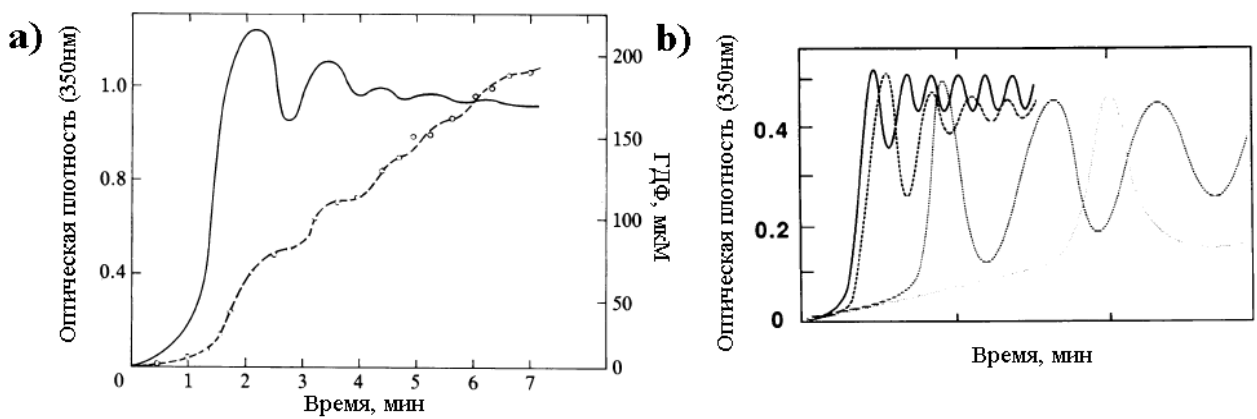

**Рис.1.11.** Периодические режимы поведения ансамбля микротрубочек. **a)** сплошная линия – зависимость от времени оптической плотности, отражающей общее количество полимеризованного тубулина в системе. Пунктирная линия – количество ГДФ в системе. Рисунок приведен из работы [Carlier et al., 1987]. **b)** Изменение колебаний оптической плотности при различных концентрациях пируват-киназной системы, восстанавливающей ГТФ [Melki et al., 1988]. Уменьшение частоты колебаний соответствует уменьшению концентрации киназной системы и ее меньшей восстановительной способности.

Нестационарные апериодические режимы, типа изображенного на рис. 1.12, имеют вид, характерный для релаксационных хаотических колебаний [Анищенко, 2000; Кузнецов, 2001; Данилов, 2001; Трубецков, 2004]. Особенностью этих колебаний являются отчетливо выраженные моменты переключения микротрубочек со стадий роста на деполимеризацию и наоборот. В биологической литературе именно эти «поворотные» моменты в динамике микротрубочек принято называть «катастрофами» и «спасениями».

Под термином «динамическая нестабильность» микротрубочек в биологической литературе понимается круг явлений, связанных с

«катастофами» и «спасениями» [Alberts et al., 2002]. Во избежание недоразумений в дальнейшем под термином «катастрофа» будет подразумеваться динамическая неустойчивость роста микротрубочек, а под термином «спасение» – неустойчивость процесса деполимеризации.

Как мы видели в §§1.2.5, 1.2.6, в аксиоматических подходах к описанию динамики микротрубочек явления переключения режимов непосредственно постулируются. Этого оказывается вполне достаточно для феноменологического описания, пока не ставится вопрос о природе самих механизмов переключения.

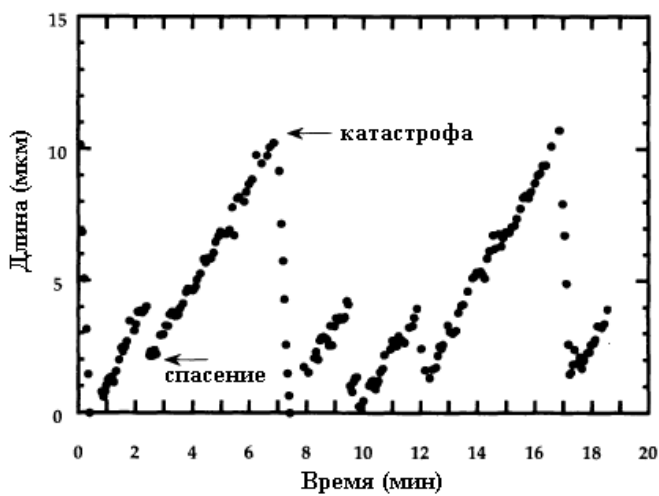

**Рис.1.12.** Изменение длины индивидуальной микротрубочки во времени, согласно работе [Fygenson et al., 1994].

По нашему мнению, природа механизмов, лежащих в основании процессов динамических неустойчивостей, кроется в кинетических особенностях физико-химических взаимодействий, свойственных процессам агрегации/дисагрегации молекул тубулина при участии молекул ГТФ. Нашей целью является вывод условий нестабильности динамики микротрубочки, опирающийся на естественные представления о физико-химических процессах, свойственных росту и деполимеризации микротрубочек. При этом свойственные динамике микротрубочек процессы будут трактоваться, как разновидность агрегационных/дисагрегационных переходов в неравновесной реакционно-диффузионно-преципитационной системе.

## § 1.4 Анализ характерных времен кинетических процессов в динамике микротрубочек

На основании изложенного в предыдущих разделах видно, что дискуссия о природе динамических нестабильностей микротрубочек активно продолжается уже длительное время. Наиболее острыми остаются четыре группы вопросов:

- природа пороговых эффектов роста микротрубочки, проявляющихся при изменении концентрации тубулина в растворе;
- возможное влияние гидролиза ассоциированной с тубулином молекулы ГТФ, происходящего после присоединения тубулина к микротрубочке, на стабильность микротрубочки (в частности, гипотеза ГТФ-крышки);
- обусловленность нестабильностей роста микротрубочек накоплением микродефектов в структуре;
- степень влияния концентрационных градиентов молекул тубулина в растворе на условия роста и деполимеризации микротрубочек.

Взаимная сочетаемость или непротиворечивость разных моделей не может быть в настоящее время признана удовлетворительной. Каждая из вышеперечисленных проблем анализируется в литературе в рамках частных ситуативных моделей. На основании имеющихся в литературе данных и выдвинутых гипотез до сих пор исследователям не удалось сделать окончательный вывод о том, обуславливаются ли динамические нестабильности преимущественно внутренними – структурными или же внешними – кинетическими причинами.

На наш взгляд, вышеперечисленные противоречия вызваны недоучетом того обстоятельства, что в случае с реконструируемыми системами *in vitro*, мы имеем дело с <u>двуфазными</u> системами вблизи границы потери ими устойчивости. Наблюдаемые при этом эффекты, свойственные динамике микротрубочки, являются по сути внешними (частными) проявлениями нестационарного агрегационного перехода, в ходе определенных стадий которого молекулы тубулина конденсируются из раствора в «твердую» фазу, т.е. в микротрубочки.

С физической точки зрения, проблема коренится в том, как соотносятся характерные времена и пространственные масштабы реакционных и диффузионных процессов в рассматриваемой системе. Для выявления условий, при которых определяющую роль в развитии динамической нестабильности микротрубочек играют структурные (внутренние) неустойчивости, а не внешние факторы (обусловленные реакционно-диффузионными процессами в растворе), остановимся подробнее на анализе основных эффектов, связанных с нестабильностью динамики микротрубочки, с кинетической точки зрения.

В таблице 1.2 приведены характерные оценки параметров основных процессов: сорбции (десорбции) тубулина на плюс-концы, восстановления тубулина-ГТФ, диффузии тубулина в растворе и крупномасштабных катастроф, известные из литературы.

Данные, приведенные в таблице 1.2, позволяют в ряде случаев оценить характерные кинетические параметры, характеризующие процессы молекулярной сорбции и десорбции как in vitro, так и in vivo (см. табл.1.3). В качестве оценки характерного времени роста микротрубочки $\tau_{gr}$ в работе используется отношение ее длины к скорости полимеризации. Так как скорость роста микротрубочки зависит от концентрации тубулина в растворе, которая сама по себе может варьироваться в достаточно широких пределах (см. табл. 1.2), диапазон изменений характерного времени роста достаточно широк. Характерным временем деполимеризации микротрубочки – $\tau_{dep}$, является отношение длины микротрубочки к скорости ее укорочения. Известно, что последняя варьируется в широких пределах [Pedigo and Williams, 2002].

**Таблица 1.2.** Количественные оценки параметров системы микротрубочек *in vitro*.

| Величина | Обозначение | Диапазон значений | Литературный источник |
|---|---|---|---|
| Характерная длина микротрубочек in vitro | $L_{vitro}$ | 100 мкм | Fygenson et al., 1994 |
| Характерная длина микротрубочек in vivo | $L_{vivo}$ | 3 мкм | Vorobjev et al., 1999 |
| Радиус микротрубочки | $R_{MT}$ | 12 нм | Nogales et al., 1998 |
| Размер тубулинового димера | $l$ | 8 нм | Nogales et al., 1998 |
| Расстояние между микротрубочками | $s$ | 100÷1000 нм | Dogterom et al., 1995 |
| Коэффициент диффузии тубулина в растворе in vitro | $D_{vitro}$ | 0.06 мкм²/с | Salmon et al., 1984 |
| Коэффициент диффузии тубулина in vivo | $D_{vivo}$ | 0.6 мкм²/с | Salmon et al., 1984 |
| Концентрация тубулина в растворе | $[Tu]$ | 3÷150 мкМ | Carlier et al., 1987; Walker et al., 1988 |
| Скорость роста микротрубочки | $V_{gr}$ | 0.2÷25 мкм/мин | Mitchison and Kirschner, 1984; Walker et al., 1988 |
| Скорость деполимеризации микротрубочки | $V_{dep}$ | 5÷70 мкм/мин | Walker et al., 1988; Pedigo and Williams, 2002 |
| Константа скорости восстановления тубулина-ГТФ in vitro | $k^{res}_{vitro}$ | 100 М⁻¹с⁻¹ | Melki et al., 1988 |
| Константа скорости восстановления тубулина-ГТФ in vivo | $k^{res}_{vivo}$ | 1000 М⁻¹с⁻¹ | Brylawski and Caplow, 1982 |
| Концентрация ГТФ в растворе in vitro | $[ГТФ]_{vitro}$ | 200-2000 мкМ | Melki et al., 1988 |
| Концентрация ГТФ in vivo | $[ГТФ]_{vivo}$ | 20-200 мкМ | Brylawski and Caplow, 1982 |

Характерное время диффузии $\tau_D$, для двумерного случая (именно в такой постановке есть возможность определить коэффициент диффузии из данных микроскопии [Salmon et al., 1984]) определяется, как отношение средней площади, на которой происходят диффузионные процессы, к коэффициенту диффузии тубулина в растворе. При этом в качестве средней площади использовалось произведение длины микротрубочки на характерное расстояние между соседними микротрубочками. Данное расстояние зависит от количества микротрубочек, одновременно растущих от одного нуклеационного центра. В качестве оценки времени реакции восстановления молекул тубулина-ГТФ из тубулина-ГДФ в растворе $\tau_{res}$, нами использовалась величина $1/(k_1[\text{ГТФ}])$, где $k_1$ – константа скорости реакции восстановления, а [ГТФ] – концентрация молекул ГТФ (см. табл.1.3).

**Таблица 1.3.** Диапазоны изменения масштабов характерных времен процессов сорбции/десорбции молекул тубулина

| Процесс, обозначение времени | Рост, $\tau_{gr}$, с | Деполимеризация, $\tau_{dep}$, с | Диффузия, $\tau_D$, с | Время восстановления тубулина, $\tau_{res}$, с |
|---|---|---|---|---|
| Формула оценки | $\tau_{gr} = L/V_{gr}$ | $\tau_{dep} = L/V_{dep}$ | $\tau_D = (L*s)/D$ | $\tau_{res} = 1/(k_1[\text{ГТФ}])$ |
| Значение in vitro | $120 \div 3000$ | $60 \div 1200$ | $160 \div 1600$ | $20 \div 200$ |
| Значение in vivo | $4 \div 100$ | $2 \div 40$ | $5 \div 50$ | $20 \div 200$ |

Сопоставляя характерные масштабы времен из таблицы 1.3, можно выделить представляющие интерес следующие предельные случаи:

А) $\tau_{gr} \gg \tau_{dep}$, $\tau_{gr} \gg \tau_D$, $\tau_{dep} \gg \tau_D$, $\tau_{res} \gg \tau_{gr}$. Этот случай соответствует отсутствию выраженного роста микротрубочек, то есть полному преобладанию процесса деполимеризации над ростом. Так как восстановление тубулина-ГТФ происходит медленно, а перемешивание в растворе – быстро, рост микротрубочек затруднен.

Б) $\tau_{gr} \gg \tau_D$, $\tau_{dep} \gg \tau_D$, $\tau_{gr} \gg \tau_{res}$, $\tau_{dep} \gg \tau_{res}$. В этом предельном случае «жидкую» систему (т.е. раствор) можно рассматривать, как полностью однородную и состоящую преимущественно из тубулина-ГТФ. Динамические нестабильности обусловлены исключительно внутренними причинами, что соответствует приближению, использующемуся в работах [Hill and Chen, 1983; Dogterom and Leibler 1993; Flyvbjerg et al., 1996; Hammele, Zimmermann; 2003]).

В) $\tau_{dep} \ll \tau_D$, $\tau_{res} \ll \tau_D$, $\tau_{gr} \ll \tau_D$. При выполнении данных неравенств динамика микротрубочек определяется исключительно диффузионной частью системы микротрубочки-раствор. Данный предельный случай характерен для работ ([Dogterom et al., 1995; Odde, 1997]).

Г) $\tau_{gr} \gg \tau_D$, $\tau_{dep} \gg \tau_D$, $\tau_{gr} \gg \tau_{res}$, $\tau_{dep} \gg \tau_{res}$, $\tau_{gr} \sim \tau_{dep}$. В этом предельном случае раствор полагается полностью однородным и динамика микротрубочек определяется исключительно процессами сорбции/десорбции на плюс-конце [Vorobjev et al., 1997; Vorobjev et al., 1999; Воробьев и др., 2000; Komarova et al., 2002]

Д) $\tau_{res} \gg \tau_{gr}$, $\tau_{res} \gg \tau_{dep}$, $\tau_{res} \gg \tau_D$. В данном предельном случае динамика системы микротрубочки-раствор будет определяться исключительно процессами энергообмена в системе [Катруха и Гурия, 2006].

Таким образом, кинетический подход к анализу данных позволил сгруппировать и расклассифицировать опубликованные в печати работы. Важным представляется и тот факт, что «кинетический взгляд» позволил выявить лакуны в существующих подходах. В рамках настоящей диссертационной работы ставилась цель проанализировать наиболее подробно ситуации, которые следует отнести к предельному случаю «Д». Предпринимается попытка построения физико-математического подхода, позволяющего трактовать проблемы нестабильностей в динамике микротрубочек единообразно, с позиций теории неравновесных систем. При этом во второй главе основное внимание уделяется анализу внутренних структурных факторов динамических нестабильностей. В последующих главах внимание концентрируется на эффектах, связанных с процессами сорбции (десорбции) тубулина из раствора.

# Глава 2. Стохастические аспекты структурной динамики тубулиновых волокон

В основе развиваемого главе 2 подхода, рассматривающего внутренние причины нестабильности микротрубочек, лежит допущение о ключевой роли дефектов структуры в потере устойчивости роста микротрубочек. В первой части главы (§§ 2.1 и 2.2) представлена математическая модель, в которой используются методы теории клеточных автоматов для описания динамики отдельных тубулиновых димеров. Оценки величин вероятностей одночастичных событий приближенно определяются из величин термодинамических характеристик системы. Для рассматриваемой системы постулируется ряд интегральных критических условий, приводящих к коллективным переходам. Проводится сопоставление экспериментов с результатами компьютерного моделирования. В § 2.3 представлена упрощенная феноменологическая модель, в рамках которой построены точные аналитические решения.

## § 2.1. Описание динамики димеров тубулина и вакансий

Модель рассматривает рост отдельной микротрубочке в растворе, содержащем димеры тубулина и необходимые для полимеризации компоненты (буфер, ГТФ). В отношении системы, в которой рассматривается рост микротрубочки, делаются следующие предположения. Процесс нуклеации (то есть зарождения) нити проходит за время, которое пренебрежимо мало по сравнению с характерными временами остальных процессов, свойственных динамике микротрубочки [Moritz et al., 1995; Jobs et al., 2003]. Быстрая нуклеация в реальной системе соответствует тому, что рост микротрубочки инициируется на центросомах, которые помещаются в систему *in vitro*. Динамика закрепленных минус-концов далее рассматриваться не будет.

Полагается, что концентрации веществ, оказывающих влияние на константы скоростей протекающих реакций (ионы $Ca^{2+}$, $Mg^{2+}$, ГТФ)

поддерживаются на постоянном уровне и не изменяются во времени и пространстве, а какие-либо белки MAP (microtubule associated proteins), оказывающие влияние на динамику микротрубочки, в растворе отсутствуют.

В рамках модели тубулиновое волокно представляет собой упорядоченный в пространстве ансамбль из N частиц, для каждой из которых определены соответствующие правила поведения в каждый момент времени. Полагается, что микротрубочка состоит из тринадцати прямых, параллельных протофиламентов, уложенных с шагом спирали, равным трем. Структура микротрубочки полагается однородной по всей длине. Таким образом, положение каждой частицы в пространстве определяется двумя целыми числами. Первое определяет номер протофиламента, в котором она находится. Второе соответствует ее порядковому номеру в протофиламенте, причем нумерация начинается от закрепленного на центросоме конца. Локус структуры микротрубочки может быть занят частицами двух типов: тубулиновыми димерами и дефектами (вакансии или дырки, т.е. вакантные локусы в структуре микротрубочки, образующиеся за счет выхода димеров из стенок в раствор).

Для всех возможных событий (процессов), происходящих с каждой отдельной частицей (переход из одного состояния в другое, рождение новой частицы, ее смещение и т.д.) определена вероятность в единицу времени, которая при усреднении по всем частицам ансамбля есть ни что иное, как эффективная константа скорости соответствующей реакции. Например, для события сорбции одного димера из раствора в торец микротрубочки (рождение новой частицы), вероятность пропорциональна концентрации димеров в растворе, площади, доступной для связывания на конце микротрубочки и некоторому стерическому фактору, вообще говоря, неодинаковому для разных частиц на краю. Точно также образование дефектов, их движение вдоль волокна и исчезновение происходят с некоторыми вероятностями, которые в конкретный момент зависят от текущего окружения частицы. Данный подход к описанию многочастичного ансамбля формально имеет много общего с концепцией клеточных автоматов, широко используемой для моделирования разного рода неравновесных физико-химических и биологических систем [Ermentrout et al., 1993; Seybold et al., 1998; Wolfram, 2002].

Полагается, что в определенный момент времени в системе существует N частиц и M возможных процессов. Если для каждого k-го процесса ($k \in [1..M]$) определить вероятность $p_{i,k}$ того, что он произойдет в единицу времени над i-ой частицей ($i \in [1...N]$), мы получим полный набор правил и законов поведения частиц. В соответствии с этим, в каждый момент времени система в целом имеет определенное состояние, зависящее от ее «истории жизни». Множество всех возможных конфигураций, или состояний системы образует дискретное пространство элементарных событий $\Omega(\omega_j)$. Так как в единицу времени с частицей может произойти с некоторой вероятностью только один какой-то процесс, то в этом пространстве можно выделить классы смежности элементов. Поясним на примере. Пусть у нас есть только один процесс – сорбция, нить толщиной в димер и состояние, в котором микротрубочка состоит из трех молекул тубулина. Обозначим этот элемент A. Состояния с четырьмя и пятью димерами обозначим соответственно B и C. Элемент B является смежным элементу A, потому что есть ненулевая вероятность P(A->B) перехода между ними в единицу времени, тогда как C – нет, ввиду P(A->C) = 0. Причем P(A->B) = $p_{сорб}$. Если теперь принять во внимание, что имеется M процессов и N частиц, то у каждого элемента $\omega_j$ пространства состояний системы будет существовать свой класс смежности $E(\omega_j) = \{ \omega_f : P(\omega_j \to \omega_f) \neq 0\}$. Где по определению $P(\omega_j \to \omega_f) = \sum_{i=1}^{N}\sum_{k=1}^{M} p_{i,k}$, в суммировании берутся только те слагаемые, которые означают процесс k над частицей i при переходе из состояния $\omega_j$ в $\omega_f$. Должно выполняться условие нормировки $\sum_f P(\omega_j \to \omega_f) = 1$.

В итоге, в таком представлении реальный рост нити выглядит, как «блуждание» или переход между элементами пространства состояний $\Omega(\omega_j)$, с учетом классов смежности для каждого из них. Обратимся теперь к подробному рассмотрению каждого процесса и проведем его детальное описание.

*Сорбция димеров из раствора.* Полагается, что в растворе существует нуклеационная «затравка» - стабильный, неразрушаемый сегмент нити, с которого начинается рост. На «затравку» сорбируются молекулы тубулина из раствора. Это первый процесс, и, как уже отмечалось выше, эффективная константа скорости или, иными словами, вероятность присоединения молекулы в секунду пропорциональна концентрации димеров в растворе, площади конца микротрубочки и некоторому стерическому фактору. Для исключения влияния внешних факторов полагается, что в растворе поддерживается постоянная концентрация димеров, а рост микротрубочки существенно ее не изменяет. Площадь активной сорбирующей поверхности трубки варьируется во времени благодаря «выростам», отдельным более длинным протофиламентам, которые образуются за счет флуктуаций процесса сорбции. Поэтому стерический фактор, стоящий в предэкспоненте константы скорости сорбции (при записи ее в виде классического уравнения Аррениуса [Eyring, 1935; Мелвин-Хьюз Э.-А., 1962]), вообще говоря, является функцией положения соседних димеров на торце (см. рис.2.1, **A, B**).

Полагается, что энергия присоединения для различных мест на торце тубулинового волокна не является одинаковой. Это объясняется из следующих соображений. Молекулы тубулина в трубке стабилизированы боковыми взаимодействиями друг с другом, иначе микротрубочка просто-напросто развалилась бы на отдельные полоски (протофиламенты) [VanBuren et al., 2002]. Исходя из принципа минимума энергии, величина порога присоединения меньше у локуса, для которого площадь соприкосновения с боковыми соседями больше, поэтому отдельные «выросты» будут сглаживаться. В среднем же край растущей микротрубочки будет двигаться с постоянной средней скоростью, которая и регистрируется в экспериментах in vitro.

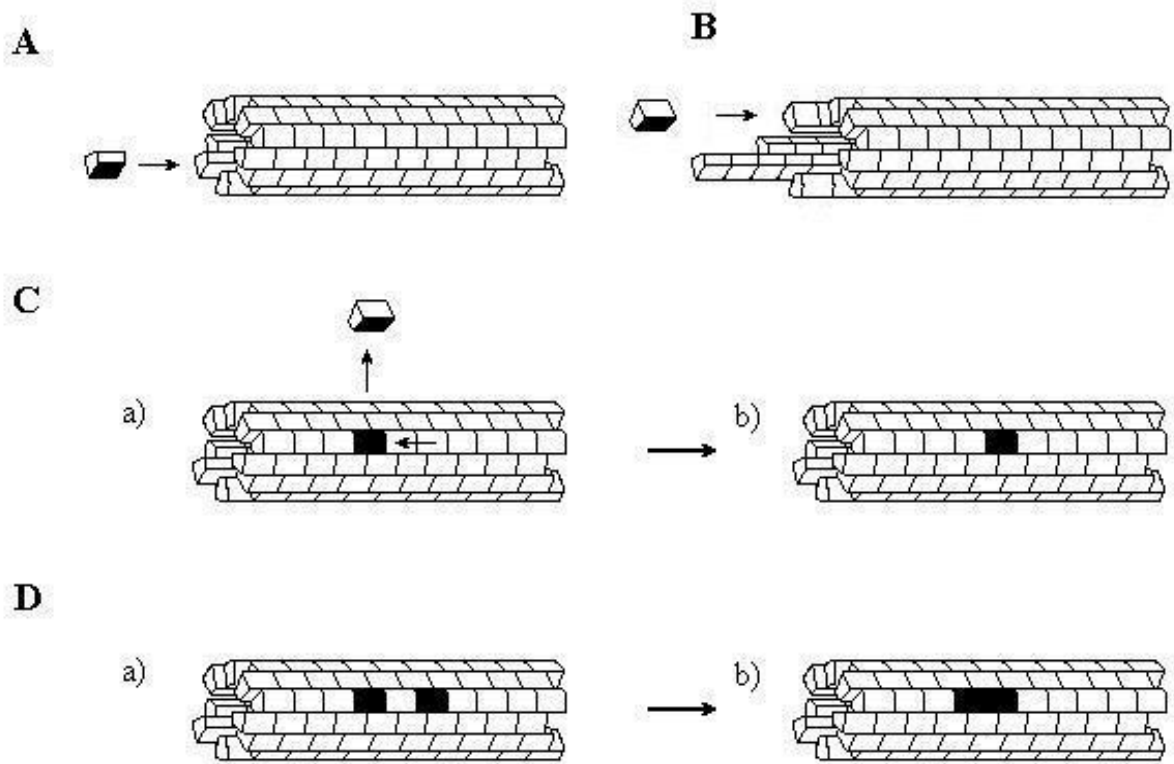

**Рис.2.1.** Основные типы взаимодействий в микротрубочке. **A, B** – сорбция димеров тубулина из раствора. В случае **A** величина стерического фактора и вероятности сорбции меньше, т.к. меньше соответствующая площадь сорбирующей поверхности. **C, D** – образование и продольное движение дырок.

Одновременно с сорбцией проходит обратный процесс – десорбция димеров тубулина. В случае, если отсоединение происходит непосредственно с сорбирующей поверхности (плюс-конца, свободного края микротрубочки), димер возвращается в раствор. Если же его успевает «прикрыть» следующий слой, и только потом происходит десорбция, то на его месте возникает свободный локус, то есть вакансия. Таким образом, происходит вток дырок с края в объем растущей микротрубочки.

*Образование и движение вакансий.* Ключевым процессом в описываемом подходе является образование дырок в стенках и на краю микротрубочки, т.е возвращение димеров в раствор. Возможность их существования объясняется следующими предпосылками. Своей упорядоченной периодичной структурой и наличием оси симметрии волокна во многом напоминают кристаллы, и такая аналогия не является случайной. Количественную оценку энергии, затраченной на присоединение тубулинового димера в стенки микротрубочки можно произвести исходя из величины энергии гидролиза ГТФ, что дает величину порядка 13 ккал/моль. Очевидно, что поверхностная энергия полимера также является периодической функцией длины микротрубочки с периодом, равным размеру молекулы димера [VanBuren et al., 2002]. Т.к. образование вакансий (дефектов) возможно в периодической кристаллической решетке, где величины энергии связи атома порядка 40 ккал/моль [Френкель, 1945], естественно предположить, что вероятность выхода тубулина из стенки микротрубочки в раствор также не равна нулю.

Полагается, что полимеризовавшаяся фаза является устойчивой и для внесения какого-либо искажения в ее структуру необходимо затратить энергию. Поэтому предполагается, что существует потенциальный барьер, препятствующий выходу тубулина из стенок микротрубочки в раствор. Его величина, и, следовательно, вероятность выхода зависит от наличия соседей, как продольных, так и боковых.

Обозначив энергию продольной $\mathbf{E}_\text{П}$, а боковой – $\mathbf{E}_\text{Б}$, то, следуя Гиббсу, вероятность покинуть стенку микротрубочки у молекулы тубулина с шестью соседями (2 продольных и 4 боковых в смежных протофиламентах) будет пропорциональна:

$$\omega_\text{esc} \sim \exp(-\frac{2\mathbf{E}_\text{П} + 4\mathbf{E}_\text{Б}}{\mathbf{T}}) \qquad (2.1)$$

где $\omega_\text{esc}$ – вероятность выхода, $\mathbf{T}$ – температура системы. Образование дефекта увеличивает вероятность возникновения рядом с ним еще одной вакансии, т.к. величина работы выхода для соседней молекулы тубулина становится меньше и упорядоченная структура микротрубочек становится более «разрыхленной». Таким образом, уместно говорить об автокаталитическом механизме образования дефектов.

Появление вакансий de novo конкурирует со вставкой димеров из раствора внутрь стенок микротрубочки. Присоединение молекул свободного тубулина из раствора может осуществляться по механизму, аналогичному сорбции димеров в торец. Различие состоит в величине стерического фактора, который пропорционален телесному углу. Для одного дефекта он будет незначительно отличаться от нуля. Но если вакансии расположены рядом или образуют своеобразный агрегат, то присоединение становится достаточно активным и необходимо учитывать его вклад в динамику дефектов.

Микротрубочки, в отличие от гибких актиновых волокон, довольно жесткие структуры. Из клеточных органелл – наиболее жесткие, их персистентная длина порядка 1 мм [Janson and Dogterom, 2004], что значительно превышает характерный размер отдельного тубулинового димера – 8 нм. В частности микротрубочки проявляют такое свойство макротел, как упругость. Они используются клеткой везде, где необходим крепкий и прочный каркас. Сам принцип трубки (известный всем создателям опорных конструкций) состоит в максимуме жесткости при минимуме исходного материала. Поэтому по аналогии с твердыми телами можно предположить наличие в них дальнего порядка. Исходя из этого полагается, что вакансия способна перемещаться вдоль волокна, подобно движению дефектов в кристалле. В некоторой степени этот процесс напоминает продольную диффузию. Смещение вакансии на одну позицию вправо соответствует «перескоку» правой молекулы влево, на свободное место. Поперечный перескок димеров, из-за спиральной структуры микротрубочек, маловероятен и рассматриваться в дальнейшем не будет. Вероятность передвижения дефекта в разные стороны вообще говоря, различна и зависит от положения других вакансий.

Свойство макроскопической упругости выражается в наличии дальних корреляций. Поэтому полагается, что движение вакансий не хаотично, а коррелированно, подобно образованию трещин в стекле. Другими словами, каждая из вакансий «чувствует» распределение соседних вакансий на расстояниях, много больших ее собственного размера. Рассмотрим подробней, какие процессы приводят к таким взаимодействиям.

Если рассматривать энергию дефектов на молекулярном уровне, то комбинация из двух соседних, смежных вакансий имеет меньшую энергию, чем два дефекта, разделенные молекулами тубулина. Это объясняется тем, что вместо восьми свободных граней (продольных и поперечных) у разделенного состояния, у объединившихся вакансий их шесть. Поэтому дефект будет стремиться «слиться» с другим, или, иными словами, притягиваться к другим дефектам (см. рис. 2.1, **D** a,b), причем как расположенным в данном протофиламенте, так и в соседних с ним. Вследствие этого перескоки вакансии влево и вправо не являются равновероятными событиями, их вероятности будут зависеть от положения других, смежных дефектов.

В свою очередь, локальное увеличение концентрации дефектов приводит к уменьшению энтропии системы, что противоречит второму началу термодинамики. Согласно закону роста энтропии более предпочтительным является равномерное распределение дефектов. Для того, чтобы учесть оба этих факта, запишем свободную энергию системы частиц:

$$\mathbf{F} = \mathbf{E} - \mathbf{TS} \qquad (2.2)$$

Перескок вакансии – это переход всей системы из определенного начального состояния в состояние, где вакансия смещена на один локус. При условии, что никаких других событий не произошло, вероятность перескока будет пропорциональна:

$$\omega_{mov} \sim \exp\left(-\frac{\Delta \mathbf{F}}{\mathbf{T}}\right) = \exp\left(-\frac{\Delta \mathbf{E}}{\mathbf{T}} + \Delta \mathbf{S}\right) \qquad (2.3)$$

где $\Delta \mathbf{F}$, $\Delta \mathbf{E}$ и $\Delta \mathbf{S}$ – изменения соответственно свободной энергии, внутренней энергии и энтропии, а $\omega_{mov}$ – вероятность перескока. Из-за отсутствия точных экспериментальных данных трудно дать количественную оценку вклада, осуществляемого за счет внутренней энергии. Ввиду малости энергии перескока по сравнению с тепловой, она предположительно будет давать небольшой вклад, причем лишь на сравнительно малых расстояниях. Изменение же энтропии системы можно оценить из следующих соображений.

Мысленно разделим микротрубочку на совокупность составляющих ее подсистем, в каждой из которых находится M частиц (дефектов и димеров). Допустим, i-тая подсистема содержит $m_i$ дефектов и, соответственно, $n_i = M - m_i$ молекул тубулина. Ее энтропия будет равна (с точностью до аддитивной постоянной)

$$S_i = \ln\left(C_M^{m_i}\right) = \ln\left(\frac{M!}{m_i!(M-m_i)!}\right) \qquad (2.4)$$

т.е. является функцией $m_i$. Аналогичную формулу можно написать для j-ой, соседней с первой подсистемы. Предположим, что вакансии могут свободно пересекать лишь проходящую между ними границу. Тогда энтропия двух подсистем в силу своей аддитивности дается выражением

$$S_{i,j} = S_i + S_j = S_{i,j}(m_i, m_j) \qquad (2.5)$$

и будет максимальной, если $m_i = m_j$, т.е. наиболее вероятным будет состояние с однородным распределением вакансий. Таким образом, можно оценить энтропийный вклад в вероятность перескока отдельной вакансии, ведь он не будет затрагивать остальные подсистемы. Перенося данные рассуждения на всю систему в целом, и окончательно подводя итог рассуждениям, получим, что движение вакансий действительно самокоррелированно и "управляется" с одной стороны энергией, которая вынуждает их притягиваться друг к другу, минимизируя «открытую» площадь, а с другой – энтропией, вызывающей взаимное отталкивание.

*Критические условия катастрофы.* Обратимся к кооперативным процессам, которые вызывают деполимеризацию микротрубочки, а точнее – причинам потери устойчивости состояния монотонного роста микротрубочки. Наличие искажений-дыр в упорядоченной структуре микротрубочки приводит к неизбежному возникновению напряжений и выходу системы из состояния энергетического минимума. Средняя скорость роста волокна велика по сравнению с частотой появления вакансий, но образование хотя бы одной из них приводит к "разрыхлению" окружающей ее зоны и увеличению вероятности появления других дефектов в окрестности. Кроме того, вакансии предположительно будут конденсироваться за счет перемещений, уменьшая свою энергию и формируя "дырочные рои", а это также способно привести к «испарению» микротрубочки – ее деполимеризации.

Согласно принимаемым в данной работе предположениям динамические нестабильности могут происходить вследствие кластеризации структурных дефектов в микротрубочках. При этом такого рода критические кластеры могут возникать как в окрестности плюс-конца, так и в теле микротрубочки. Размер критического кластера $R_C$ определяется кинетическими параметрами, характеризующими процессы рождения, гибели и миграции вакансий в микротрубочке.

Переходы микротрубочек от роста к деполимеризации могут происходить по двум основным механизмам. Согласно первому механизму, если в каком-то месте микротрубочки вакансии образуют перколяционный кластер, такой, что его характерный размер $R_C$ сопоставим с диаметром микротрубочки, в системе происходит «поперечная» перколяция. Это приведет к отсоединению от микротрубочки целого ее фрагмента, расположенного между дырочным кластером и плюс-концом. Тем самым, в таких ситуациях мы имеем дело с фрагментацией микротрубочек, как формой ее крупноблочной деполимеризации.

Согласно второму механизму, достигнув в некоторый момент критического среднего «разрыхления»[1], вся микротрубочка приобретет способность пороговым образом утрачивать устойчивость, как целое. При дальнейшем увеличении степени разрыхления с какого-то момента микротрубочка окажется в абсолютно неустойчивом состоянии. Для запуска стремительной деполимеризации достаточно любой сколь угодно малой флуктуации. Характерный размер перколяционного кластера $R_C$ при этом сопоставим с длиной микротрубочки. Результатом потери устойчивости при этом является «объемное вскипание» (или «расплетание» филаментов), которое будет продолжаться до тех пор, пока отношение числа дефектов к общему числу тубулиновых димеров в микротрубочке не уменьшится до подпорогового уровня. Резкое уменьшение длины микротрубочки воспринимается, как проявление динамической нестабильности процесса роста. Таким образом, необходимым и достаточным условием катастрофы по второму сценарию является превышение средним разрыхлением микротрубочки определенного критического уровня.

---

[1] Под средним «разрыхлением» структуры мы здесь и далее понимаем отношение общего числа вакансий к числу тубулиновых димеров в микротрубочке.

Проходит ли динамическая нестабильность микротрубочек по тому или иному механизму, зависит от значения кинетических параметров, определяющих вероятности основных элементарных случайных событий, таких, как рождение вакансий, перемещение вакансий по структуре микротрубочки и т.д.

## § 2.2. Вычислительные эксперименты

Для выявления качественных характеристик и зависимостей модели использовалась компьютерная программы, моделирующая процесс роста микротрубочек (детальное описание алгоритма и величины параметров модели см. Приложение I). Использование компьютера в данном случае обусловлено значительным количеством частиц, что требует запоминания больших объемов информации и одновременно – быстрых ресурсоемких вычислений для определения вероятностей перехода. Простейшая математическая модель микротрубочки, реализованная на ЭВМ, реализуется следующим образом. Для нити толщиной в один тубулиновый димер задается одномерный массив. Значением элемента массива является число, несущее информацию о нахождении в данном локусе либо молекулы тубулина, либо дефекта. За один шаг виртуального времени с каждой частицей с определенной вероятностью может произойти любой из описанных выше процессов. Определение того, какой именно из них произойдет, происходит путем опроса генератора случайных чисел, в который с заданными весами входят вероятности различных событий (сорбции, образования дырки и т. д.).

Одновременно работало два алгоритма обнаружения момента катастрофы. Первый, локальный, отслеживал отношение числа дырок к общему числу локусов, находящихся в пяти соседних поперечных слоях. Если данное отношение превышало указанное критическое значение, то кусок микротрубочки "обрезался" от места скопления дефектов до свободного конца. Полагалось, что в дальнейшем этот кусок полностью деполимеризуется в растворе. Второй алгоритм обнаружения катастрофы (интегральный, соответствующий второму сценарию) на каждом шаге по времени определял степень разрыхления всей микротрубочки, т.е. отношение общего числа дефектов к массе полимеризованного тубулина. Если эта величина превышала значение 5 – 10% (см. Приложение I), происходила мгновенная деполимеризация вплоть до сегмента, содержащего наибольшее число дефектов.

*Распределение деполимеризующихся фрагментов.* На основании выдвинутых выше положений было написано несколько программ на языке C++ (компилятор и оболочка Borland C++ ver.3.1, подробное описание программ см. Приложение I). Изменение параметров модели проводились в пределах указанных в Приложении I, табл.I.1, диапазонов. Качественные наблюдения показали, что при преобладании энергетического члена, в стохастическом поведении дырок явно прослеживалась тенденция образования дырочных агрегатов или «роев», что приводило к первому сценарию катастрофы. Если же с большим весом входил энтропийный член, дырки распределялись более равномерно, и преимущественно преобладало развитие катастрофы по второму сценарию. В зависимости от конкретного сценария развития системы приводился в действие один из алгоритмов обрезания.

Следующей поставленной задачей было определение характера зависимости частоты катастроф от числа тубулиновых димеров во фрагментах,

отсоединяющихся от микротрубочки при фиксированной скорости роста МТ (или концентрации тубулина). С этой целью в программу была добавлена часть, подсчитывающая число димеров до катастрофы и размер деполимеризующегося фрагмента. Типичный вид зависимости представлен на рис.2.2.

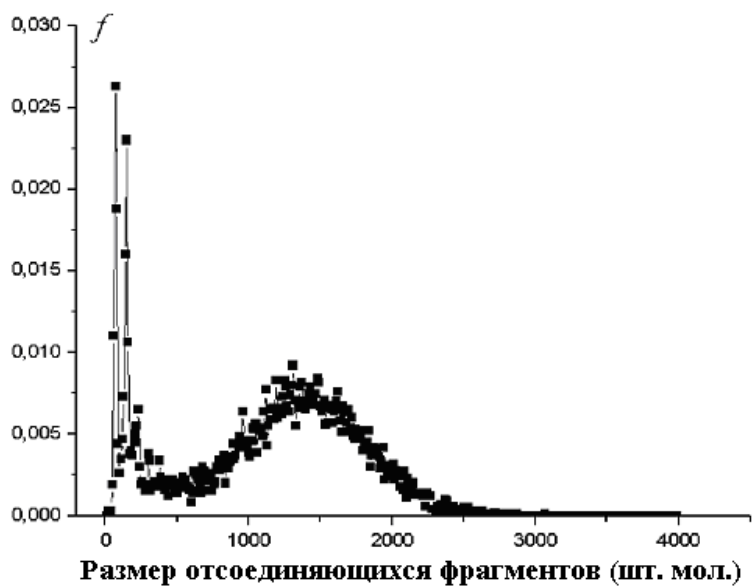

**Рис.2.2.** Характерная зависимость частоты катастроф от величины деполимеризующегося фрагмента микротрубочки. (Общее число катастроф – 10000).

Заметно наличие острого пика в районе фрагментов небольшой длины. Неожиданным оказался тот факт, что вклад в него дают оба сценария. Локальный сценарий катастрофы, как уже было указано выше (см. § 2.1), характеризуется деполимеризацией небольших фрагментов вследствие более вероятного образования вакансий у свободного конца микротрубочки. Интегральный же сценарий вносит свой вклад в левый пик распределения следующим образом. Когда объемная деполимеризация микротрубочки проходит не полностью (не достигая нуклеационной затравки), оставшаяся часть все еще содержит определенное число вакансий. Они быстро конденсируются, за счет уменьшения энергии отсоединения появляются новые вакансии и происходит небольшое укорачивание, как правило, следующее за деполимеризацией крупного фрагмента. Более широкий второй пик отражает общую закономерность, вызванную вторым сценарием. Находящийся справа пик с хорошей точностью (учитывая сдвиг по оси **M**) можно аппроксимировать функцией:

$$f = A^* \exp\left\{-(\mathbf{M} - \mathbf{M}^*)^2 / 2\sigma^2\right\} \qquad (2.6)$$

где $f$ – частота катастроф, при которых происходит отделение фрагмента, содержащего **M** тубулиновых димеров.

Отметим, что вид аппроксимационной зависимости (2.6) сохранялся при варьировании микроскопических параметров рассматриваемой модели в достаточно широких пределах (см. табл.1 приложение I).

На приведенном графике заметно, что хотя величина $\sigma$ в несколько раз меньше среднего значения количества молекул в деполимеризующемся фрагменте, ее значение существенно. Вследствие этого, не обладая достаточно длинным статистическим рядом измерений (который легко получить в компьютерном эксперименте, но непросто в реальном), о точном значении величины $\mathbf{M}^*$ судить сложно. Следовательно, наблюдая за малочисленным ансамблем микротрубочек в экспериментах *in vitro*, нельзя предсказать точный момент катастрофы, т.к. это случайное событие. Однако для большого ансамбля или для продолжительных наблюдений имеет смысл говорить о средней частоте наблюдаемых катастроф. Сложность в реальных экспериментах заключается в том, что процедура обработки каждого отдельного события трудоемка и плохо автоматизируема, поэтому статистический ряд достаточно короткий (порядка 70-100 событий, сравнительно с 10000 в вычислительном эксперименте) и величины погрешностей значительны [Walker et al., 1988; Caudron et al., 2000].

*Зависимость частоты катастроф от концентрации тубулина.* Одним из ключевых опытов, выявляющих особенности динамики микротрубочек, является эксперимент по определению спектра катастроф *in vitro*. Под спектром катастроф имеется ввиду зависимость средней частоты катастроф от концентрации свободного тубулина в растворе (либо скорости роста микротрубочек) [Walker et al., 1988]. Соотношение между экспериментальными и вычислительными характеристиками катастроф устанавливалось следующим образом. Естественно принимать за характерный среднее количество молекул тубулина в деполимеризующемся фрагменте положение пика $\mathbf{M}^*$ в формуле распределения (2.6). Зависимость характерного количества молекул тубулина в деполимеризующемся фрагменте от средней частоты катастроф может быть вычислено согласно формуле:

$$<\nu> = \frac{13}{l}\frac{<V_{gr}>}{\mathbf{M}^*} \qquad (2.7)$$

где $<\nu>$ [мин$^{-1}$] - средняя частота катастроф, $<V_{gr}>$ [мкм/мин] - усредненная скорость роста микротрубочки и $l = 8*10^{-3}$ мкм – характерный размер одной молекулы димера.

На рис. 2.3 приведено сопоставление экспериментально измеренной зависимости частоты катастроф от скорости роста микротрубочек [Walker et al., 1988] и результатов компьютерных расчетов. В качестве «ошибки измерения» в вычислительных экспериментах на рисунке отложена величина *σ*, характеризующая среднеквадратичное отклонение в распределении (2.6). Для каждой точки было вычислено 10 000 событий. Из рисунка 2.3 можно сделать вывод, что компьютерная реализация модели демонстрирует наблюдаемое уменьшение частоты катастроф при увеличении концентрации тубулина в растворе, и в пределах погрешностей результаты проведенных вычислений совпадают с данными эксперимента. Детали сопоставления параметров расчетов и реальных значений концентраций и скоростей роста приведены в Приложении I.

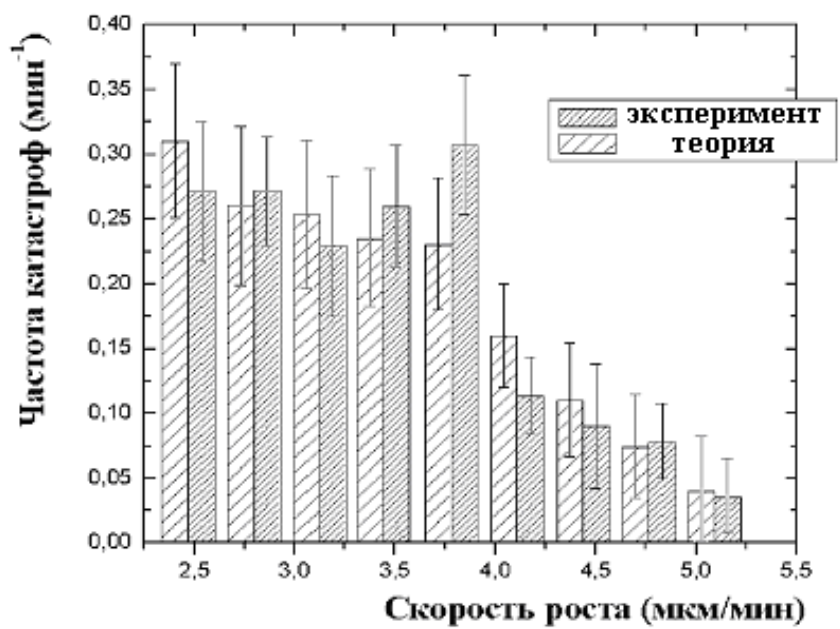

**Рис.2.3.** Сравнение зависимости частоты катастроф от скорости роста микротрубочки в эксперименте [Walker et. al, 1988] и результатов компьютерного моделирования.

При значениях скорости роста меньше ~ 2 мкм/мин (соответственно пороговая концентрация тубулина составляет около 5 мкМ), правый пик (рис.2.2) смещался в сторону уменьшения размера деполимеризущихся фрагментов и накладывался на высокий левый, поэтому определение величины **M**$^*$ становилось невозможным. Следует заметить, что отсутствие выраженного макроскопического роста при концентрации ниже пороговой действительно наблюдается в реальном эксперименте, причем ее величина совпадает с теоретической оценкой – 5 мкМ [Walker et al., 1988].

*Сопоставление с данными экспериментов по вымыванию димеров.* Развитый в данной работе подход позволяет сопоставить теоретические результаты с данными экспериментов по вымыванию димеров [Walker et al., 1991]. Вкратце их постановка сводится к следующему.

Рост микротрубочек происходил *in vitro*, в проточной кювете специальной конструкции [Walker et al., 1991]. Через определенное время, выбиравшееся заведомо большим характерного времени катастрофы при данной концентрации, из подводимого шланга в кювету под большим напором подавался буфер, не содержащий димеров. Поток вымывал из объема молекулы свободного тубулина. Большинство полимеризовавшихся микротрубочек оставалось внутри, т.к. центры нуклеации (центросомы), были специальным образом прикреплены ко дну кюветы. Измерения меченого флуоресцентными метками тубулина в растворе показали, что за время порядка 4-7 секунд концентрация свободных димеров падает ниже отметки 1 мкМ (что значительно меньше пороговой концентрации роста) даже при значительных начальных значениях. В работе [Walker et al., 1991] авторы измеряли величину времени от прекращения подачи буфера до момента катастрофы в зависимости от начальной концентрации димеров (рис.2.4) и длины микротрубочки.

В рамках излагаемого подхода не сложно объяснить природу существования задержки перед катастрофой (укорачиванием), следующей за мгновенным вымыванием свободных димеров из раствора [Walker et al., 1991; Voter et al., 1991]. Эксперименты показали, что продолжительность во времени данной задержки не зависит от концентрации тубулина, при которой происходила полимеризация. Действительно, если окружающая концентрация тубулина падает, это приводит к уменьшению скорости полимеризации, и рост микротрубочки останавливается. Однако дефекты продолжают появляться за счет выхода димеров из стенок и, более того, вероятность «вставок» уменьшается, т.к. свободные димеры вымыты. Вследствие этого происходит постепенное накопление дефектов в объеме и деполимеризация всей микротрубочки. Согласно интегральному условию катастрофы деполимеризация произойдет, когда отношение количества дефектов к общему число полимеризовавшихся молекул тубулина (т.е. фактически – к длине микротрубочки) превысит некоторое критическое значение. Поэтому величина временной задержки (времени от момента вымывания димеров до катастрофы) должна зависеть не только от скорости полимеризации (или концентрации тубулина), но и от длины микротрубочки.

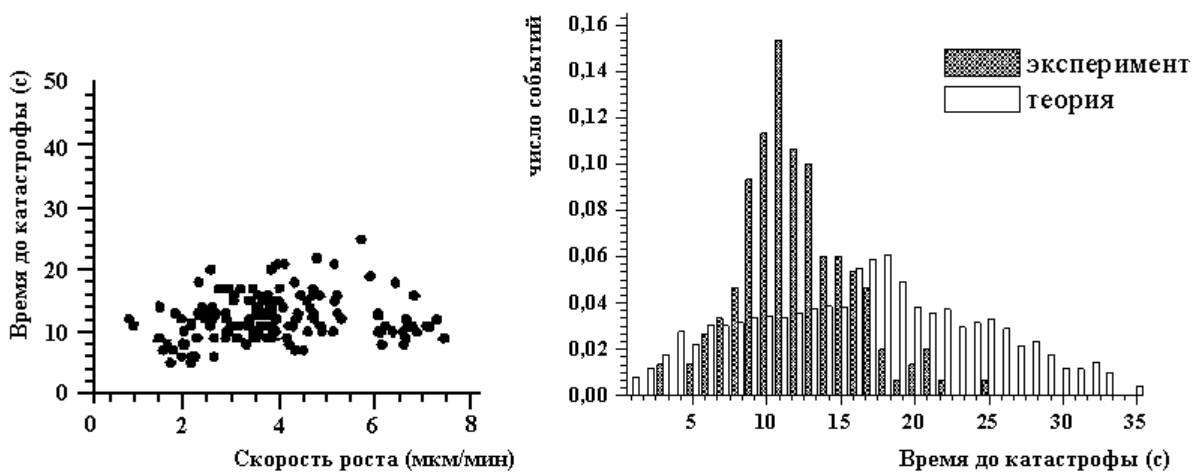

**Рис.2.4.** Зависимость времени после прекращения подачи буфера до момента катастрофы от скорости роста микротрубочки. Левый график: приведенная в работе [Walker et al. 1991] зависимость. Правый график: сопоставление результатов эксперимента и расчетов.

Трудность сопоставления данных заключается в том, что в публикации [Walker et al., 1991] зависимости времени до катастрофы от длины и скорости роста микротрубочек отложены на двух независимых графиках и не представляется возможным построить их взаимное соответствие. Для сопоставления реального эксперимента с расчетами, компьютерная эмуляция[2] проводилась следующим образом. Полагалось, что в момент поступления в кювету буфера, из-за несинхронной нуклеации распределение микротрубочек по длинам случайно. При фиксированной скорости роста микротрубочек (концентрации димеров), в случайные моменты «выключался» процесс сорбции и фиксировалось время до наступления катастрофы. Результаты показали, что его значение варьировалось в пределах от нуля до 35 секунд, что, вообще говоря, превышает экспериментальные значения (рис.2.4). Высота пика и вид распределения для расчетов и экспериментальных данных также отличаются между собой. На наш взгляд, это отчасти может быть связано с тем, что экспериментальный статистический ряд (100 точек) содержит недостаточное количество данных для корректного анализа.

---

[2] Под термином эмуляция в данном случае подразумевается программная реализация модели.

## § 2.3. Феноменологическая «огрубленная» модель

Приведенные в предыдущем параграфе результаты численного моделирования показали, что наряду с исключительно стохастическими аспектами, в динамике микротрубочки можно выделить процессы и явления, которые могут быть описаны усредненными переменными (скорость роста, частота катастроф). Данный параграф посвящен построению минимально достаточной, простой аналитической модели, которая тем не менее способна качественно и количественно характеризовать поведение микротрубочки без громоздких компьютерных вычислений. Предлагаемый в данной модели подход позволяет трактовать спектры макроскопических катастроф в динамике микротрубочек, не выходя за рамки достаточно общих феноменологических предположений, что позволяет пролить свет на природу экспериментально наблюдаемого степенного закона распределения катастроф в динамике тубулиновых волокон.

В рамках данной модели поведение ансамбля микротрубочек описывалось двумя параметрами: средней скоростью роста и усредненной частотой катастроф – т.е. числом катастроф за единицу времени. Катастрофы и спасения трактовались, как кооперативные явления в духе теории фазовых переходов [Френкель, 1945]. В отличие от подробной феноменологии, необходимой для статистического описания, данная модель содержит лишь следующие упрощенные предположения относительно динамики микротрубочки:

— скорость роста микротрубочки зависит от концентрации свободного тубулина в растворе;

— в стенках микротрубочки могут существовать микронеоднородности – своеобразные дефекты;
— вероятность образования нового дефекта в микротрубочке тем выше, чем большее число дефектов в ней уже имеется.

В соответствии с тремя сделанными выше допущениями, рост микротрубочки описывался уравнением:

$$\frac{dN}{dt} = k_g Tu - k_d \qquad (2.8)$$

где $N(t) = \frac{13}{l} L(t)$ [шт. димеров] – число димеров в микротрубочке ($L$ – [мкм] ее длина, $l$ = 8 нм – характерный размер тубулинового димера), $Tu$ [мкМ] – концентрация свободного тубулина в растворе, $k_g$ [мкМ$^{-1}$мин$^{-1}$шт.димеров] – константа скорости реакции полимеризации и $k_d$ - константа скорости реакции десорбции, не зависящая от концентрации тубулина в растворе. Уравнение, описывающее накопление дефектов, имеет вид:

$$\frac{dD}{dt} = k_{ap} + k_m D \qquad (2.9)$$

где $D(t)$ [шт.] – число дефектов в микротрубочке. Первый член в правой части уравнения (2.9) – $k_{ap}$, описывает среднее число дефектов, спонтанно образующихся в растущей микротрубочке за единицу времени как в ее объеме, так и на краю. Второй член – $k_m D$ отображает наличие кооперативности в образовании дефектов.

Модель (2.8)-(2.9) сама по себе не является замкнутой, пока не определены критические условия потери устойчивости микротрубочки, т.е. условия наступления катастроф. В данной работе полагалось, что потеря устойчивости представляет собой событие, наступающее с вероятностью единица, когда отношение числа дефектов к числу составляющих микротрубочку тубулиновых димеров превышает некоторое критическое, предельно допустимое значение $\alpha_{cr}$. Т.е. первый (локальный) сценарий развития катастрофы (см. § 2.1) не рассматривался. Критическое условие имеет вид:

$$\frac{D(t^*)}{N(t^*)} = \alpha_{cr.} \qquad (2.10)$$

где $t = t^*$ – момент катастрофы, т.е. момент достижения критического условия. Развитие катастроф по этому сценарию напоминает эффекты, наблюдаемые при объемном вскипании перегретых жидкостей [Nucleation theory and applications, 1999]. Отметим, что условие развития катастрофы (2.10) носит пороговый характер и по своей природе является достаточным условием.

В рамках модели полагалось, что катастрофа влечет за собой полную деполимеризацию микротрубочки, т.е. вероятность спасения считается пренебрежимо малой. Выбирая в качестве начального момент, когда стартует рост, далее полагалось $N(0)=N_0=0$, $D(0)=D_0=0$. С учетом этого начального условия уравнения (2.8-2.10) легко разрешаются аналитически. Кривые, соответствующие решениям уравнения (2.8) – $\alpha_{cr} N(t)$ (сплошные (**a**) и (**b**)) и уравнения (2.9) – $D(t)$ (пунктир)**,** представлены на рис.2.5. Несложный анализ показал, что уравнение (2.10) имеет лишь тривиальные решения ($t^*=0$), если концентрация свободного тубулина в растворе ниже порогового уровня $Tu \leq Tu^{th}$. При этом макроскопический рост микротрубочки невозможен.

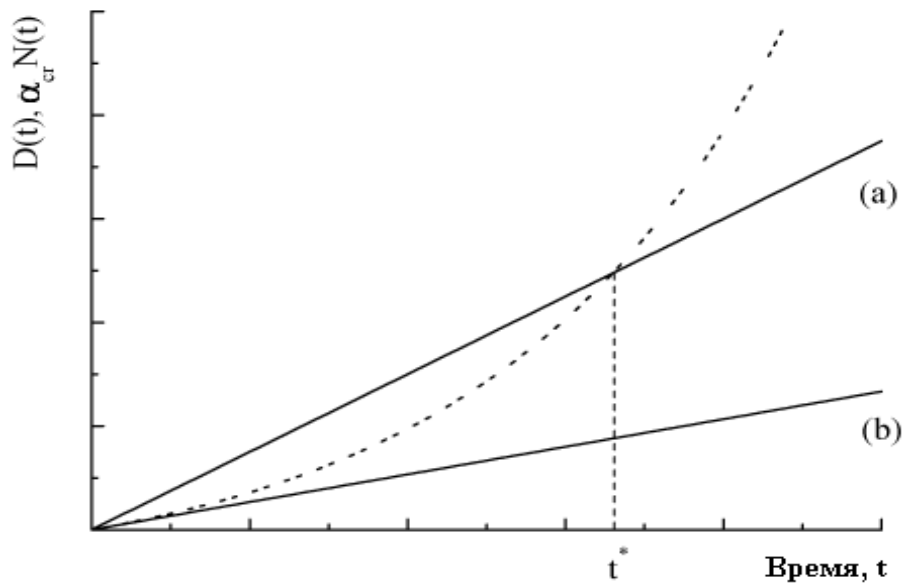

**Рис.2.5** Решения уравнений (2.8), (2.9). Число дефектов растет экспоненциально (пунктирная линия). Число димеров N на $\alpha_{cr}$ прямо пропорционально времени (сплошные линии (a),(b))**.** Катастрофа происходит в момент времени **t**$^*$. Он строго равен нулю, когда концентрация димеров Tu меньше критической $Tu^{th}$ (b). Если же $Tu>Tu^{th}$, одновременно с тривиальным решением (**t**$^*$=0) существует нетривиальное, соответствующее точке пересечения пунктирной и сплошной (a) линий.

Критическая концентрация свободного тубулина $Tu^{th}$ определяется выражением:

$$Tu^{th} = (1+\eta)Tu^*  \qquad (2.11)$$

где $\eta = k_{ap}/(\alpha_{cr}k_d)$ и $Tu^* = k_d/k_g$. Если же выполняется $Tu > Tu^{th}$, то всегда существует как минимум два решения, причем ненулевое соответствует наличию в рассматриваемой системе периодически повторяющихся катастроф с характерной частотой $f = 1/t^*$.

Было установлено, что зависимость между частотой катастроф и концентрацией свободного тубулина в растворе дается уравнением:

$$\frac{k_m}{f} = \left(\exp\left[\frac{k_m}{f}\right] - 1\right)\frac{\eta}{Tu/Tu^* - 1} \qquad (2.12)$$

Соотношение (2.12) в неявном виде задает теоретическую зависимость средней частоты катастроф *f*, как функции концентрации свободного тубулина (см. рис.2.6). Построенные в соответствии с (2.12) зависимости были сопоставлены с данными, полученными в независимых экспериментах рядом авторов (см. рис.2.6). В развитом в данном разделе подходе формально имеется всего один истинный параметр задачи – $\eta$. Его численное значения определялось путем сопоставления теоретических кривых и экспериментальных данных. Отвечающие оптимальному значению $\eta$ численные значения прочих параметров представлены в таблице 2.1.

**Таблица 2.1.** Оценка параметров модели из результатов применения фиттинга теоретической зависимости к экспериментальным данным.

| Литературный источник | Пороговая скорость (мкм/мин) | $\eta$ | $k_m \times 10^{-2}$ (мкм мкМ$^{-1}$ с$^{-1}$) |
|---|---|---|---|
| [Walker et al, 1988] | $2.02 \pm 0.15$ | $29.61 \pm 3.25$ | $2.43 \pm 0.75$ |
| [Drechsel et al., 1992] | $0.28 \pm 0.10$ | $6.15 \pm 1.37$ | $1.77 \pm 0.31$ |

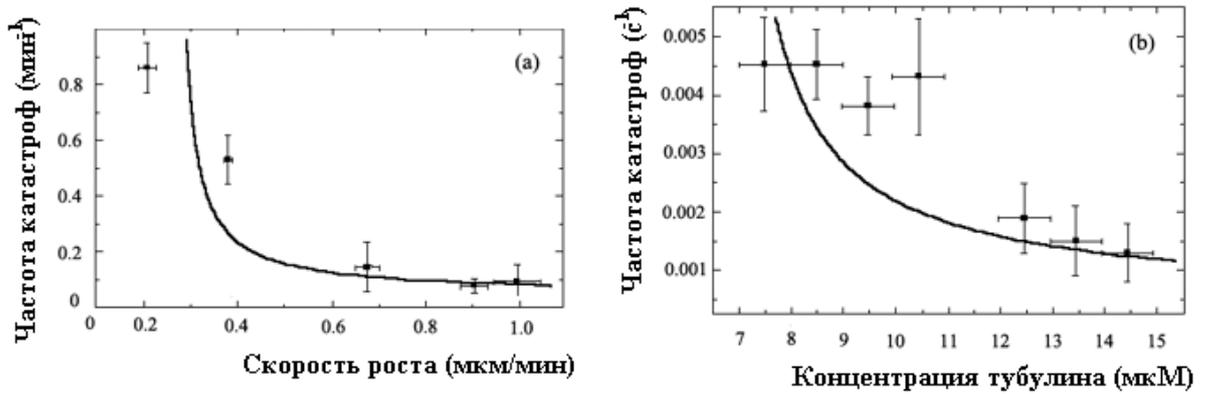

**Рис.2.6.** Частота катастроф, как функция скорости роста (a) и концентрации тубулина (b) для плюс-конца микротрубочки (согласно уравнению (2.12)). Экспериментальные точки взяты из работ [Drechsel et al. 1992] (a) и [Walker et al., 1988] (b).

Наличие качественного согласия между теоретическими кривыми и экспериментальными данными, полученными независимо разными группами экспериментаторов, свидетельствует в пользу того, что развитый в данной работе феноменологический в своей основе подход, в принципе может быть эффективно использован при интерпретации крупномасштабных критических явлений в динамике микротрубочек.

Уравнение (2.12) в предельном случае, когда $\tau = k_m t^* \ll 1$, может быть переписано в виде:

$$f = \frac{k_m}{2} \frac{1}{\left(Tu/Tu^{th}\right) - 1} \qquad (2.13)$$

Откуда вытекает, что поведение системы микротрубочек определяется значением всего двух параметров: константой $k_m$ и величиной предельной концентрации $Tu^{th}$. Каждый из них несложно измерить в эксперименте. Однако для независимого определения $k_m$, возможно, придется ожидать некоторых улучшений методов микроскопии [Arnal and Wade, 1995; Diaz et al., 1998; Schaap et al., 2004].

Отметим, что степенные зависимости типа (2.13) свойственны процессам фрагментации-ассоциации [Колмогоров, 1941], в частности, имеющим место в круге явлений самоорганизованной критичности [Bak, 1996].

Исходя из упрощенного вида зависимости (2.12), можно получить оценку максимального времени между вымыванием и катастрофой для эксперимента [Walker et al., 1991]. После подачи буфера в кювету рост прекращается и это соответствует тому, что на фазовой плоскости (рис.2.5) точка, соответствующая $N(t)$ с определенного момента отклонится от движения по наклонной и начнет двигаться параллельно оси абсцисс, пока не пересечет кривую $D(t)$, что по условию (2.10) приведет к катастрофе. Таким образом из графика становится понятным, что выражение для максимального времени задержки легко найти путем нахождения максимума функции $t_1(\alpha_{cr} N) - t_2(D)$, где $t_1$ и $t_2$ – функции, обратные к $N(t)$ и $D(t)$. После преобразований получается выражение:

$$t_{max} = \frac{1}{k_m}\left(\frac{Tu^{th}}{2Tu} + \frac{Tu}{2Tu^{th}} - 1\right) \qquad (2.14)$$

Проведя усреднение по всем концентрациям, получаем значение $\langle t_{max}\rangle = 24$с, которое соотносится с данными эксперимента (рис.2.4).

Подводя общий итог следует отметить, что по результатам вышеизложенного анализа, сопряженная с кластеризацией динамика структурных дефектов может рассматриваться, как один из ключевых механизмов развития динамических нестабильностей микротрубочек.

# Глава 3. Кинетические механизмы регуляции динамики микротрубочек

В предыдущей главе рассматривались внутренние, структурные причины, способные приводить к динамической неустойчивости микротрубочек. Глава 3 посвящена формулированию реакционно-диффузионно-преципитационной математической модели динамики роста и деполимеризации микротрубочек. Особенность данной модели состоит в предположении о том, что *a priori* не существует неустойчивостей, вызываемых особенностями строения и структуры микротрубочек. Полагается, что рост микротрубочек способен вызывать изменения в концентрационном окружении раствора, которые в свою очередь способны влиять на кинетику сорбции (агрегации, преципитации) и деполимеризации (дисагрегации). В результате основной акцент изучения ставится на рассмотрении взаимодействия концентрационных и преципитационных структур в динамике микротрубочек.

## § 3.1. Математическая модель

Следуя работе [Chen and Hill, 1987], в описываемой модели полагается, что ключевую роль в динамике микротрубочек играют следующие реакции:

$$\text{Tu-ГДФ} + \text{ГТФ} \underset{k_{-1}}{\overset{k_1}{\rightleftarrows}} \text{Tu-ГТФ} + \text{ГДФ} \qquad (3.1)$$

$$\text{MT}_n + \text{Tu-ГТФ} \xrightarrow{k_2} \text{MT}_{n+1} \qquad (3.2)$$

$$\text{MT}_n \xrightarrow{k_3} \text{MT}_{n-1} + \text{Tu-ГДФ} \qquad (3.3)$$

где $\text{MT}_n$ обозначает микротрубочку, содержащую n молекул тубулина, Tu-ГТФ и Tu-ГДФ обозначают растворенный тубулин, связанный соответственно с молекулами ГТФ и ГДФ. В ходе реакции с константой скорости $k_1$ в растворе происходит обмен связанной с тубулином молекулы ГДФ на молекулу ГТФ. Обратная реакция (с константой скорости $k_{-1}$) отображает обмен связанной с тубулином молекулы ГТФ на ГДФ, также происходящая в растворе. Реакция (3.2) описывает удлинение микротрубочки при полимеризации тубулина. Последняя реакция соответствует деполимеризации, при которой из микротрубочек в раствор высвобождается тубулин, связанный с молекулой ГДФ. Предполагается, что молекулы тубулина способны диффундировать в растворе.

Перейдем к построению кинетических уравнений, описывающих динамику микротрубочек в растворе. Полагается, что у микротрубочек свободен лишь плюс-конец, а минус-конец закреплен на центросоме. Будем полагать, что миграция плюс-концов, в силу жесткости микротрубочки, происходит радиально, от центра к периферии клетки. Система кинетических уравнений, учитывающая реакции (3.1)-(3.3), в общем случае имеет вид:

$$\frac{\partial p(\vec{r},t)}{\partial t} = \beta \ \ div\left(\frac{\vec{r}}{r}(k_3 p - k_2 u p)\right) \tag{3.4}$$

$$\frac{\partial u(\vec{r},t)}{\partial t} = D\Delta u - k_2 u p + k_1 [\text{ГТФ}] v - k_{-1}[\text{ГДФ}] u \tag{3.5}$$

$$\frac{\partial v(\vec{r},t)}{\partial t} = D\Delta v + k_3 p - k_1 [\text{ГТФ}] v + k_{-1}[\text{ГДФ}] u \tag{3.6}$$

где $u(\vec{r},t)$ – соответствует значению концентрации молекул Tu-ГТФ в точке с радиус-вектором $\vec{r}$ в момент времени $t$, $v(\vec{r},t)$ – концентрации молекул Tu-ГДФ, $p(\vec{r},t)$ – плотности вероятности нахождения плюс-конца микротрубочки, деленной на число Авогадро $N_A$. $k_1$ [М$^{-1}$с$^{-1}$], $k_{-1}$ [М$^{-1}$с$^{-1}$], $k_2$ [М$^{-1}$с$^{-1}$], $k_3$ [с$^{-1}$] обозначают константы скоростей соответствующих реакций (3.1)-(3.3), β [нм] – стерический фактор, равный приращению длины микротрубочки при присоединении к ней одной молекулы тубулина [Li et al., 2002]. Концентрации молекул ГТФ и ГДФ полагаются однородными и выступают в качестве параметров.

Также необходимо выполнение дополнительных условий:

$$N_A \cdot \int_V p(\vec{r},t) dV = 1 \tag{3.7}$$

$$\int_V \left[ u(\vec{r},t) + v(\vec{r},t) + \frac{|\vec{r}|}{\beta} p(\vec{r},t) \right] dV = \frac{N_{tot}}{N_A} \tag{3.8}$$

где $N_{tot}$ обозначает общее количество молекул тубулина в системе, а $V$ – общий объем системы. Уравнение (3.7) отображает полноту рассматриваемой системы событий. Соотношение (3.8) отражает сохранение в ходе реакций (3.1)-(3.3) общего количества молекул тубулина. Инициация полимеризации предполагается происходящей на затравке, расположенной в начале координат, т.е. при $\vec{r}=0$. Рост микротрубочек происходит в радиальном направлении. Оказывается удобным для дальнейшего рассмотрения ввести также дополнительный параметр $\mathrm{Tu} = \frac{N_{tot}}{V N_A}$, равный общему количеству молекул тубулина, деленному на число Авогадро и объем системы и по смыслу соответствующий "общей молярной концентрации" молекул тубулина в растворе.

Наиболее простым представляется псевдоодномерное приближение, рассматриваемое в области, параллельной оси абсцисс, сечением $S$ и длиной L. Рост микротрубочек происходит в положительном направлении вдоль оси абсцисс. Правая граница $x=\mathrm{L}$ соответствует местоположению клеточной стенки, которая считается непроницаемой. После дискретизации по пространственной переменной (см. Приложение II), система кинетических уравнений (3.4)-(3.8) в псевдоодномерном приближении принимает вид:

$$\frac{dp_0}{dt} = k_3^0 p_1 - k_2^0 u_1 p_0 \tag{3.9}$$

$$\frac{dp_1}{dt} = \frac{\beta}{h}\left(k_3 p_2 - k_3^0 p_1 - k_2 u_1 p_1 + k_2^0 u_1 p_0\right) \tag{3.10}$$

$$\frac{dp_m}{dt} = \frac{\beta}{h}\left(k_3(p_{m+1} - p_m) + k_2(u_{m-1}p_{m-1} - p_m u_m)\right) \qquad m \in \{2, M-1\} \tag{3.11}$$

$$\frac{dp_M}{dt} = \frac{\beta}{h}\left(k_2 u_{M-1} p_{M-1} - k_3 p_M\right) \tag{3.12}$$

$$\frac{du_1}{dt} = D\frac{u_2 - u_1}{h^2} - k_2^0 u_1 p_0 - k_2 p_1 u_1 + k_1[\text{ГТФ}]v_1 - k_{-1}[\text{ГДФ}]u_1 \qquad (3.13)$$

$$\frac{du_m}{dt} = D\frac{u_{m-1} - 2u_m + u_{m+1}}{h^2} - k_2 p_m u_m + k_1[\text{ГТФ}]v_m - k_{-1}[\text{ГДФ}]u_m,$$

$$m \in \{2, M-1\} \qquad (3.14)$$

$$\frac{du_M}{dt} = D\frac{u_{M-1} - u_M}{h^2} + k_1[\text{ГТФ}]v_M - k_{-1}[\text{ГДФ}]u_M \qquad (3.15)$$

$$\frac{dv_1}{dt} = D\frac{v_2 - v_1}{h^2} + k_3^0 p_1 - k_1[\text{ГТФ}]v_1 + k_{-1}[\text{ГДФ}]u_1 \qquad (3.16)$$

$$\frac{dv_m}{dt} = D\frac{v_{m-1} - 2v_m + v_{m+1}}{h^2} + k_3 p_m - k_1[\text{ГТФ}]v_m + k_{-1}[\text{ГДФ}]u_m,$$

$$m \in \{2, M-1\} \qquad (3.17)$$

$$\frac{dv_M}{dt} = D\frac{v_{M-1} - v_M}{h^2} + k_3 p_M - k_1[\text{ГТФ}]v_M + k_{-1}[\text{ГДФ}]u_M \qquad (3.18)$$

$$\sum_{m=0}^{M} w_m = w_0 + hSN_A \sum_{m=1}^{M} p_m = 1 \qquad (3.19)$$

$$w_0 + hSN_A \sum_{m=1}^{M}\left(u_m + v_m + \frac{mh + \beta}{\beta}p_m\right) = N_0 \qquad (3.20)$$

где $u_m$ – соответствует значению концентрации молекул Tu-ГТФ в интервале $((m-1)h, mh]$, $m \in \{1, M\}$, $v_m$ – концентрации молекул Tu-ГДФ, $p_m = \frac{w_m}{hSN_A}$ – плотности вероятности нахождения плюс-конца микротрубочки в интервале $((m-1)h, mh]$, $m \in \{1, M\}$, деленной на число Авогадро $N_A$. Вероятность нахождения плюс-конца на затравке обозначается через $w_0$, при этом $p_0 = \frac{w_0}{\beta SN_A}$. $k_1$, $k_{-1}$, $k_2$, $k_3$ обозначают константы скоростей соответствующих реакциям (3.1)-(3.3), константы $k_2^0, k_3^0$ отражают реакции первичной сорбции и десорбции тубулина на затравке, β – стерический фактор, равный приращению длины микротрубочки при присоединении к ней одной молекулы тубулина [Nogales et al., 1998]. Концентрации молекул ГТФ и ГДФ выступают в качестве параметров.

Введем новые, безразмерные переменные, причем их связь со старыми выражается формулами:

$$u_m = \frac{k_3}{k_2}\tilde{u}_m, \quad v_m = \frac{k_3}{k_2}\tilde{v}_m, \quad p_m = \frac{k_1[\text{ГТФ}]}{k_2}\tilde{p}_m$$

$$t = \frac{\tilde{t}}{k_1}, \quad h = \sqrt{\frac{D}{k_1}}\tilde{h}, \quad L = \sqrt{\frac{D}{k_1}}\tilde{L}, \tag{3.21}$$

$$\gamma = \frac{k_{-1}[\text{ГДФ}]}{k_1[\text{ГТФ}]}, \quad \theta = \frac{\beta}{\sqrt{Dk_1[\text{ГТФ}]}}\,k_3, \quad \varphi = \frac{k_2}{k_1[\text{ГТФ}]hSN_A}, \quad Tu^0 = \frac{k_2}{k_3}Tu \tag{3.22}$$

$$\varepsilon_2 = \frac{k_2^0}{k_2}, \quad \varepsilon_3 = \frac{k_3^0}{k_3}, \quad \eta = \frac{k_1[\text{ГТФ}]}{k_3}$$

В новых, безразмерных переменных, система (3.9)-(3.20) принимает вид:

$$\frac{d\tilde{p}_0}{d\tilde{t}} = \frac{\varepsilon_3}{\eta}\tilde{p}_1 - \frac{\varepsilon_2}{\eta}\tilde{u}_1\tilde{p}_0 \tag{3.23}$$

$$\frac{d\tilde{p}_1}{d\tilde{t}} = \frac{\theta}{\tilde{h}}\left(\tilde{p}_2 - \varepsilon_3\tilde{p}_1 - \tilde{u}_1\tilde{p}_1 + \varepsilon_2\tilde{u}_1\tilde{p}_0\right) \tag{3.24}$$

$$\frac{d\tilde{p}_m}{d\tilde{t}} = \frac{\theta}{\tilde{h}}\left(\tilde{p}_{m+1} - \tilde{p}_m + \tilde{u}_{m-1}\tilde{p}_{m-1} - \tilde{p}_m\tilde{u}_m\right) \qquad m \in \{2, M-1\} \tag{3.25}$$

$$\frac{d\tilde{p}_M}{d\tilde{t}} = \frac{\theta}{\tilde{h}}\left(\tilde{u}_{M-1}\tilde{p}_{M-1} - \tilde{p}_M\right) \tag{3.26}$$

$$\frac{d\tilde{u}_1}{d\tilde{t}} = \frac{\tilde{u}_2 - \tilde{u}_1}{\tilde{h}^2} - \varepsilon_2\tilde{u}_1\tilde{p}_0 - \tilde{p}_1\tilde{u}_1 + \tilde{v}_1 - \gamma\tilde{u}_1 \tag{3.27}$$

$$\frac{d\tilde{u}_m}{d\tilde{t}} = \frac{\tilde{u}_{m-1} - 2\tilde{u}_m + \tilde{u}_{m+1}}{\tilde{h}^2} - \tilde{p}_m\tilde{u}_m + \tilde{v}_m - \gamma\tilde{u}_m, \quad m \in \{2, M-1\} \tag{3.28}$$

$$\frac{d\tilde{u}_M}{d\tilde{t}} = \frac{\tilde{u}_{M-1} - \tilde{u}_M}{\tilde{h}^2} + \tilde{v}_M - \gamma\tilde{u}_M \tag{3.29}$$

$$\frac{d\tilde{v}_1}{d\tilde{t}} = \frac{\tilde{v}_2 - \tilde{v}_1}{\tilde{h}^2} + \varepsilon_3\tilde{p}_1 - \tilde{v}_1 + \gamma\tilde{u}_1 \tag{3.30}$$

$$\frac{d\widetilde{v}_m}{d\widetilde{t}} = \frac{\widetilde{v}_{m-1} - 2\widetilde{v}_m + \widetilde{v}_{m+1}}{\widetilde{h}^2} + \widetilde{p}_m - \widetilde{v}_m + \gamma\widetilde{u}_m, \quad m \in \{2, M-1\} \quad (3.31)$$

$$\frac{d\widetilde{v}_M}{d\widetilde{t}} = \frac{\widetilde{v}_{M-1} - \widetilde{v}_M}{\widetilde{h}^2} + \widetilde{p}_M - \widetilde{v}_M + \gamma\widetilde{u}_M \quad (3.32)$$

$$\frac{\beta}{h}\widetilde{p}_0 + \sum_{m=1}^{M}\widetilde{p}_m = \varphi \quad (3.33)$$

$$\eta\frac{\beta}{h}\widetilde{p}_0 + \sum_{m=1}^{M}\left(\widetilde{u}_m + \widetilde{v}_m + \left(\frac{m\widetilde{h}}{\theta} + \eta\right)\widetilde{p}_m\right) = Tu^0 M \quad (3.34)$$

## § 3.2. Параметрическая диаграмма состояния

Решение системы уравнений (3.23)-(3.34) проводилось численно, методом Розенброка для жестких систем [Press et al., 2002]. Значения параметров и диапазон их изменений указаны в таблице 3.1.

**Таблица 3.1.** Значения параметров модели и диапазон их изменений.

| Параметр модели, единицы измерения | $k_1$,[a] $M^{-1}c^{-1}$ | $k_{-1}$,[a] $M^{-1}c^{-1}$ | $k_2$,[b] $мкM^{-1}c^{-1}$ | $k_3$,[b] $c^{-1}$ | $k_2^0$, $M^{-1}c^{-1}$ | $k_3^0$, $c^{-1}$ | D,[c] $нм^2/c$ |
|---|---|---|---|---|---|---|---|
| Значение | 0,02 | 0,02 | 18 | 20 | 3000 | 20 | 4000 |

| Параметр модели, единицы измерения | L,[a] мкм | M | β,[d] нм | S, $мкм^2$ | [ГТФ],[a] мкМ | [ГДФ],[a] мкМ | $Tu^0$,[a] мкМ |
|---|---|---|---|---|---|---|---|
| Значение | 90 | 90 | 0,165 | 40 | 50 | $2 \div 300$ | $0 \div 400$ |

Величины параметров модели взяты из работ: [a] [Melki et al., 1988]; [b] [Mitchison and Kirschner, 1984]; [c] [Salmon et al., 1984]; [d] [Nogales et al., 1998]

Результаты расчетов представлены на диаграмме, изображенной на рис. 3.1. По оси абсцисс отложена безразмерная общая концентрация молекул тубулина в системе:

$$Tu^O = \frac{k_2}{k_3} Tu \qquad (3.35)$$

а по оси ординат – величина:

$$\gamma = \frac{k_{-1}[\text{ГДФ}]}{k_1[\text{ГТФ}]} \qquad (3.36)$$

Выяснилось, что параметрическая плоскость $(Tu^0, \gamma)$ (см. рис. 3.1) разбивается семейством найденных нами кривых на пять областей, каждой из которых соответствует свой динамический режим. Области значений параметров, обозначенной на диаграмме символом **"0"**, соответствует наличие в системе единственного решения в виде унимодального устойчивого распределения микротрубочек по длинам, с выраженным максимумом на левой границе, т.е. в области инициации полимеризации.

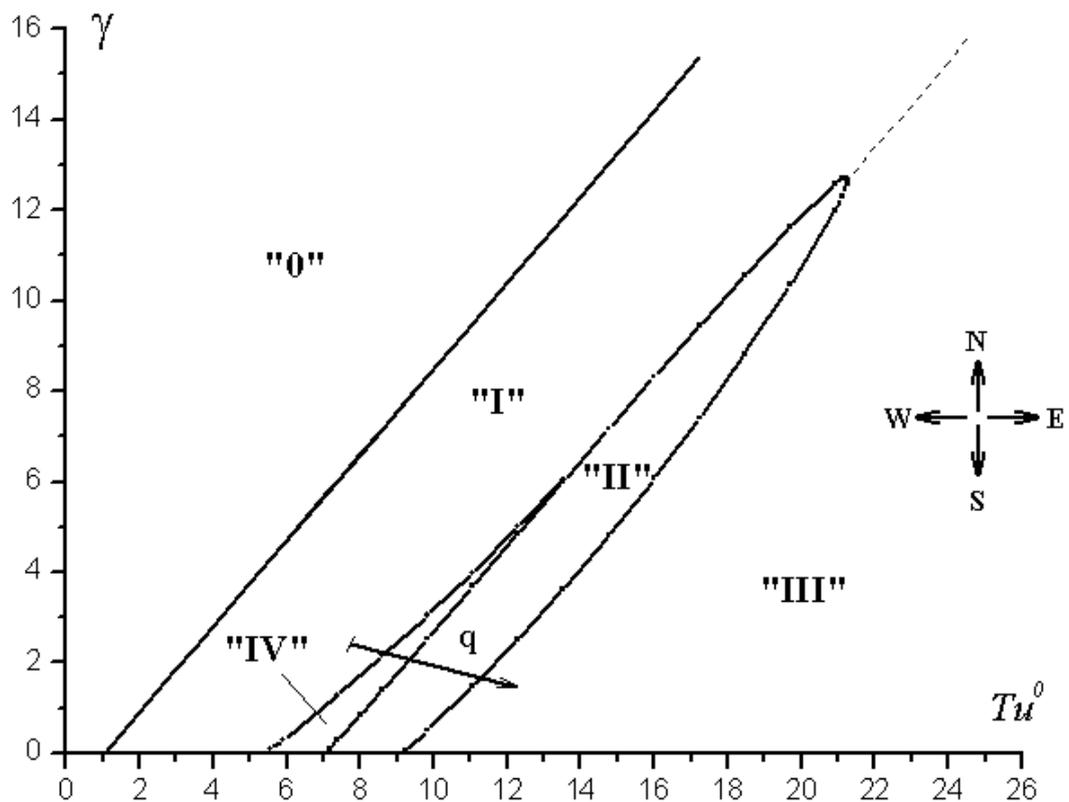

**Рис. 3.1.** Диаграмма состояний, построенная в соответствии с системой уравнений (3.23)-(3.34). Линии границ разбивают параметрическую плоскость на пять областей: зоны стационарных режимов **"0"**, **"I"** и **"III"**, зону нестационарных режимов **"II"**, а также зону **"IV"**, которая соответствует как метастабильным, так и нестационарным состояниям (см. пояснение в тексте).

Областям решений "**I**" и "**III**" соответствуют стационарные устойчивые решения (см. рис. 3.2а). В области значений параметров "**I**" распределение микротрубочек по размерам характеризуется наличием отчетливо выраженного пика на левом крае и локального максимума, расположенного на существенном удалении от левого края (см. рис. 3.2а).

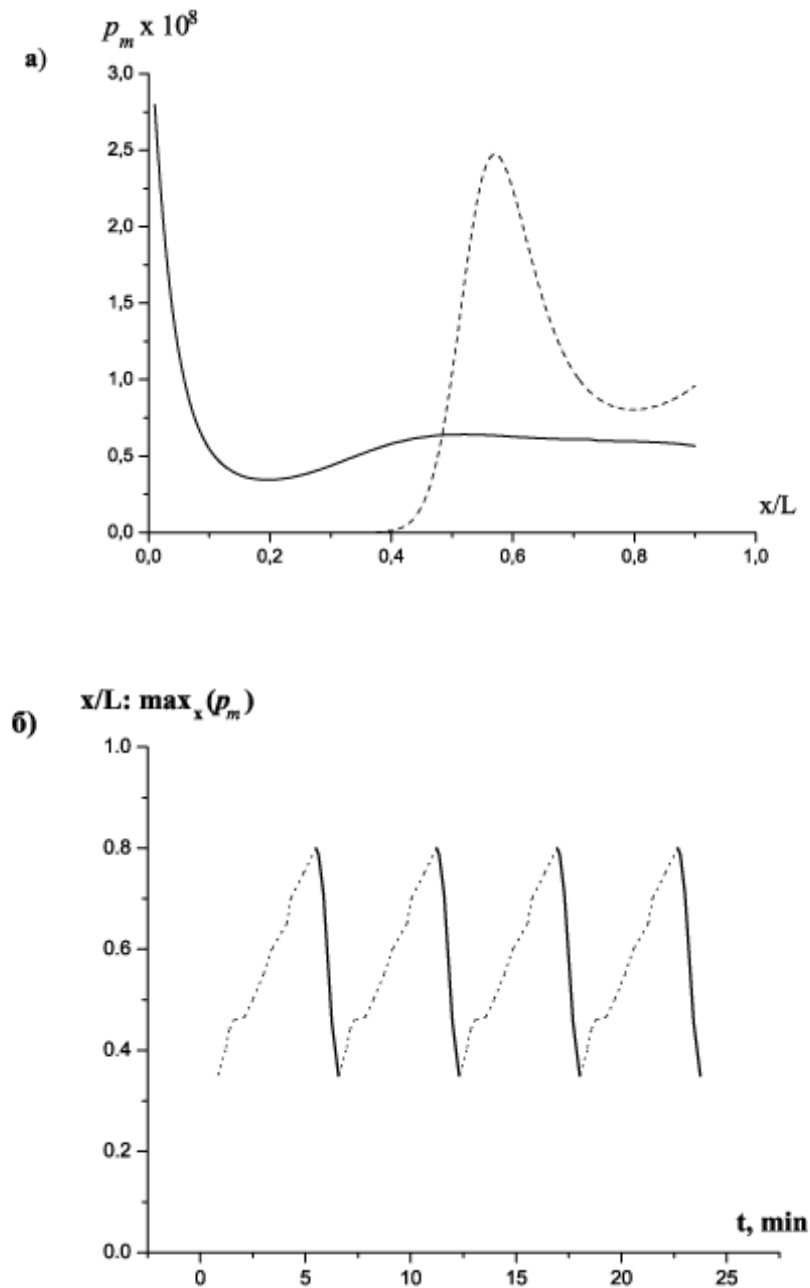

**Рис. 3.2. а)** интерполяционно сглаженные решения, соответствующие областям "**I**" (сплошная линия) и "**III**" (пунктирная линия) на диаграмме состояния. **б)** изменение во времени положения максимума нестационарного распределения плотности плюс-концов микротрубочек, характерное для режимов, соответствующих области "**II**". Периоды медленной полимеризации (пунктирная линия) чередуются с волнами быстрой деполимеризации (сплошная линия).

В области параметров **"III"** характерные распределения микротрубочек по размерам, как видно из рис. 3.2а, имеют минимум на левой границе (x=0), экстремум примерно в центре рассматриваемой области и локальный максимум на границе x=L. Это говорит о том, что при значениях параметров из данной области в системе должны доминировать протяженные микротрубочки, сопоставимые по размерам с размером клетки.

Область **"II"** соответствует нестационарным режимам, проявляющимся в виде периодически бегущих справа налево волн деполимеризации микротрубочек, чередующихся с фазами их медленного полимеризационного роста (рис. 3.2б). На границе области **"II"** имеет место параметрическая неустойчивость стационарных решений.

Представляется, что отвечающие данной области параметров режимы наблюдаются в динамике реальных микротрубочек как in vitro [Mitchison and Kirschner, 1984; Walker et al., 1988], так и in vivo [Sammak and Borisy, 1988; Воробьев и др., 2000].

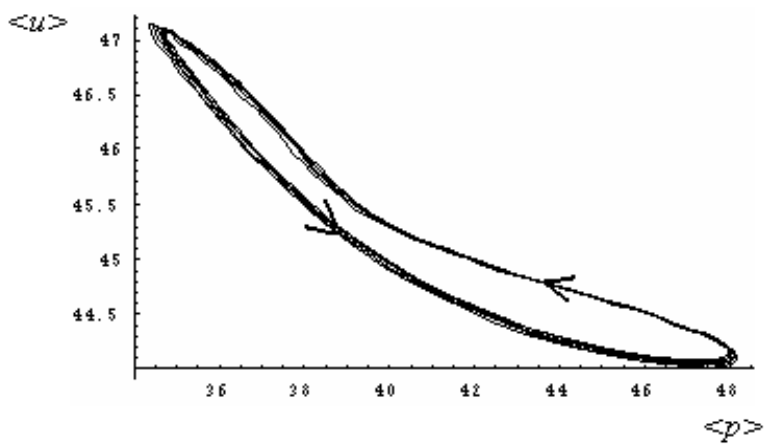

**Рис. 3.3.** Фазовая траектория для "центров масс" концентрации тубулина-ГТФ и плотности плюс-концов при величинах параметров, соответствующих зоне **"II"**.

Для параметров из области **"II"** рассмотрим поведение изображающей систему точки на плоскости, где по оси абсцисс отложен "центр масс" плотности вероятности нахождения плюс-концов, а по оси ординат – концентрации тубулина-ГТФ (рис. 3.3). Видно, что на указанной плоскости имеет место предельный цикл Пуанкаре. То есть в системе явным образом реализуются автоколебания! Более тщательный анализ показывает, что по мере роста микротрубочек (движение по нижней ветви цикла) "центр масс" концентрации тубулина-ГТФ начинает смещаться к центру нуклеации (центросоме). Это соответствует обеднению концентрации способного к полимеризации тубулина в области правой границы, образованию там "провала" за счет диффузионного размытия. Поэтому в момент, когда микротрубочки достигают правой границы, локальное пресыщение в области плюс-конца приводит к тому, что запускается процесс деполимеризации (верхняя ветвь на рис.3.3).

В области **"IV"** рассматриваемая система имеет два типа решений. Наряду со стационарными метастабильными решениями, имеют место и нестационарные решения, схожие с наблюдаемыми в области **"II"**. В зависимости от истории эволюции значений параметров в системе может установиться либо стационарный, либо нестационарный режим. Если параметры ($Tu^0$, $\gamma$) будут медленно изменяться так, что на диаграмме состояния (рис. 3.1) изображающая точка будет плавно перемещаться из области **"I"** в область **"IV"**, то система будет оставаться в стационарном метастабильном состоянии до тех пор, пока параметры не пересекут границу между областями **"IV"** и **"II"**. Пересечение последней повлечет за собой "жесткое" возбуждение пространственных осцилляций и волн, характерных для всей области **"II"**. Примечательно, что при изменении параметров в обратном направлении, то есть движении изображающей точки из зоны **"II"** в зону **"IV"**, система сохранит нестационарное поведение вплоть до границы между зоной **"IV"** и зоной **"I"**. Иными словами, в области значений параметров **"IV"** система демонстрирует отчетливо выраженное гистерезисное поведение.

Наличие целой области метастабильных состояний у системы регуляции микротубулинового цитоскелета может играть важную биологическую роль. При определенных условиях микротубулиновый цитоскелет находится в покоящемся, но "взведенном" состоянии [Potapova et al., 2006]. Небольшого внешнего воздействия, иногда укола клетки иглой, достаточно для мгновенного перехода цитоскелета к масштабной перестройке.

## § 3.3. Динамические характеристики автоколебательных решений

Качественно поведение решений системы уравнений (3.23)-(3.24) при пересечении изображающей точкой границ областей "**I**", "**II**" "**III**" и "**IV**" показано на рис. 3.4а. Вдоль оси абсцисс отложено изменение параметров $Tu^0$ и $\gamma$ вдоль вектора $\vec{q}$, проходящего через четыре разных области на диаграмме состояний (см. рис. 3.1). По оси ординат отложен средний радиус кривизны $<R>$ решения системы (3.23)-(3.34) $p_m(t)$, $u_m(t)$, $v_m(t)$, $m \in \{1, M\}$, представляющего однопараметрическую кривую в 3М-мерном пространстве [Ефимов и Розендорн, 1970]. Для режимов, характерных для областей параметров "**I**" и "**III**" на диаграмме состояний, справедливо $<R> = 0$. Было обнаружено, что при проходе из области "**III**" в область параметров "**II**" в системе мягко возбуждаются автоколебания. О величине амплитуды соответствующего предельного цикла можно судить по среднему значению радиуса кривизны $<R>$ за период. При антипараллельном вектору $\vec{q}$ движении из зоны "**III**" в зону "**II**", амплитуда $<R>$ растет, как квадратичный корень величины закритичности (расстояния от $\vec{q} = \vec{q}^*$, т.е. границы областей "**II**" и "**III**", до текущей изображающей точки на диаграмме состояний). Иными словами, при переходе системы через границу областей "**III**" и "**II**" имеет место закритическая бифуркация Андронова-Хопфа [Андронов и др., 1959].

При антипараллельном вектору $\vec{q}$ изменении параметров, в точке $\vec{q} = \vec{q}_1$ происходит бифуркация удвоения цикла (рис. 3.4б). Однако перехода к каскаду удвоений Фейгенбаума не наблюдается [Feigenbaum, 1978]. Так как далее, при $\vec{q} = \vec{q}_2$ наблюдается бифуркация схлопывания "удвоившегося ранее цикла" к однопериодичному решению (уменьшение периода вдвое), которое продолжает существовать при дальнейшем изменении параметров вплоть до границы областей **"IV"** и **"I"**.

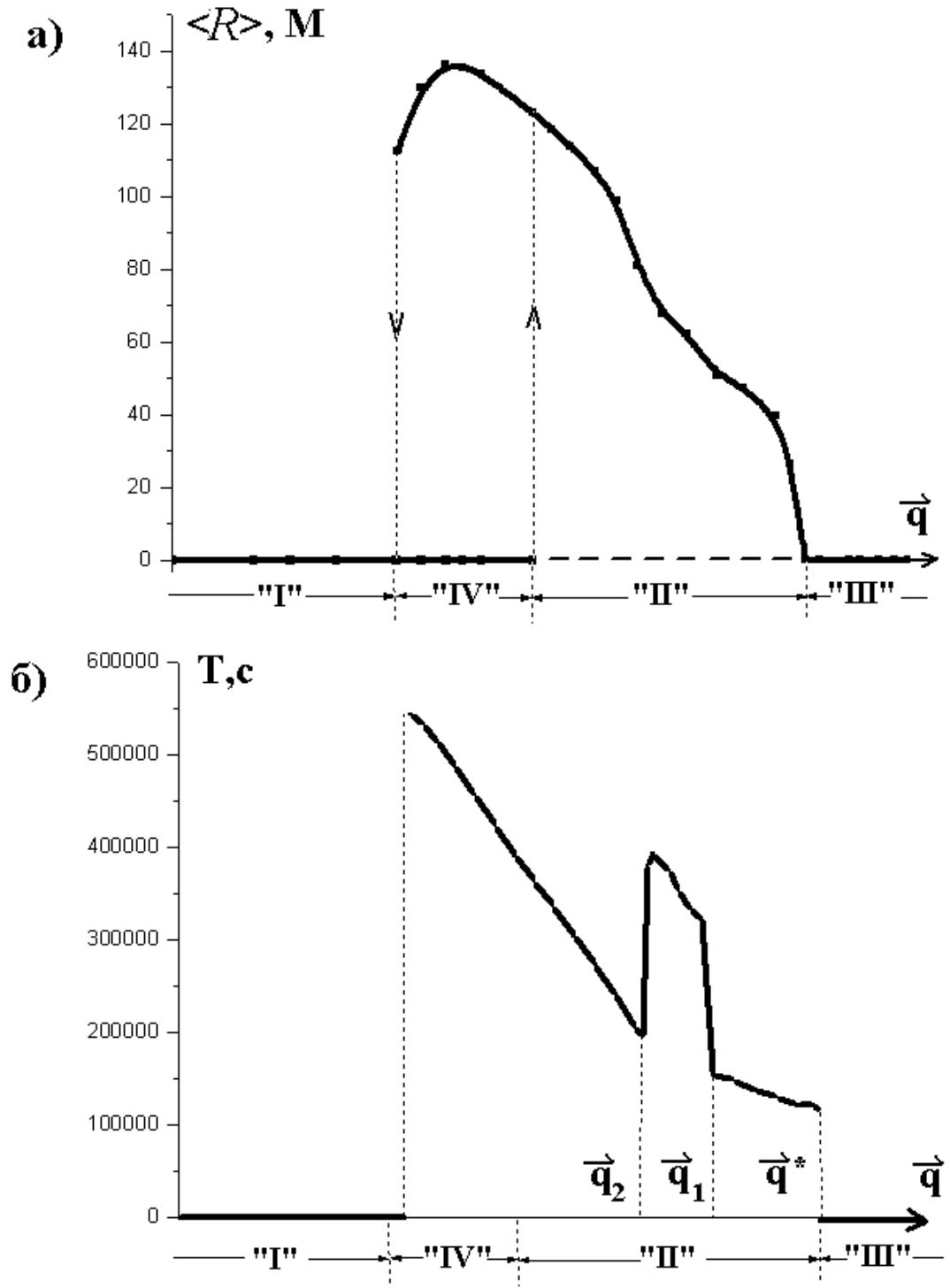

**Рис. 3.4. а)** Зависимость среднего радиуса кривизны <*R*> траектории решения системы (3.23)-(3.34) и **б)** периода предельного цикла **T** при изменении параметров $Tu^0$ и γ вдоль вектора $\vec{q}$ (см. рис. 3.1).

Из рис. 3.4а видно, что в области **"I"** величина <*R*> равняется нулю, что соответствует устойчивому стационарному решению. Дальнейшее медленное изменение параметров вдоль вектора q̄ оставляет решение системы на нижней ветви <*R*> = 0, вплоть до границы с областью **"II"**. Как уже отмечалось, пересечение этой границы приводит к жесткому возбуждению колебаний, что выражается в скачкообразном увеличении величины <*R*>. Система переходит на верхнюю ветвь (см. рис. 3.4а), отвечающую ненулевым значениям <*R*>. Отметим, что во всей области значений параметров **"II"** стационарное решение системы (3.23)-(3.34) существует, но оно неустойчиво (обозначено горизонтальной пунктирной линией на рис. 3.4а.

При изменении параметров из области **"II"** в противоположном вектору q̄ направлении, система остается на верхней ветви до тех пор, пока изображающая точка не пересечет границу областей **"IV"** и **"I"**. При пересечении этой границы происходит потеря устойчивости нестационарного решения, и система возвращается на нижнюю ветвь. Таким образом, история изменения параметров определяет ветвь, на которой будет находиться система при значении параметров из области **"IV"**.

Подводя итог, отметим, что построенная параметрическая диаграмма состояния, содержит области, отвечающие как устойчивым стационарным, так и нестационарным состояниям. При этом найдена область, отвечающая метастабильным состояниям. В результате бифуркационного анализа показано, что потеря устойчивости стационарных состояний при изменении параметров системы, сопровождается бифуркацией рождения предельного цикла по механизму Андронова-Хопфа. Обнаружена вторичная бифуркация удвоения предельного цикла.

# Глава 4. Классификация регуляторных воздействий на тубулиновый цитоскелет

Построенная диаграмма состояния может быть использована для качественного анализа механизмов влияния различных биохимических агентов на динамику цитоскелета. С кинетической точки зрения, их действие на рассматриваемую систему наиболее часто проявляется в изменении эффективных констант скоростей реакций [Wilson et al., 1999]. В рамках развиваемого подхода изменение констант, в свою очередь, влечет за собой изменение параметров модели и сдвиг изображающей систему точки на диаграмме состояний. В таблице 4.1 представлены сведения об изменениях ключевых параметров модели $Tu^0$ и $\gamma$ при воздействии на систему ряда биохимических регуляторных факторов.

**Таблица 4.1.** Изменение параметров модели $Tu^0$ и $\gamma$ при воздействии на систему биохимических факторов. Стрелка вверх соответствует увеличению значения параметра при действии соответствующего фактора, вниз – уменьшению.

| Название фактора | $k_1$[ГТФ] | $k_2$ | $k_3$ | $k_{-1}$[ГДФ] | Tu | [ГТФ]/[ГДФ] | $Tu^0$ | $\gamma$ | Класс |
|---|---|---|---|---|---|---|---|---|---|
| Таксол, доцетаксел [a] | — | ↑ | ↓ | — | — | — | ↑ | — | **E** |
| Tau, MAP2, MAP4 [b),c),d)] | — | ↑ | ↓ | — | — | — | ↑ | — | **E** |
| Колхицин, винбластин, нокодазол [a] | — | ↓ | ↑ | — | — | — | ↓ | — | **W** |
| Op18/статмин [e] | — | — | ↑ | — | ↓ | — | ↓ | — | **W** |
| Митохондриальные факторы | ↑ | — | — | ↓ | — | ↑ | — | ↓ | **S** |
| Факторы ГТФ-ГДФ обмена (GEFs) [f] | ↑ | — | — | ↓ | — | ↑ | — | ↓ | **S** |
| Малые ГТФазы (rho, ran, rac) [g), h)] | ↓ | — | — | ↑ | — | ↓ | — | ↑ | **N** |
| Активирующие ГТФазы белки (GAPs) [g] | ↓ | — | — | ↑ | — | ↓ | — | ↑ | **N** |

Сведения о влиянии факторов приведены по публикациям: [a)] Wilson et al., 1999; Grigoriev et al., 1999 [b)] Drechsel et al., 1992; [c)] Itoh and Hotani, 1994; [d)] Nguyuen et al., 1999; [e)] Andersen, 2000; [f)] Schimdt and Hall, 2002; [g)] Caudron et al., 2005; [h)] Grigoriev et al., 2006.

Из таблицы 4.1 видно, что рассмотренные нами факторы могут быть разбиты на две группы. Одни оказывают влияние на параметр $Tu^0$, а другие – на $\gamma$. Каждая группа в свою очередь разбивается на два класса, в зависимости от того, к увеличению или уменьшению соответствующего параметра приводит действие входящих в данный класс агентов. В рамках модели действие агентов сводится к смещению изображающей точки на диаграмме состояний в направлении, параллельном осям, поэтому удобно соотнести обозначения указанных четырех классов с направлениями, принятыми для указания сторон света (см. рис. 3.1). К первому классу, обозначаемому буквой **E**, относятся вещества (таксол, доцетаксел, см. табл. 4.1), вызывающие смещение изображающей точки на восток. Аналогичным образом колхицин и винбластин могут быть отнесены к классу **W**, а митохондриальные факторы и факторы ГТФ-ГДФ обмена – к классу **S**.

На стадии деления тубулиновый цитоскелет претерпевает значительные пространственно-временные трансформации. Поэтому мы предполагаем, что по крайней мере, на отдельных стадиях митоза клеточный скелет должен находиться в состояниях, отвечающих зоне **"II"** на построенной диаграмме (см. рис. 3.1).

Известно, что вещества, оказывающие цитостатическое действие, нарушают нормальное прохождение клеткой митоза [Jordan, 2002; Honore et al., 2005]. Эти агенты подавляют высокую динамическую лабильность микротрубочек, необходимую для пространственного разделения хромосом. Таким образом, цитостатики блокируют ту фазу митоза, в которой клетка находилась бы в области параметров, отвечающей нестационарным режимам, т.е. области **"II"**.

Действие агентов класса **E** сопровождается удлинением микротрубочек, с последующей остановкой пульсаций. В терминах нашей модели эти вещества переводят систему на диаграмме состояния в область **"III"**, например, из области **"II"**. Поэтому к классу **W** относятся цитостатики, приводящие к существенной деполимеризации микротрубочек с последующим подавлением осцилляций в их динамике. Аналогично, вещества из класса **W** переводят систему в области **"0"** и **"I"** на диаграмме состояния.

Например, добавление к ансамблю микротрубочек таких цитостатиков, как таксол (или доцетаксел), выражается, как известно, в ускорении сорбции тубулина из раствора с последующей стабилизацией длины микротрубочек [Wilson et al., 1999; Grigoriev et al., 1999]. В рамках развиваемого нами подхода ускорение скорости сорбции молекул тубулина на плюс концы микротрубочек формально учитывается за счет увеличения константы $k_2$. Замедление же реакций десорбции – уменьшением константы $k_3$ (См. таблицу 4.1). Изменение обеих констант, в силу (3.35), эквивалентно увеличению величины ключевого безразмерного параметра системы – $Tu^0$. Иными словами, действие таксола (и его аналогов) должно вызывать на диаграмме состояния (см. рис. 3.1) сдвиг изображающей точки вправо по оси $Tu^0$. В частности, это должно приводить к переходу из области нестационарных решений **"II"** в область **"III"**, отвечающую стационарным распределениям плюс-концов у внешнего края клетки (см. рис. 3.1).

Следует отметить, что добавление таксола (или доцетаксела) к клеткам, находящимся на стадии деления, действительно приводит к остановке деления [Wilson et al., 1999; Jordan, 2002]. При этом распределение микротрубочек по длине становится статичным с ярко выраженным пиком у внешней мембраны клетки.

Известно, что цитостатики, дестабилизирующие микротрубочки, (такие, как колхицин, винбластин, нокодазол) увеличивают скорость реакции деполимеризации [Vasques et al., 1997; Wilson et al., 1999]. Из уравнения (3.35) следует, что увеличение константы этой реакции – $k_3$ в рамках нашей модели равносильно уменьшению значения параметра $Tu^0$. Следовательно, действие цитостатиков из **W** класса должно проявляться в виде перемещения изображающей точки влево на диаграмме состояний, например, из области **"II"** в область стационарных режимов, соответствующих зонам **"0"** и **"I"**.

Предложенную классификацию механизмов регуляции динамики микротрубочек можно применить и к ряду эндогенных белков, отвечающих за регуляцию цитоскелета в разных фазах клеточного цикла (табл. 4.1). К примеру, действие белков семейств MAP2 и Tau (табл. 4.1), аналогично таксолу вызывает уменьшение скорости элементарной реакции деполимеризации (константа $k_3$) и одновременно увеличивает скорость реакции полимеризации микротрубочек (константа $k_2$) [Drechsel et al., 1992; Itoh and Hotani, 1994]. Поэтому в рамках изложенной классификации они относятся к ранее описанному классу **E**. Активация этих белков в клетке должна приводить к стабилизации распределений микротрубочек по длине с преобладанием протяженных экземпляров.

Статмин (табл. 4.1), как было показано в ряде экспериментов *in vitro* и *in vivo* [Andersen, 2000; Rubin and Atweh, 2004], связывается с молекулами свободного тубулина в растворе и образует неспособный к полимеризации комплекс. В рамках нашего подхода взаимодействие тубулина со статмином снижает эффективную концентрацию молекул свободного тубулина, способных участвовать в реакциях формирования микротрубочек. В терминах нашей модели это соответствует уменьшению параметра $Tu^0$, что должно проявляться на диаграмме состояний в смещении изображающей точки системы влево по оси $Tu^0$. Таким образом статмин, как и колхицин и винбластин, в рамках предложенной классификации относится к веществам из **W**-класса.

Кроме белков, непосредственно взаимодействующих с тубулином, динамика ансамбля микротрубочек зависит также и от концентраций молекул ГТФ и ГДФ. Уровень последних в клетке определяется ферментными системами, ответственными за энергетический метаболизм [Janmey, 1998; Alberts et al., 2002; Caudron et al., 2005; Grigoriev et al., 2006]. Среди ряда факторов, регулирующих уровень гуанозин-фосфатов, наиболее изучены малые ГТФазы (семейства Rho, Ran, Rac) и активирующие их факторы обмена между ГТФ и ГДФ (guanine nucleotide exchange factors – GEFs) (табл. 4.1) [Schmidt and Hall, 2002; Grigoriev et al., 2006; Narumiya and Yasuda, 2006]. Активация GEF-белков приводит к увеличению концентрации ГТФ в клетке, что в рамках нашего подхода эквивалентно в силу (3.36) уменьшению значения параметра $\gamma$. Таким образом, воздействие GEF-белков должно проявляться на диаграмме состояния в виде сдвига изображающей точки системы вниз по оси $\gamma$. Следовательно, эта группа факторов относится к классу **S**. Если исходно изображающая точка находилась в области стационарных решений **"0"** или **"I"**, то значительное уменьшение параметра $\gamma$ может привести к переходу рассматриваемой системы в область нестационарных решений **"II"**. Аналогично, при исходном нахождении системы в области **"II"**, воздействие данного фактора может вызвать смещение изображающей точки на диаграмме состояний в область **"III"**, соответствующую стационарным решениям.

Действие малых ГТФаз и GAP-белков (GTPase activating proteins) приводит к гидролизу ГТФ и увеличению в цитоплазме уровня концентрации молекул ГДФ [Wilde and Zheng, 1999; Caudron et al., 2005]. Это приводит к увеличению параметра $\gamma$ и вызывает сдвиг изображающей точки системы вверх на диаграмме состояний. То есть в данном случае мы имеем дело с действием, прямо противоположным воздействию GEF-белков. Это позволяет причислить малые ГТФазы к веществам из **N** класса.

В естественных условиях вышеперечисленные клеточные факторы участвуют в регуляции динамики микротрубочек. За счет создания пространственных градиентов вышеперечисленных клеточных регуляторов имеет место образование локальных областей "стабильных" и "нестабильных" микротрубочек в клетке [Niethammer et al., 2004; Cauldron et al., 2005].

В рамках предложенного подхода зоны "**II**" и "**IV**" соотносятся с состояниями, в которых тубулиновый цитоскелет демонстрирует макроскопические трансформации. Выше отмечалось, что клетка на определенных стадиях митоза должна проходить через области нестационарного поведения. Вследствие этого понятно, что более эффективными следует считать те цитостатические препараты (либо их сочетания), которые при прочих равных условиях вызывают большее смещение изображающей точки на диаграмме состояния из областей нестационарного поведения микротрубочек.

**Таблица 4.2.** Сочетаемость препаратов согласно предложенной классификации. Знаком "+" обозначены сочетания, при которых имеет место эффект взаимного усиления воздействия препаратов. Прочерк "–" обозначает отсутствие эффекта.

|   | N | S | E | W |
|---|---|---|---|---|
| **N** | + | — | — | + |
| **S** | — | + | + | — |
| **E** | — | + | + | — |
| **W** | + | — | — | + |

Кратчайшим путем является смещение вдоль нормали к оси области "**II**" (см. рис. 3.1). Например, за счет одновременного смещения вверх и влево (что характерно для совместного действия веществ из классов **W** и **N**), или же вниз и вправо (при сочетании веществ **S** и **E** классов). Это означает, что при использовании указанных веществ в приведенных сочетаниях, требуемый цитостатический эффект будет достигаться при меньших дозировках. Откуда непосредственно следует, что сопутствующая токсичность противоопухолевых цитостатических препаратов при одновременном их применении также будет снижена. В таблице 4.2 представлены данные о совместном действии веществ, принадлежащих различным из перечисленных классов.

Необходимо отметить, что взаимное усиление (синергизм) действия различных цитостатических агентов достигается лишь в тех случаях, когда имеет место влияние одного агента на "тубулиновую" составляющую (т.е. на изменение скорости реакций полимеризации/деполимеризации), а другого – на "энергетическую", т.е. на баланс между концентрациями молекул ГТФ и ГТФ.

# ЗАКЛЮЧЕНИЕ

В связи с биологическими и медицинскими приложениями, проблема регуляции динамических трансформаций микротубулинового цитоскелета представляет большой научный и практический интерес [Jordan and Wilson, 2004; Honore et al., 2005]. К настоящему времени исследован широкий класс биохимических веществ, обладающих цитостатическим регуляторным действием [Wilson et al., 1999; Jordan, 2002]. Развитый нами подход позволил расклассифицировать регуляторные воздействия на четыре основных типа. При этом открылась возможность, опираясь на диаграмму состояний микротубулинового цитоскелета, судить об эффективности совместного действия разных регуляторов (см. табл. 4.2, глава 4).

Микротубулиновый цитоскелет клетки высокочувствителен к условиям проведения эксперимента. Наибольшие экспериментальные затруднения возникают при отыскании условий, при которых проявляется динамическая нестабильность микротрубочек [Carlier et al., 1987; Fygenson et al., 1994]. В свете проведенного анализа эти наблюдения становятся понятными. Дело в том, что области, отвечающие нестационарным режимам на построенной нами диаграмме состояний, представляют собой узкие полосы (см. рис. Z.1). Поэтому случайное попадание параметров в эти обладающие малой площадью области маловероятно.

Намного проще направленная постановка опытов по регистрации нестабильностей в динамике микротрубочек в системах *in vitro*. Достаточно задать высокую начальную концентрацию тубулина и ГТФ и дождаться возбуждения колебаний в ходе естественного понижения уровня ГТФ [Carlier et al., 1987; Melki et al., 1988]. Отметим, что снижение уровня ГТФ со временем на диаграмме состояний (рис. Z.1) соответствует движению изображающей точки сверху вниз. При этом, как несложно видеть, система микротрубочек на границе между зонами "**III**" и "**II**" должна претерпевать переход из стационарного состояния в нестационарное. Таким образом ясно, что в системах *in vitro* появление нестабильностей в динамике микротрубочек обусловлено рождением предельного цикла (см. рис. 3.3). В терминах теории катастроф, при переходе системы через границу областей "**III**" и "**II**" имеет место закритическая бифуркация Андронова-Хопфа [Андронов и др., 1959].

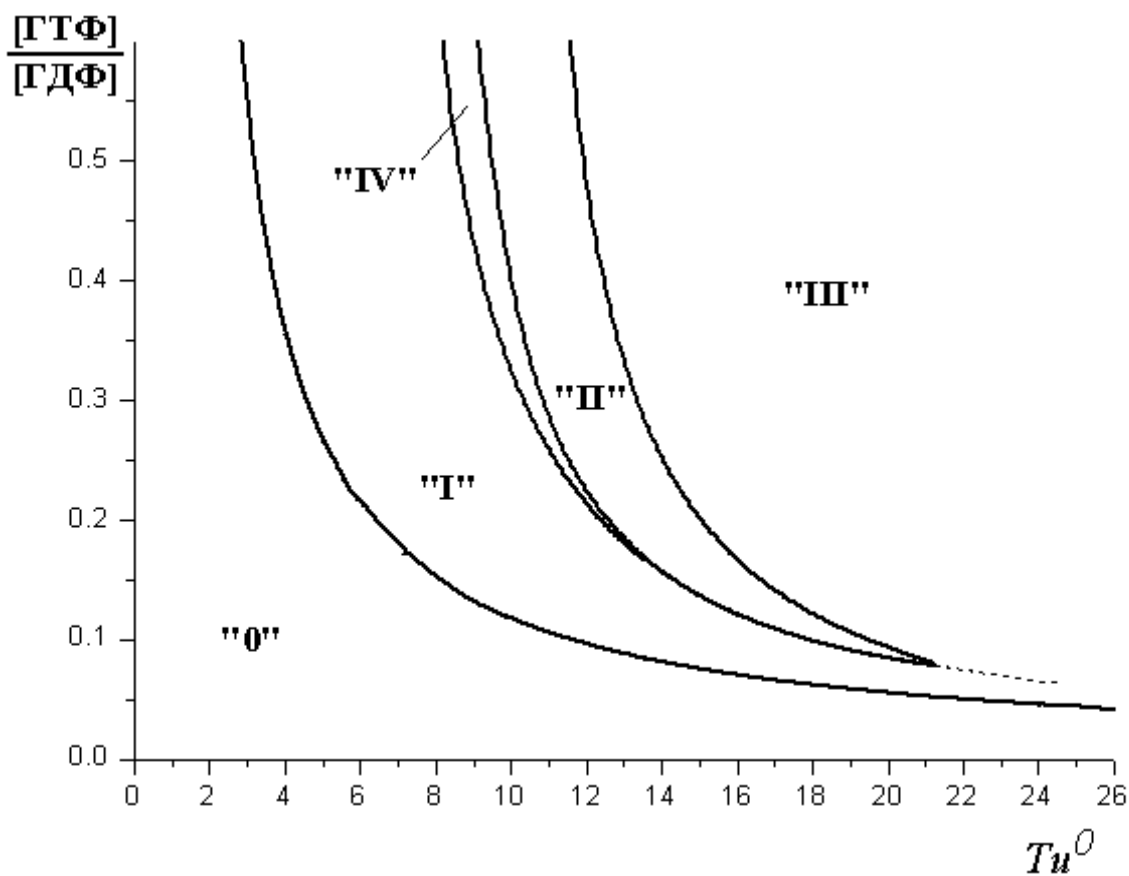

**Рис. Z.1.** Диаграмма состояний (см. рис 3.1), в которой по оси ординат отложено отношение концентрации ГТФ к концентрации ГДФ в растворе, характеризующее степень энергообеспеченности системы.

Природа нестационарных пульсаций микротрубочек, при которых стадии полимеризационного роста внезапно сменяются стадиями деполимеризации, интенсивно обсуждается более 20 лет [Mitchison and Kirschner, 1984]. Наиболее широко дебатируется гипотеза о существовании в окрестности плюс конца микротрубочки так называемой ГТФ-крышки (GTP-cap). Согласно данной гипотезе, гидролиз присоединенной к тубулину молекулы ГТФ происходит не сразу после присоединения очередного димера к микротрубочке, а спустя некоторое время [Chen and Hill, 1983; Chen and Hill, 1984; Stewart et al., 1990; Flyvbjerg et al., 1996]. Образующийся слой молекул тубулина-ГТФ в окрестности плюс-конца предотвращает деполимеризацию и стабилизирует всю микротрубочку [Bayley et al., 1989; Baylay et al., 1990; Stewart et al., 1990]. Переход от роста микротрубочки к ее деполимеризации связан с исчезновением ГТФ-крышки.

Нельзя исключить, что в определенных экспериментальных условиях задержка гидролиза ГТФ в принципе способна вызвать нестабильность в динамике микротрубочек [Carlier et al., 1987; Melki et al., 1988]. Однако проведенный выше анализ показал, что, и не прибегая к этой гипотезе (опираясь только на реакционно-диффузионные представления в рамках модели, изложенной в главе 3), удается получить решения, соответствующие динамическим нестабильностям тубулиновых микротрубочек. Попутно заметим, что в рамках нашего подхода сам собой снимается активно обсуждаемый в альтернативных подходах вопрос о виде частотного спектра случайных событий "переключения" микротрубочек с роста на деполимеризацию и обратно [Dogterom et al., 1995; Houchmandzadeh and Vallade, 1996; Hammele and Zimmermann, 2003].

В рамках развитого нами подхода динамика плюс конца микротрубочки определяется исключительно реакциями сорбции/десорбции молекул тубулина. При этом скорость реакции сорбции зависит от локальной концентрации молекул тубулина, связанных с ГТФ. В этом смысле изменение размера микротрубочки регулируется реакционно-диффузионными процессами в окружающей системе. Проведенный анализ показал, что рост микротрубочки связан с распространением концентрационных автоволн молекул тубулина в растворе. Бегущая слева направо автоволна ГТФ вызывает рост микротрубочек, а распространяющаяся справа налево волна ГДФ сопровождает деполимеризацию. По аналогии с упоминавшимися ранее диффузионно-лимитированными процессами (DLA [Vicsek, 1992]) можно сказать, что критические явления в динамике микротрубочек относятся к энергетически лимитированным процессам агрегации-дисагрегации (ELAD).

Вопрос о роли концентрационных автоволн в динамике микротрубочек, насколько нам известно, не рассматривался ранее теоретически. Однако в пользу влияния концентрационных градиентов на динамику микротрубочек *in vivo* свидетельствуют недавно опубликованные экспериментальные данные [Caudron et al., 2005; Pearson et al., 2006]. Это позволяет надеяться, что полученные нами теоретические результаты могут оказаться небезынтересными специалистам в области клеточной биологии.

Стоит заметить, что в предпринятой в данной работе попытке описания критических явлений в динамике микротрубочек развиваются общие физико-химические положения теории диссипативных структур [Николис и Пригожин, 1979]. Распределение реагентов в растворе, в котором формируются микротрубочки, описывается в рамках реакционно-диффузионного приближения. А сами микротубулиновые нити трактуются, как "преципитат" из жидкой фазы, то есть, как выпавшее в осадок вещество. В рамках такого подхода проблема роста микротрубочек по своей физико-химической сути предстает, как проблема конденсации молекул тубулина из одной фазы в другую. Несмотря на то, что построенная нами кинетическая реакционно-диффузионно-преципитационная (РДП) модель, вне всякого сомнения, является сильно упрощенной и ухватывает лишь часть из наиболее характерных черт реальной системы, на ее основе удалось построить диаграмму состояний рассматриваемой системы.

В физике переход вещества из одной формы пространственного упорядочения в другую принято рассматривать в рамках теории фазовых переходов, с термодинамических позиций [Ландау и Лившиц, 1964]. Традиционный взгляд на явления смены агрегатного состояния связаны с выяснением условий термодинамической устойчивости фаз [Gibbs, 1928]. Сосуществование двух фаз возможно лишь в случае, если на границе раздела фаз химические потенциалы равны. О спектре агрегатных состояний системы и возможных переходах между ними при изменении термодинамических параметров принято судить по виду "диаграммы состояний" [Gibbs, 1928]. Поэтому построение таких диаграмм является одной из ключевых задач в физике конденсированного состояния. Надо сказать, что диаграммы состояний широко используются и в физической химии. Например, при изучении термодинамических условий выпадения в осадок растворенных в жидкой фазе солей [Мэлвин-Хьюз, 1962]. Применительно к рассмотренной нами ELAD системе, уместно говорить о неравновесных переходах, поскольку в ходе формирования и роста микротрубочек имеет место энергопотребление – расходуются молекулы ГТФ.

Ввиду того, что тубулиновые микротрубочки способны формироваться и в реконструированных биохимических системах, нам представлялось естественным рассматривать их с позиций современной неравновесной физической химии [Ван Кампен, 1990; Пригожин и Кондепуди, 2002]. С этой точки зрения реконструированная система может рассматриваться, как состоящая их двух взаимодействующих между собой подсистем (см. рис. Z.2). В "концентрационной" подсистеме молекулы тубулина, ассоциированные с ГТФ или ГДФ, способны индивидуально перемещаться в пространстве. Напротив, в "преципитационной" подсистеме отдельные молекулы тубулина плотно упакованы в цилиндрическую 3-хзаходную спиральную структуру (см. рис. 1.2, глава 1). В этом смысле тубулиновая микротрубочка представляет собой элемент конденсированной фазы.

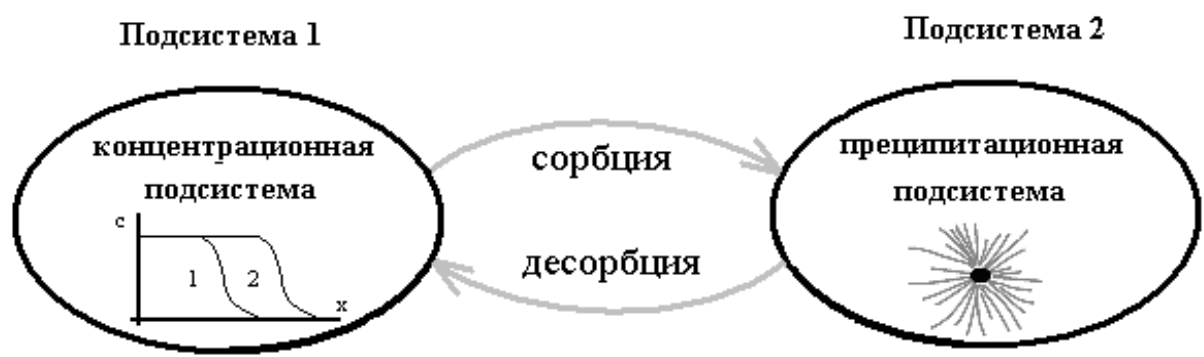

**Рис.Z.2.** Принципиальная схема рассмотренной реакционно-диффузионно-преципитационной системы.

При этом условная "граница раздела фаз" представляет собой кольцо внешним размером 24 нм и внутренним 13 нм, лежащее в торцевом сечении микротрубочки, в зоне местоположения ее плюс-конца. Понятно, что указанная "граница раздела фаз" должна быть в равновесной ситуации неподвижной (испытывать только флуктуационное дрожание), так как при этом скорость прямых и обратных процессов (сорбции/десорбции) должны быть равны.

Однако взаимное действие указанных выше подсистем в неравновесных условиях может быть, вообще говоря, и нескомпенсированным. Например, в условиях, когда доминируют процессы сорбции молекул тубулина, будет иметь место рост микротрубочки. Как следствие, будет смещаться граница раздела фаз. При доминировании процессов десорбции, трубочки будут укорачиваться, а концентрационная фаза насыщаться молекулами тубулина-ГДФ. Действие подсистем друг на друга, как выяснилось, не всегда бывает компенсаторно сбалансированным в каждый момент времени. Наряду с режимами, которым отвечают стационарные распределения микротрубочек по длинам, имеющими место в зонах "**0**", "**I**" и "**III**" (см. рис. Z.1), возможны и нестационарные режимы, которым отвечают области "**II**" и "**IV**" на параметрической диаграмме состояний (см. рис. Z.1). Как мы видели, в последнем случае поведение системы определяется наличием в фазовом пространстве предельного цикла А. Пуанкаре. Представляется важным подчеркнуть, что при этом масштабные периодические изменения длины микротрубочек оказываются сопряженными с концентрационными автоволнами в растворе. В физике широко известно, что при приближении к критическим состояниям, имеет место нарастание крупномасштабных флуктуаций и увеличение радиуса корреляции (второго корреляционного момента) [Балеску, 1978; Ma, 1980]. В рассмотренной нами неравновесной системе увеличение пространственно-временной корреляции в критических условиях обеспечивается за счет автоволновых процессов и сопряженных с ними процессов полимеризации-деполимеризации.

В определенной области параметров (см. "**II**","**IV**", рис. Z.1) имеет место своеобразная взаимная синхронизация в поведении подсистем, которая проявляется в виде взаимообусловленной пространственно-временной ритмики. Рождение предельного цикла и связанных с ним пространственно-временных форм трансформации элементов тубулинового цитоскелета идет в полном соответствии с открытым А.А. Андроновым механизмом (см. рис. 3.4а).

Таким образом, исследованная реакционно-диффузионно-преципитационная система, которая в рамках развитого подхода описывает поведение микротрубочек в условиях in vitro, позволяет эффективно трактовать широкий круг динамических нестабильностей, свойственных данной неравновесной двуфазной системе.

# ВЫВОДЫ

1. Показано, что динамические нестабильности роста микротрубочек могут развиваться как вследствие кластеризации внутренних структурных микродефектов, так и вследствие лимитирования роста за счет исчерпания в растворе молекул тубулина и/или ГТФ.

2. Построена феноменологическая модель динамических неустойчивостей, обусловленных коллективной кластеризацией микродефектов. Модель позволила объяснить степенную зависимость частоты катастроф от скорости роста микротрубочек.

3. Построена реакционно-диффузионно-преципитационная модель динамики микротрубочек. Показано, что масштабные периодические изменения длины микротрубочек сопряжены с распространением в растворе концентрационных тубулиновых автоволн.

4. Построена параметрическая диаграмма состояний тубулинового цитоскелета, содержащая области, отвечающие устойчивым стационарным и нестационарным состояниям. Выделена область, отвечающая метастабильным состояниям.

5. Показано, что потеря устойчивости стационарных состояний при изменении параметров системы, сопровождается бифуркацией рождения предельного цикла по механизму Андронова-Хопфа. Обнаружена вторичная бифуркация удвоения предельного цикла.

6. Исходя из вида диаграммы состояний, предложена классификация регулирующих динамику микротрубочек биохимических агентов (клеточных факторов и цитостатических препаратов) на четыре класса. Проанализирована проблема взаимной сочетаемости цитостатических воздействий разных классов.

# БЛАГОДАРНОСТИ

# СПИСОК ЛИТЕРАТУРЫ

1. Андронов А.А., Витт А.А., Хайкин С.Э. (1959) Теория колебаний, 2-е изд. М.: Физматгиз. – 926 с.

2. Анищенко В.С. (2000) Знакомство с нелинейной динамикой. Саратов, Из-во Гос УНЦ "Колледж". – 180 с.

3. Аносов Д.В., Арнольд В.И. (1985) Динамические системы – 1. Современные проблемы математики. Фундаментальные направления. М.: ВИНИТИ. – 260 с.

4. Балеску Р. (1978) Равновесная и неравновесная статистическая механика. В двух томах. Т.1. М.: Мир. – 350 с.

5. Буравцев В.Н. (1982) Периодический фазовый переход в растворе аммиака. // Ж. Физ. Химии. Т. 57. с.1822-1824.

6. Васильев В.А., Романовский Ю.М., Яхно В.Г. (1987) Автоволновые процессы. М.: Наука. – 240 с.

7. Волькенштейн М.В.(1988) Биофизика. М.: Наука. – 592 с.

8. Ван Кампен Н.Г. (1990) Стохастические процессы в физике и химии. М.: Высшая школа, – 375 с.

9. Воробьев И.А., Григорьев И.С. (2003) Динамика и жизненный цикл микротрубочек в клетке. Цитология и генетика. Т.37. с.22-38.

10. Воробьев И.А., Григорьев И.С., Бориси Г.Г. (2000) Динамика микротрубочек в культивируемых клетках. Онтогенез. Т. 31. с. 420-428.

11. Гегузин Я.И. (1981) Живой кристалл. М.: Наука. – 192 с.

12. Данилов Ю.А. (2001) Лекции по нелинейной динамике. Элементарное введение. М.: Постмаркет. – 184 с.

13. Ефимов Н.В., Розендорн Э.Р. (1970) Линейная алгебра и многомерная геометрия. М.: Наукаю – 528 с.

14. Жаботинский А.М. (1974) Концентрационные колебания. М.: Наука. – 120 с.

15. Канцерогенез. Ред. Д.Г. Заридзе М.: Медицина, 2004. – 576 с.

16. Катруха Е.А., Гурия Г.Т. (2006) Динамические нестабильности тубулинового цитоскелета. Диаграмма состояния. Биофизика, 2006, Т.51, с.885-893.

17. Коваленко О.В. (2005) Формирование радиальной системы микротрубочек в интерфазных клетках: роль динеина и протеинкиназы LOSK. Дисс. на соискание научной степени к.б.н., МГУ

18. Колебания и бегущие волны в химических системах. Ред. Р.Филд, М. Бургер. М.: Мир, 1988. – 720 с.

19. Колмогоров А.Н. (1941) О логарифмически-нормальном законе распределения частиц при дроблении. ДАН СССР, Т. XXXI, Вып. 2, с.99-101.

20. Кузнецов С.П. (2001) Динамический хаос. М.: Из-во физ-мат. Литературы. – 296 с.

21. Ландау Л.Д., Лившиц Е.М. (1964) Статистическая физика. М.: Наука. – 568 с.

22. Ма Ш. (1980) Современная теория критических явлений. М.: Мир, – 299 с.

23. Мелвин-Хьюз Э.-А. (1962) Физическая химия. (в двух томах) М.: Изд-во Иностран. Литературы.

24. Николис Г., Пригожин И. (1979) Самоорганизация в неравновесных системах. М.: Мир.

25. Пригожин И., Кондепуди Д. (2002) Современная термодинамика. От тепловых двигателей до диссипативных структур. М.: Мир. – 464 с.

26. Рёпке Г. (1990) Неравновесная статистическая механика. М.:Мир.

27. Трубецков Д.И. (2004) Введение в синергетику. Хаос и структуры. М. УРС – 240 с.

28. Франк-Каменецкий Д.А. (1947) Диффузия и теплопередача в химической кинетике. М.: Изд. АН СССР.

29. Френкель Я.И. (1945) Кинетическая теория жидкостей. М.-Л.: АН СССР.

30. Эбелинг В. (1979) Образование структур при необратимых процессах. Введение в теорию диссипативных структур. М.: Мир. – 279 с.

31. Alberts B., Johnson A., Lewis J., Raff M., Roberts K., Walter P. (2002) Molecular Biology of Cell. New York: Garland Science.

32. Alieva I.B., Vorobjev I.A. (2000) Interphase microtubules in cultured cells: long or short? Membr. Cell Biol., V.14 P. 57-67.

33. Amos L.A., van den Ent F., Lowe J. (2004) Structural/functional homology between the bacterial and eukaryotic cytoskeletons. Curr Opin Cell Biol. V. 16 P.24-31.

34. Andersen S.S.L. (2000) Spindle assembly and the art of regulating microtubule dynamics by MAPs and Stathmin/Op18. Trends Cell Biol. V. 10 P. 261-267.

35. Anesti V., Scorrano L. (2006) The relation between mitochondrial shape and function and cytoskeleton. Biochim. Biophys Acta. V.1757. P.692-699.

36. Arnal I., Wade R.H. (1995) How does taxol stabilize microtubules? Curr. Biol., V.5. P.900-908.

37. Bak P. (1996) How nature works. Copernicus Press, New York.

38. Bayley P.M., Schilstra M.J., Martin S.R. (1989) A lateral cap model of microtubule dynamic instability. FEBS Lett. V. 259, P.181-184.

39. Bayley P.M., Schilstra M.J., Martin S.R. (1990) Microtubule dynamic instability: numerical simulation of microtubule transition properties using a Lateral Cap model. J. Cell Sci. V.95 P.33-48.

40. Bergen L.J., Borisy G.G. (1980) Head-to-tail polymerization of microtubules in vitro. Electron microscopy analysis of seeded assembly. J. Cell. Biol. V.84, P.141-150

41. Bolterauer H., Limbach H.J., Tuszynski J.A. (1999) Models of assembly and dissasembly of individual microtubules: stochastic and averages equations. J. Biol. Phys. V.25 P.1-22.

42. Borisy G.G., Olmsted J.B. (1972) Nucleated assembly of microtubules in porcine brain extracts. Science V.177. P.1196-1197.

43. Brandt R., Lee G. (1994) Orientation, assembly, and stability of microtubule bundles induced by a fragment of tau protein. Cell Motil. Cytoskeleton. V. 28(2) P.143-154.

44. Caplow M., Fee L. (2003) Concerning the chemical nature of tubulin subunits that cap and stabilize microtubules. Biochemistry V.42 P.2122-2126.

45. Carlier M.F., Hill T.L., Chen Y. (1984) Interference of GTP hydrolysis in the mechanism of microtubule assembly: An experimental study. PNAS, V.81, P.771-775.

46. Carlier M.F., Melki R., Pantaloni D., Hill T.L., Chen Y. (1987) Synchronous oscillations in microtubule polymerization. PNAS, V.84, P.5257-5261.

47. Carlier M.F., Pantaloni D. (1981) Kinetic analysis of guanosine 5'-triphosphate hydrolysis associated with tubulin polymerization. Biochemistry. V. 20 P. 1918–1924.

48. Carlier M.F. (1982) Guanosine-5'-triphosphate hydrolysis and tubulin polymerization. Review article. Mol/ Cell Biochem. V. 47 P. 97–113.

49. Carlson J.F. (1952) Microdissection studies of the dividing neuroblast of. the grasshopper, Chortophaga viridifasciata (De Greer). Chromosoma (Berlin). V.5 P.200-220.

50. Cassimeris L. (1999) Accessory protein regulation of microtubule dynamics throughout the cell cycle. Curr. Opin. Cell Biol. V. 11 P.134–141.

51. Cassimeris L., Pryer N.K., Salmon E.D. (1988) Real-time observation of microtubule dynamic instability in living cells. J. Cell Biol. V.107. P.2223-2231.

52. Caudron M., Bunt G., Bastiaens P., Karsenti E. (2005) Spatial coordination of spindle assembly by chromosome-mediated signaling gradients. Science, V. 309. P.1373-1376.

53. Caudron N., Valiron O., Usson Y., Valiron P., Job D. (2000) A reassessment of the factors affecting microtubule assembly and disassembly in vitro. J.Mol.Biol. V. 297. P.211-220.

54. Caviston J.P., Holzbaur E.L. (2006) Microtubule motors at the intersection of trafficking and transport. Trends Cell Biol. V.16(10) P.530-537.

55. Chen Y., Hill T.L. (1983) Use of Monte Carlo calculations in the study of microtubule subunit kinetics. PNAS. V.80. P.7520-7523.

56. Chen Y., Hill T.L. (1984) Theoretical treatment of microtubules disappearing in solution. PNAS. V.82. P.4127-4131.

57. Chen Y., Hill T.L. (1987) Theoretical studies on oscillations in microtubule polymerization. PNAS. V.84. P.9418-8423.

58. Chretien D., Metoz F., Verde F., Karsenti E., Wade R.H. (1992) Lattice defects in microtubules: protofilament numbers vary within individual microtubules. J Cell Biol. V.117(5) P.1031-1040.

59. Chrétien D., Fuller S.D., Karsenti E. (1995) Structure of growing microtubule ends: two dimensional sheets close into tubes at variable rates. J Cell Biol V. 129 P.1311–1328.

60. Davis L.J., Odde D.J., Block S.M., Gross S.P. (2002) The importance of lattice defects in katanin-mediated microtubule severing in vitro. Biophys. J. V. 82. P.2916-2927.

61. Desai A., Mitchison T.J. (1997) Microtubule polymerization dynamics. Annu. Rev. Cell Dev. Biol., V. 13, p.83-117.

62. Diaz J.F., Valpuesta J.M., Chacon P., Diakun G., Andreu J.M. (1998) Changes in microtubule protofilament number induced by taxol binding to an easily accessible site. J. Biol. Chem. V. 273 P.33803–33810.

63. Dogterom M., Maggs A.C., Leibler S. (1995) Diffusion and formation of microtubule asters: physical processes versus biochemical regulation. PNAS V.92 P.6683-6688.

64. Dogterom M., Leibler S. (1993) Physical aspects of the growth and regulation of microtubule structures. Phys. Rev. Lett. V. 70. P.1347-1350.

65. Drechsel D.N., Kirschner M.W. (1994) The minimum GTP cap required to stabilize microtubules. Current Biology, V. 4 P.1053-1061.

66. Drechsel D.N., Hyman A.A., Cobb M.H., Kirschner M.W. (1992) Modulation of the dynamic instability of tubulin assembly by the microtubule associated protein tau. Mol Biol Cell V.3, P.1141-1154.

67. Dustin P. (1984) Microtubules. 2nd edn. New York: Springer.

68. Erickson H.P., O'Brien E.T. (1992) Microtubule dynamic instability and GTP hydrolysis. Ann. Rev. Bioph. Biomol. Struct. Vol. 21, P.145-166.

69. Ermentrout C.B., Edelstein-Keshet L. (1993) Cellular automata approaches to biological modeling. J.theor.Biol., V. 160, P.97-133.

70. Eyring H. (1935) The activated complex in chemical reactions. J. Chem. Phys. V.3. P.107-115.

71. Feigenbaum M.J. (1978) Quantitative universality for a class of nonlinear transformations.J. Stat. Phys., V.19, N 1, P.25-52.

72. Flyvbjerg H., Holy T.E., Leibler S. (1994) Stochastic dynamics of microtubules: a model for caps and catastrophes. Phys. Rev. Lett, V.73, P.2372-2375.

73. Flyvbjerg H., Holy T.E., Leibler S. (1996) Microtubule dynamics: caps, catastrophes, and coupled hydrolysis. Phys. Rev. E, V.54, P.5538-5560.

74. Fokker A.D. (1914) Ann. Physik V.43 P.810

75. Fygenson D.K., Braun E., Libchaber A. (1994) Phase diagram of microtubules. Phys. Rev. E. V.50. P.1579-1588.

76. Gibbs J.W. (1928) The Collected Works. Thermodynamics. New York: Longmans, Green and Co., V. 1.

77. Grigoriev I.S., Chernobelskaya A.A., Vorobjev I.A. (1999) Nocodazole, vinblasine and taxol at low concentrations affect fibroblast locomotion and salutatory movements of organelles. Membr. Cell. Biol. V.13 P.23-48.

78. Grigoriev I., Borisy G., Vorobjev I. (2006) Regulation of microtubule dynamics in 3T3 fibroblasts by Rho family GTPases. Cell Motil. Cytoskel. 2006. V. 63. P. 29-40.

79. Gurret J.-P. (1995) Modelling the mitotic apparatus: from the discovery of bipolar spindle to modern concepts. Acta Biotheor. V. 43, P. 127-142.

80. Haken H. (1977) Synergetics, an introduction. Nonequilibrium Phase-Transitions and Self-Organization in Physics, Chemistry and Biology. Springer. (русский перевод: Хакен Г. Синергетика, М.: Мир, 1980.)

81. Hammele M., Zimmermann W. (2003) Modeling oscillatory microtubule polymerization. Phys. Rev. E. V. 67. P.21903-19

82. Hill T.L. (1984) Introductory analysis of the GTP-cap phase-change kinetics at the end of a microtubule. V.81 P.6728-6732.

83. Hill T.L., Carlier M.F. (1983) Steady-state theory of the interference of GTP hydrolysis in the mechanism of microtubule assembly. V.80 P.7234-7238.

84. Hill T.L., Chen Y. (1984) Phase changes at the end of a microtubule with GTP cap. V.81 P.5772-5776.

85. Hirokawa N. (1994) Microtubule organization and dynamics dependent on microtubule-associated proteins. Curr. Opin. Cell Biol. V.6 P.74–81.

86. Holmfeldt P., Brattsand G., Gullberg M. (2002) MAP4 counteracts microtubule catastrophe promotion but not tubulin-sequestering activity in intact cells. Curr Biol. V. 12(12) P. 1034-1039.

87. Holy T.E., Leibler S. (1994) Dynamic instability of microtubules as an efficient way to search in space. PNAS, V.91 P.5682-5685.

88. Honore S., Pasquier E., Braguer D. (2005) Understanding microtubule dynamics for improved cancer therapy. Cell. Mol. Life Sci., V. 62. P. 3039-3056.

89. Houchmandzadeh B., Vallade M. (1996) Collective oscillations in micro-tubule growth. Phys. Rev. E. 1996. V. 53. P.6320-6324.

90. Howard J., Hyman A.A. (2003) Dynamics and mechanics of the microtubule plus end. Nature, V. 422, P.752-758.

91. Howard J., Hyman A.A. (2007) Microtubule polymerases and depolymerases. Curr. Opin. Cell Biol, V. 19, P.31-35.

92. Inoue S. (1960) On the physical properties of the mitotic spindle. Ann. N. Y. Acad. Sci. V. 90, P. 529-530.

93. Inoue S., Ritter H. (1975) Dynamics of mitotic spindle organization and function. in "Molecules and Cells Movement" (eds. Inoue S., Stephens R.E.) Raven Press, New York. P.3-30.

94. Inoue S., Sato H. (1967) Cell motility by labile association of molecules. The nature of the mitotic spindle fibers and their role in chromosome movement. J. Gen. Physiol. V. 50 P. 259-292.

95. Inoue S. (1981) Cell division and the mitotic spindle. J. Cell Biol. V. 97. P. 131-147.

96. Inoue, S. (1952) The effect of colchicine on the microscopic and submicroscopic structure of the mitotic spindle. Expl. Cell Res. (Suppl.) V. 2 P.305-319.

97. Inoue, S. (1959) Motility of cilia and the mechanism of mitosis. Rev. mod. Phys. V. 31 P. 402-408.

98. Itoh T.J., Hotani H. (1994) Microtubule-stabilizing activity of Microtubule-Associated Proteins (MAPs) is due to increase in frequency of rescue in dynamic instability: shortening length decreases with binding of MAPs onto microtubules. Cell Struct. And Func.V.19. P.279-290.

99. Janmey P.A. (1998) The cytoskeleton and cell signaling: component localization and mechanical coupling. Physiol. Rev. V.78 P.763-781.

100. Janosi I.M., Chretien D., Flyvbjerg H. (1998) Modeling elastic properties of microtubule tips and walls. Eur Biophys J, V. 27 P.501–513.

101. Janosi I.M., Chretien D., Flyvbjerg H. (2002) Structural microtubule cap: stability, catastrophe, rescue, and third state. Biophys J., V. 83, P.1317-1330.

102. Janson M.E., Dogterom M. (2004) A bending mode analysis for growing microtubules. Biophysical J. V.87, P.2723-2736.

103. Janulevicius A., Van Pelt J., Van Ooyen A. (2006) Compartment volume influences Microtubule dynamic instability: a model study. Biophysical J., V.90, p.788-798.

104. Jordan M.A. (2002) Mechanism of action of antitumor drugs that interact with microtubules and tubulin. Curr. Med. Chem. Anti-Canc. Agents V.2. P. 1–17

105. Jobs D., Valiron O., Oakley B. (2003) Microtubule nucleation. Curr.Opin.Cell Biol., V.15, P.111-117.

106. Jobs E., Wolf D.E., Flyvbjerg H. (1997) Modeling microtubule oscillations Phys.Rev.Lett. V.79, P. 519-522.

107. Karsenti E., Vernos I. (2001) The mitotic spindle: a self-made machine. Science. V. 294. P.543-547.

108. Koch A.J., Meinhardt H. (1994) Biological pattern formation - from basic mechanisms to complex structures. Rev. Modern Physics. V. 66. P.1481-1507.

109. Komarova Y.A., Vorobjev I.A., Borisy G.G. (2002) Life cycle of MTs: persistent growth in the cell interior, asymmetric transition frequencies and effects of the cell boundary. J Cell Sci. 2002 V.115 P.3527-3539.

110. Koshland D.E., Mitchison T.J., Kirschner M.W. (1988) Polewards chromosome movement driven by microtubule depolymerization in vitro. Nature V.331, P. 499-504.

111. Kowalski R.J., Williams R.C. (1993) Microtubule-associated protein 2 alters the dynamic properties of microtubule assembly and disassembly. J. Biol Chem. V. 268 P. 9847-9855.

112. Li H., DeRosier D.J., Nicholson W.V., Nogales E., Downing K.H (2002) Microtubule structure at 8 A resolution. Structure. V. 10. P.1317-1328.

113. Maddox P., Chin E., Mallavarapu A., Yeh E., Salmon E.D., Bloom K. (1999) Astral microtubule and spindle dynamics in mating and the first zygotic division in the budding yeast Saccharomyces cerevisiae. J. Cell Biol. V.144, P.977-987.

114. Maly I.V. (2002) Diffusion approximation of the stochastic process of microtubule assembly. Bull. Math. Biol. V.64. P.213-238.

115. Mandelbrot B.B. (2004) Fractals and Chaos. The Mandelbrot Set and Beyond. Springer, 308 P.

116. Mandelkow E.M., Lange G., Jagla A., Spann U., Mandelkow E. (1988) Dynamics of the microtubule oscillator: role of nucleotides and tubulin -MAP interactions. EMBO J. V.7 P.357-365.

117. Margolis R.L., Wilson L. (1978) Opposite end assembly and disassembly of microtubules at steady state in vitro. Cell, V. 13. P. 1-8.

118. McNally F.J. (1996) Modulation of microtubule dynamics during the cell cycle. Curr. Opin. Cell Biol. V.8 P.23–29

119. Melki R., Carlier M.F., Pantaloni D. (1988) Oscillations in microtubule polymerization: the rate of GTP regeneration on tubulin controls the period. EMBO J., V.7, P. 2653-2659.

120. Melki R., Carlier M.F., Pantaloni D. (1990) Direct evidence for GTP and GDP-Pi intermediates in microtubule assembly. Biochemistry V.29, P.8921-8932.

121. Mitchison T.J., Salmon E.D. (2001) Mitosis: a history of division. Nature Cell Biology, V.3, P. E17-E21.

122. Mitchison T.J., Kirschner M.W. (1984) Dynamic instability of microtubule growth. Nature, V. 312. P.237-242.

123. Molodtsov M.I., Ermakova E.A., Shnol E.E., Grishuk E.L., McIntosh J.R., Ataullakhanov F.I. (2005) A molecular-mechanical model of microtubule. Biophysical J. V.88 P.3167-3179.

124. Moritz M., Braunfeld M.B., Sedat J.W., Alberts B., Agard D.A. (1995) Microtubule nucleation by gamma-tubulin-containing rings in the centrosome. Nature, V. 378, P.638-640.

125. Muller-Reichert T., Chretien D., Severin F., Hyman A.A. (1998) Structural changes at microtubule ends accompanying hydrolysis: Information from slowly hydrolysable analogue of GTP, guanylyn (α, β) methylenediphosphonate. PNAS, V.95. P.3661-3666.

126. Narumiya S., Yasuda S. (2002) Rho GTPases in animal cell mitosis. Curr Opin Cell Biol. V.18. P.199-205.

127. Nguyen et al., (1999) Microtubule associated protein 4 (MAP4) regulates assembly, protomer-polymer partitioning and synthesis of tubulin in cultured cells. J. Cell Sci. V.112, P.1813-1824.

128. Niethammer P., Bastiaens P., Karsenti E. (2004) Stathmin-tubulin interaction gradients in motile and mitotic cells. Science. V. 303. P. 1862-1866.

129. Nogales E. (2001) Structural insights into microtubule function. Annu. Rev. Biophys. Biomol. Struct. V.30 P.397-420.

130. Nogales E., Wolf S., Downing K.H. (1998) Structure of the alfa-beta tubulin dimer by electron crystallography. Nature V.391, P.199-203.

131. Nucleation theory and applications. (1999) edited by J. W. P. Schmelzer, G. Ropke, V. R. Priezzhev, JINR, Dubna.

132. O'Brien E.T., Voter W.A., Erickson H.P. (1987) GTP hydrolysis during microtubule assembly. Biochemistry V.26(13) P.4148-4156.

133. O'Brien E.T., Salmon E.D., Walker R.A., Erickson H.P. (1990) Effects of magnesium on the dynamic instability of individual microtubules. Biochemistry, V. 29, P.6648-6656.

134. Odde D.J., Cassimeris L., Buetten H.M. (1995) Kinetics of microtubule catastrophe assessed by probabilistic analysis. Biophys. J. V.69. P.796-802.

135. Odde D.J. (1997) Estimation of the diffusion-limited rate of microtubule assembly. Biophys J. V.73(1) P:88-96.

136. Olmsted J.B., Borisy G.G. (1975). Ionic and nucleotide requirements for microtubule polymerization in vitro. Biochemistry V. 14. P.2996-3005.

137. Oosawa F., Asakura S. (1975) Thermodynamics of the polymerization of protein. New York, Academic Press.

138. Wilson L., Panda D., Jordan M.A. (1999) Modulation of microtubule dynamics by drugs: a paradigm for the actions of cellular regulator. Cell Struct. Function V.24. P.329-335.

139. Panda D., Miller H.P., Wilson L. (2002) Determination of the size and chemical nature of the stabilizing "cap" at microtubule ends using modulators of polymerization dynamics. Biochemistry V.41 P.1609-1617.

140. Pearson C.G., Gardner M.K., Paliulis L.V., Salmon E.D., Odde D.J., Bloom K. (2006) Measuring nanometer scale gradients in spindle microtubule dynamics using model convolution microscopy. Mol Biol Cell.V.17. P.4069-4079.

141. Pedigo S., Williams R.C. (2002) Concentration Dependence of Variability in Growth Rates of Microtubules. Biophysical J. V. 83 P.1809–1819

142. Plank M. (1917) Sitzungsber. Preuss Acad. Wissens. P.324.

143. Potapova T.A., Daum J.R., Pittman B.D., Hudson J.R., Jones T.N., Satinover D.L., Stukenberg P.T., Gorbsky G.J. (2006) The reversibility of mitotic exit in vertebrate cells. Nature. V. 440. P.954-958.

144. Press W.H., Teukolsky S.A., Vetterling W.T., Flannery B.P. (2002) Numerical recipes in C. Cambridge University Press. 995 P.

145. Pryer N.K., Walker R.A., Skeen V.P., Bourns B.D., Soboeiro M.F., Salmon E.D. (1992) Microtubule-associated protein 2 alters the dynamic properties of microtubule assembly and disassembly. J. Cell Sci. V.103. P.965-976.

146. Ross J., Muller S., Vidal C. (1988) Chemical waves. Science. V. 240. P.460-465.

147. Rubin C.I., Atweh G.F.(2004) The role of stathmin in the regulation of the cell cycle. J Cell Biochem. 2004, V.93. P. 242-250.

148. Salmon E.D., Saxton W.M., Leslie R.J., Karow M.L., McIntosh J.R. (1984) Diffusion coefficient of fluorescein-labeled tubulin in the cytoplasm of embryonic cells of a sea urchin: video image analysis of fluorescence redistribution after photobleaching. J. Cell Biol. V.99. P.2157-2164.

149. Sammak P.J., Borisy G.G. (1988) Direct observation of microtubule dynamics in living cells. Nature. V. 331 P.724-726.

150. Schaap I.A., de Pablo P.J., Schmidt C.F. (2004) Resolving the molecular structure of microtubules under physiological conditions with scanning force microscopy. Eur. Biophys. J., V.33(5). P.462-467.

151. Schilstra M.J, Martin S.R., Bayley P.M. (1987) On the relationship between nucleotide hydrolysis and microtubule assembly: studies with a GTP-regenerating system. Biochem. Biophys. Res. Commun. V.147 P.588-595.

152. Schmidt A., Hall A. (2002) Guanine nucleotide exchange factors for Rho GTPases: turning on the switch. Genes and Development, V. 16, P.1587-1609.

153. Scholey J.M., Brust-Mascher I., Mogilner A. (2003) Cell division. Nature. V.422. P.746-752.

154. Semenov M.V. (1996) New concept of microtubule dynamics and microtubule motor movement and new model of chromosome movement in mitosis. J. theor. Biol. V.179, P.91-117.

155. Sept D., Baker N.A., MacCammon J.A. (2003) The physical basis of microtubule structure and stability. Protein Sci. V.12. P.2257-2261.

156. Sept D., Limbach H.J., Bolterauer, Tuszynski J.A. (1999) A chemical kinetics model for microtubule oscillations. J. Theor. Biol. V.197. P.77-88.

157. Seybold P.G., Kier L.B., Cheng C.K. (1998) Stochastic cellular automata models of molecular excited-state dynamics. J. Phys.Chem. V. 102. P.886-891.
158. Shelanski M.L., Gaskin F., Cantor C.R. (1973) Microtubule assembly in the absence of added nucleotides V.70, P.765-768
159. Slautterback D.B. (1963) Cytoplasmic microtubules. J. Cell Biol. V. 18 P.367-388.
160. Stamenovic D., Mijailovich S.M., Tolic-Norrelykke I.M., Chen J., Wang N. (2002) Cell prestress. II. Contribution of microtubules Am. J. Physiol. Cell Physiol. V. 282 P. C617–C624.
161. Stephens R.E. (1971) Microtubules. in "Subunits and Biological Systems" (eds. S.N. Timasheff and G.D.Fasman) Marcel Dekker Inc., New York. P. 355-391.
162. Stewart R.J., Farrell K.W., Wilson L. (1990) Role of GTP hydrolysis in microtubule polymerization: evidence for a coupled hydrolysis mechanism. Biochemistry V.29, p.6489-6498.
163. Thom R. (1982) Mathematical models of morphogenesis. New York, Halsted Press. (русский перевод: Том Р. (2006) Математические модели морфогенеза. М.: РХД, 2006. 136 с.)
164. Tran P.R., R.A.Walker, E.D.Salmon (1997) A metastable intermediate state of microtubule dynamic instability that differs significantly between plus and minus ends. J.Cell Biol. V. 138, P.105-117.
165. Turing A.M. (1952) The chemical basis of morphogenesis. Phil. Trans. R. Soc. London. V.237. P.37-72.
166. Valiron O., Caudron N., Job D. (2001) Microtubule dynamics. Cell. Mol. Life Sci. V.58. P.2069-2084.

167. VanBuren V., Cassimeris L., Odde D.J. (2005) Mechanochemical model of microtubule structure and self-assembly kinetics. Biophysical J. V.89. P.2911-2926.

168. VanBuren V., Odde D.J., Cassimeris L. (2002) Estimates of lateral and longitudinal bond energies within the microtubule lattice. PNAS V.99 P.6035-6040.

169. Vasques R.J., Howell B., Yvon A.C., Wadsworth P., Cassimeris L. (1997) Nanomolar concentration of nocodazole alter microtubule dynamic instability in vivo and in vitro. Mol. Biol. Cell V.8. P.973-985.

170. Verde F., Dogterom M., Stelzer E., Karsenti E., Leibler S. (1992) Control of microtubule dynamics and length by cyclin-A and cyclin-B dependent kinases in Xenopus egg extracts. V. 118 P. 1097-1108.

171. Vicsek T. (1992) Fractal growth phenomena. World Scientific. 400 P.

172. Vorobjev I.A., Rodionov V.I., Maly I.V., Borisy G.G. (1999) Contribution of plus and minus end pathways to microtubule turnover. J. Cell Sci. V. 112 P.2277–2289.

173. Vorobjev I.A., Svitkina T.M., Borisy G.G. (1997) Cytoplasmic assembly of microtubules in cultured cells. J.Cell Sci. V.110 P. 2635-2645.

174. Voter W.A., O'Brien E.T., Erickson H.P. (1991) Dilution-induced disassembly of microtubules: relation to dynamic instability and the GTP cap. Cell Motil. Cytoskeleton. V. 18 P.55-62.

175. Walker R.A., O'Brien E.T., Pryer N.K., Soboeiro M., Voter W.A., Erickson H.P., Salmon E.D. (1988) Dynamic instability of individual microtubules analyzed by video light microscopy: rate constants and transition frequencies. J.Cell.Biol, V. 107, P.1437-1448.

176. Walker R.A., Pryer N.K., Salmon E.D. (1991) Dilution of individual microtubules observed in real-time in vitro: evidence that cap size is small and independent of elongation rate. J. Cell. Biol.V.107, P.1437-1448.

177. Wang H.W., Nogales E. (2005) Nucleotide-dependent bending flexibility of tubulin regulates microtubule assembly. Nature. V.435. P.911-915.

178. Watanabe T., Noritake J., Kaibuchi K. (2005) Regulation of microtubules in cell migration. Trends Cell Biol. V. 15(2). P. 76-83.

179. Weisenberg R. (1972) Microtubule formation in vitro in solutions containing low calcium concentrations. Science, V. 177 P. 1104-1105.

180. Wilde A., Zheng Y. (1999) Stimulation of microtubule aster formation and Spindle assembly by the small GTPase Ran. Science, V.284, P.1359-1362

181. Wolfram S. (1986) Theory and applications of cellular automata. Singapore: World Publishing Co.

182. Wolfram S. (2002) A new kind of science. Wolfram Media 1192 P.

183. Wolpert L. (1998) Principles of development. Oxford. University Press, 1998.

184. Yu R.X., Holmgren E. (2007) Endpoints for agents that slow tumor growth. Contemporary Clinical Trials, V.28. P.18-24.

# ПРИЛОЖЕНИЕ I

## § I.1. Описание компьютерного моделирования динамики димеров тубулина и вакансий

В рамках проведенного компьютерного моделирования микротрубочка представляла собой двумерный массив $\{i,j\}$, где $i \in \{1..13\}$ соответствует номеру протофиламента в микротрубочке, а $j \in \{1..M\}$, где величина M способна увеличиваться, либо уменьшаться по ходу выполнения программы. Каждый из элементов массива содержал число, обозначающее, в каком состоянии находится данный локус, то есть что за частица находится в месте с координатами $\{i,j\}$. Всего использовалось три состояния: тубулиновый димер, пограничный элемент либо вакансия (дефект).

На каждом шаге по времени для каждого из процессов (см. § 2.1 главы 2) определялась вероятность того, что он произойдет с элементами массива. Далее генератор сообщал случайное число в диапазоне от 0 до 1 и если оно оказывалось меньше вычисленной вероятности, то процесс считался прошедшим над данной частицей и она либо переходила в другое состояние, либо перемещалась по массиву. Элементы, для которых $i = 1$ полагались постоянно находящимися в состоянии «тубулиновых димеров», выступая в качестве нуклеационной затравки.

Для каждого локуса $\{i,j\}$ существовало два продольных соседа (соответственно $j-1$ и $j+1$) и, с учетом продольного сдвига протофиламентов относительно друг друга – четыре поперечных (или, что то же самое – боковых, находящихся в смежных протофиламентах). Микротрубочка полагалась замкнутым цилиндром, поэтому для элементов, у которых $i=1$, поперечными продольными соседями оказывались элементы с $i=2$ и $i=13$, тогда как у элементов с координатой $i=13$, соответственно, $i=12$ и $i=1$.

Общая вероятность процесса сорбции, который происходит с элементами, находящимися на растущем краю массива, обозначается через $p_{sorb}$. Ее значение для каждого конкретного элемента на краю меняется в зависимости от количества соседних поперечных элементов, которые заняты тубулиновыми димерами. Вклад оценивался исходя из формулы:

$$p_{sorb} = p_{sorb}^0 (p_{sorb}^{\parallel} + p_{sorb}^{=} \frac{n_{=}}{4}) \qquad (I.1)$$

где $n_{=}$ – количество соседних с данным поперечных элементов, занятых тубулиновыми димерами, $p_{sorb}^{=}$ – доля вклада в вероятность, осуществляемая их присутствием, $p_{sorb}^{\parallel}$ - доля вклада в вероятность при отсутствии поперечных соседей и $p_{sorb}^0$ – общая вероятность сорбции. Изменение $p_{sorb}^0$ соответствовало изменению концентрации тубулина в растворе.

Вероятность образования вакансии на месте тубулинового димера в объеме микротрубочки вычислялась согласно формуле:

$$p_{hole} = p_{hole}^0 (p_{hole}^{\parallel} \frac{m_{=}}{2} + p_{hole}^{=} \frac{m_{=}}{4}) \qquad (I.2)$$

где $m_{\parallel}, m_{=}$ – количество соседних с данным соответственно продольных и поперечных элементов, занятых вакансиями, $p_{hole}^{\parallel}, p_{hole}^{=}$ – доли вклада в вероятность образования вакансии, обусловленные наличием соседних продольных и поперечных вакансий соответственно и $p_{hole}^0$ – общая вероятность образования дефекта.

Вероятность вставки тубулинового димера из раствора на место, занятое вакансией, была пропорциональна вероятности сорбции с коэффициэнтом $k_{ins}$, то есть, концентрации тубулина в растворе.

Вероятность перескока дырки в одну из сторон определялась формулой:

$$p_{mov} = p_{mov}^0 p_E p_S \qquad (I.3)$$

где $p_{mov}^{0}$ – общая вероятность перескока, $p_E$ – энергетический вклад в вероятность и $p_S$ - энтропийный. При этом $p_E \sim \exp\{-E_{sum}/kT\}$, где $E_{sum} = \sum_{i,j} E_r(m_{i,j}, r)$, сумма вкладов в энергию вакансии от соседних $m_{i,j}$ вакансий, находящихся на расстоянии r. Полагалось, что функция $E_r$ имеет вид параболы. Величина энтропийного вклада $p_S \sim \exp\{-S/k\}$ рассчитывалась исходя из изменения энтропии вакансии при ее вероятном перескоке, согласно формуле (2.4) главы 2.

Одновременно работало два алгоритма проверки выполнения критических условий катастрофы. Первый, локальный, отслеживал отношение числа вакансий к числу димеров, находящихся в пяти соседних поперечных слоях. Если оно превышало критическое значение $R_{loc}$, то этот фрагмент деполимеризовался до места скопления вакансий, а его дальнейшая судьба нас не интересовала. Второй (интегральный, соответствующий второму сценарию §2.1 главы 2) через каждую секунду виртуального времени определял степень «разрыхления» всей микротрубочки, т.е. отношение общего числа дефектов к общему количеству полимеризованных молекул тубулина. Если эта величина превышала критическое значение $R_{int}$, то происходила деполимеризация до тех пор, пока отношение числа дефектов к числу тубулиновых димеров не становилось меньше критического значения.

Характерные величины параметров, использованные при моделировании процессов, представлены в таблице I.1.

Время, за которое осуществляется один полный перебор всех элементов массива, полагалось равным одной виртуальной секунде. Отметим, что последняя не равна секунде реального времени. Виртуальная секунда была отнормирована с реальным, «физическим» временем. В качестве нормировочной величины была использована средняя скорость роста микротрубочки, то есть движение переднего края массива.

Пересчет производился следующим образом. Каждый димер увеличивает длину микротрубочки на dx = 8 нм/13 = 0.6 нм. Если скорость роста равна, допустим, 0,6 мкм/мин = 10 нм/с, значит, должно присоединяться $v$ = 10/dx ~ 16,6 молекул тубулина в секунду. За одну прогонку сканируется все 13 строчек массива, поэтому если положить две виртуальные секунды равными одной эффективной, вероятность присоединения отдельного кирпича будет равна $p_s = v/(n*13)$ = 16.6/(2*13) ~ 0,64, где $n$ – число виртуальных секунд, равных одной «реальной».

**Таблица.I.1** Вероятности характерных процессов частиц системы в рамках компьютерной модели.

|   | Обозначение | Вероятность процесса (частица / вирт. секунда) | Диапазон изменений |
|---|---|---|---|
| **сорбция** | $p_{sorb}^0$ | 0.8 | 0.2 – 1 |
| продольная часть | $p_{sorb}^{\parallel}$ | 0.25 | |
| поперечная часть | $p_{sorb}^{=}$ | 0.75 | |
| **образование вакансий** | $p_{hole}^0$ | $15*10^{-3}$ | $(2 – 60)*10^{-3}$ |
| продольная часть | $p_{hole}^{\parallel}$ | 0.8 | |
| поперечная часть | $p_{hole}^{=}$ | 0.2 | |
| **вставка** коэф. пропорцион. сорбции | $k_{ins}$ | 0.08 | 0.03 – 0.1 |
| **движение дырок** | $p_{mov}^0$ | 0.5 | 0.2 – 0.7 |
| энергетический вклад | $p_E$ | 0.3 | 0.1 – 1 |
| энтропийный вклад | $p_S$ | 0.7 | 0.1 – 1 |
| **крит. коэффициенты** локальный | $R_{loc}$ | 0.7 | 0.5 – 0.9 |
| интегральный | $R_{int}$ | 0.07 | 0.02 – 0.1 |

Следует оговорить, что <$V_{gr}$> обсчитывалась отдельно для каждого компьютерного эксперимента, как средняя скорость перемещения края микротрубочки. При вычислении спектра катастроф (см. рис. 2.3, глава 2), определение значений параметров программы осуществлялось следующим образом. Выбирались две некоторые отправные калибровочные точки с известными из эксперимента значениями скорости роста и частоты катастроф. Параметры модели (см. табл. I.1) подбирались таким образом, что они демонстрировали приблизительно те же величины <$V_{gr}$>, $\sigma$ и <$v$> (см. рис.2.2 и выражение (2.7), глава 2), то есть удовлетворяли обеим точкам. Расчет для всех значений скорости роста проводился при данных фиксированных параметрах. Результаты моделирования показали, что даже при произвольном выборе двух калибровочных точек и при варьировании параметров в пределах 10-15%, спектр не меняется, что свидетельствует об устойчивости найденных значений.

## § I.2. Блок-схема основного алгоритма

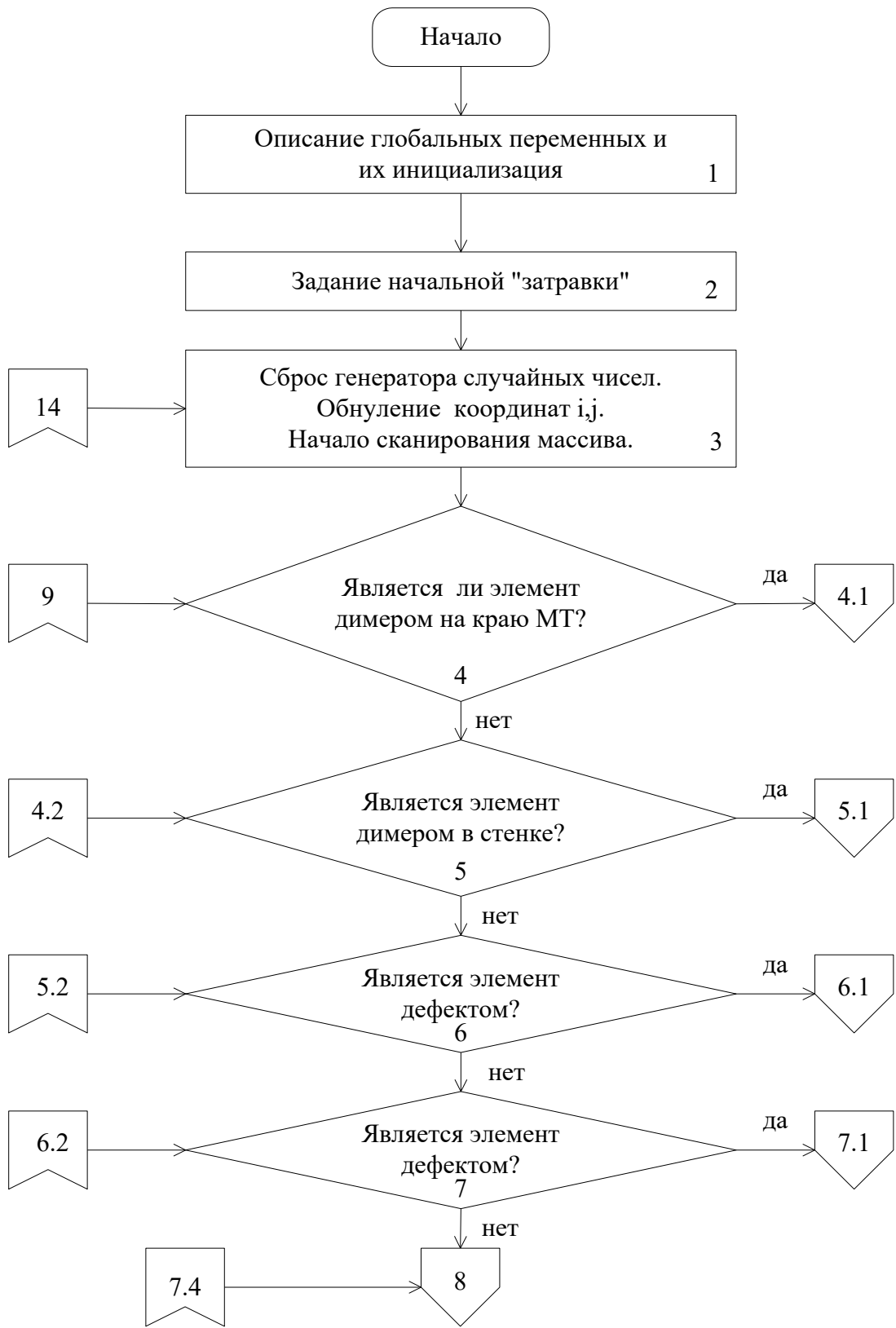

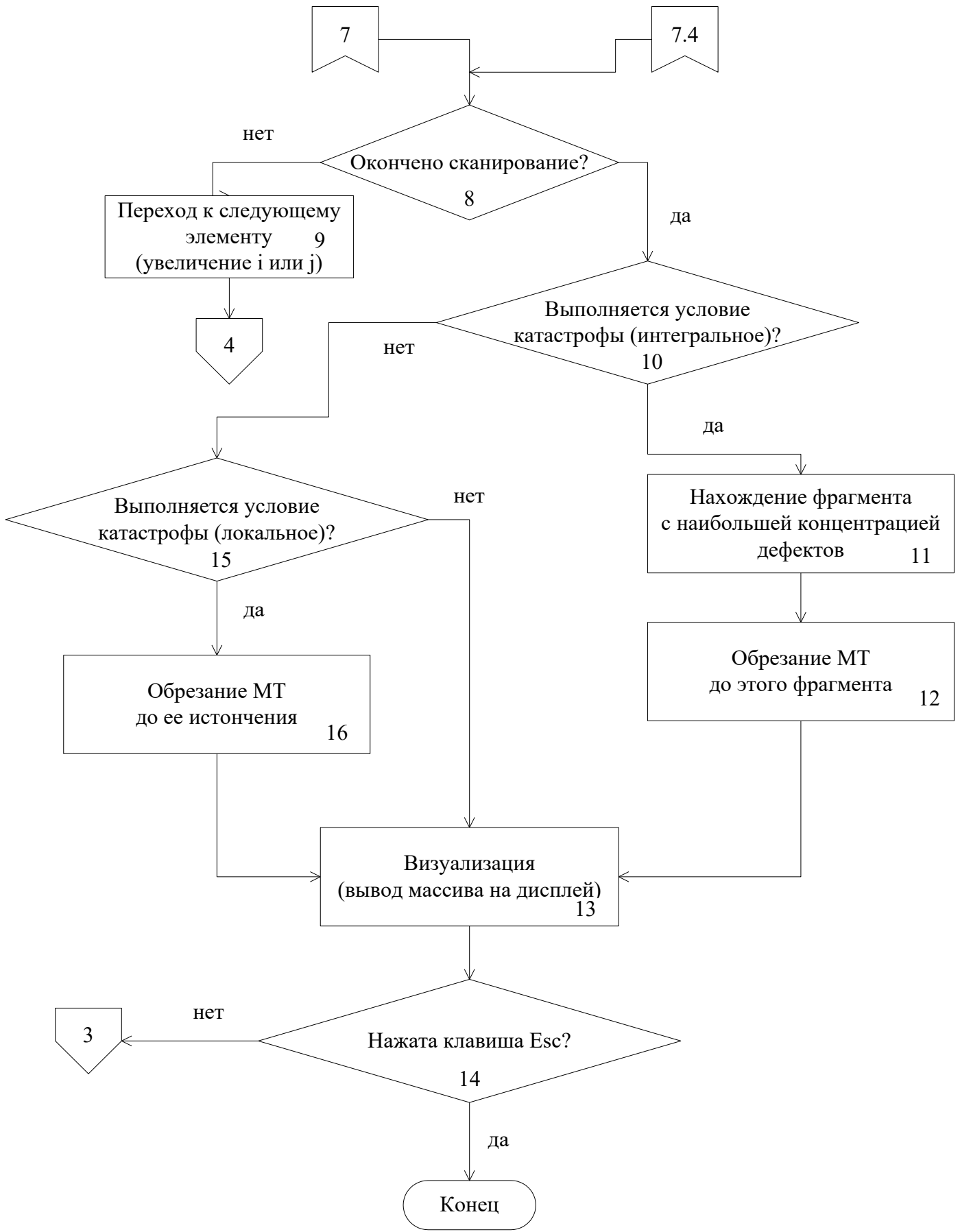

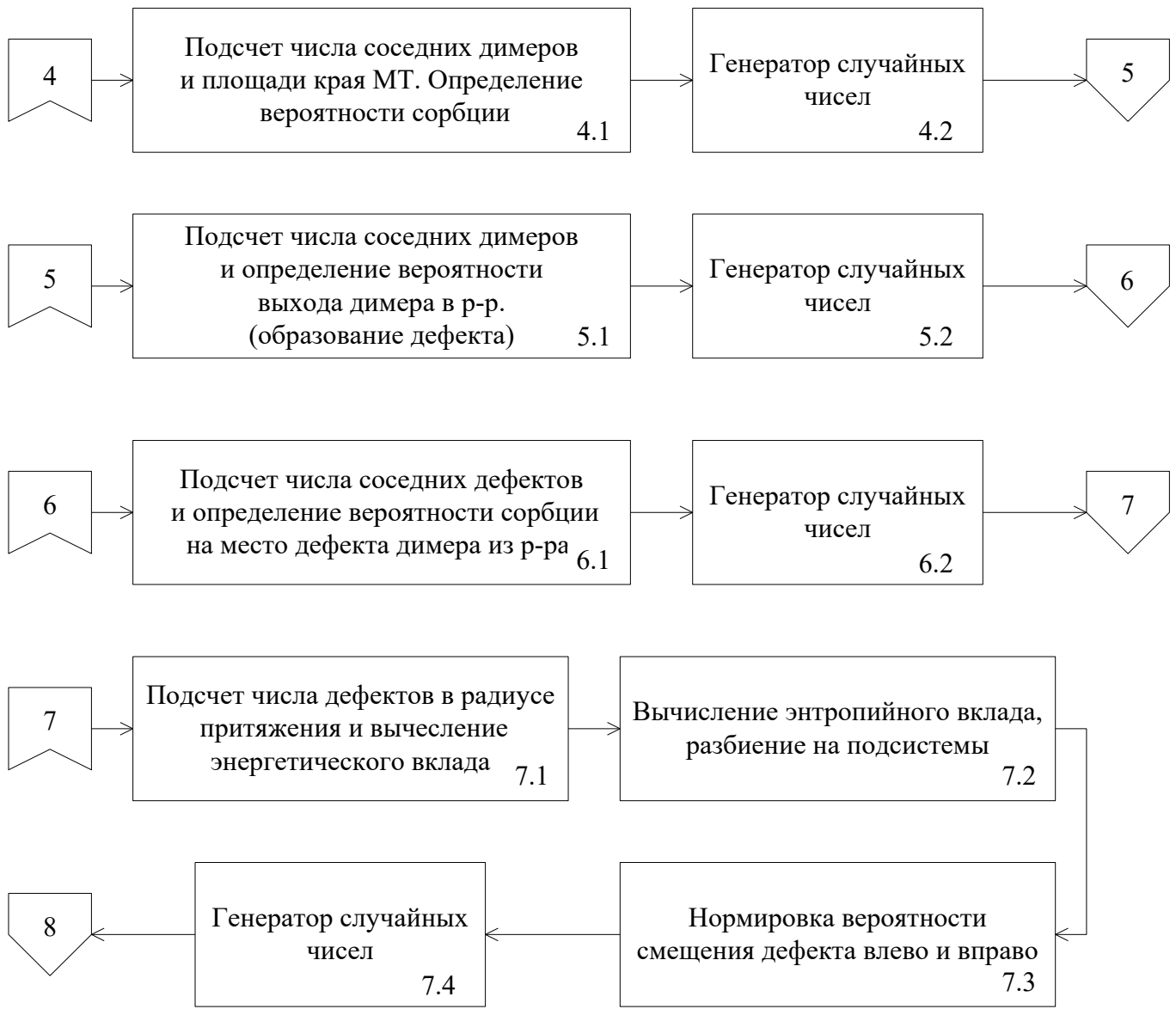

# ПРИЛОЖЕНИЕ II

**Нуклеация. Вывод дискретного аналога и безразмерный вид кинетических уравнений псевдоодномерного приближения**

Кинетические уравнения на непрерывные функций распределения концентраций реагентов и плотность вероятности плюс-концов микротрубочек для псевдоодномерного приближения выглядят следующим образом:

$$\frac{\partial p(x,t)}{\partial t} = \beta \frac{\partial}{\partial x}(k_3 p - k_2 u p) \quad \text{(II.1)}$$

$$\frac{\partial u(x,t)}{\partial t} = D \frac{\partial^2 u}{\partial x^2} - k_2 u p + k_1 [\text{ГТФ}] v - k_{-1} [\text{ГДФ}] u \quad \text{(II.2)}$$

$$\frac{\partial v(x,t)}{\partial t} = D \frac{\partial^2 v}{\partial x^2} + k_3 p - k_1 [\text{ГТФ}] v + k_{-1} [\text{ГДФ}] u \quad \text{(II.3)}$$

$$N_A S \cdot \int_0^L p(x,t) dx = 1 \quad \text{(II.4)}$$

$$\int_0^L \left[ u(x,t) + v(x,t) + \frac{x}{\beta} p(x,t) \right] dx = \frac{N_{tot}}{N_A} \quad \text{(II.5)}$$

где $u(x,t)$ соответствует значению концентрации молекул Tu-ГТФ в точке с координатой $x$ в момент времени $t$, $v(x,t)$ – концентрации молекул Tu-ГДФ, $p(x,t) = \frac{w(x,t)}{V N_A}$ – плотности вероятности нахождения плюс-конца микротрубочки, деленной на число Авогадро $N_A$. Система рассматривается на отрезке $[0, L]$ оси $x$.

Для численного исследования, представим систему (II.1)-(II.5) в дискретном виде по пространственной переменной. Разобъем рассматриваемый отрезок на $M+1$ интервал следующим образом:

$$[0,L] = [0,\beta] \cup (\beta, \beta + h] \cup (\beta + h, \beta + 2h] \cup \ldots \cup (\beta + (M-1)h, \beta + Mh] \quad \text{(II.6)}$$

Интервал $[0,\beta]$ длины $\beta$ будем обозначать индексом 0. Для остальных интервалов сетки длины $h$ будем использовать индексы от 1 до $M$, нумеруя последовательно по мере продвижения вдоль оси *x*. Для нахождения приближенного решения поставим в соответствие непрерывным функциям $u(x,t)$, $v(x,t)$ сеточные функции $\{u_m(t), v_m(t)\}$, $m \subset \{1, M\}$, определенные на каждом из интервалов (II.6), кроме нулевого. Функции же $w(x,t)$ (или же $p(x,t)$) будет соответствовать сеточная функция $\{w_m(t)\}$, $m \subset \{0, M\}$, где $w_0$ обозначает вероятность нахождения свободного плюс-конца на затравке (в нулевом интервале, см. рис.II.1). Причем будем полагать, что $p_m(t) = \dfrac{w_m(t)}{hSN_A}$, $p_0(t) = \dfrac{w_0(t)}{\beta SN_A}$.

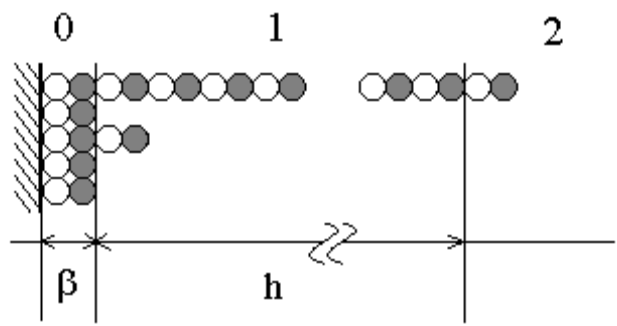

**Рис.II.1.** Пространственное разбиение отрезка [0,L]. Нулевой интервал длины β соответствует «затравке».

Построение дискретного аналога системы уравнений (II.1)-(II.5) начнем с написания условий нормировки вероятности (II.4) и сохранения материального баланса (II.5) в дискретном виде. Т.к. исходная система консервативна, то и для ее дискретного аналога необходимо выполнение условия сохранения количества вещества на каждом шаге. Дискретные аналоги уравнений (II.4)-(II.5) выглядят следующим образом:

$$\sum_{m=0}^{M} w_m = w_0 + hSN_A \sum_{m=1}^{M} p_m = 1 \qquad (II.6)$$

$$w_0 + hSN_A \sum_{m=1}^{M} \left( u_m + v_m + \frac{\beta + mh}{\beta} p_m \right) = N_0 \qquad (II.7)$$

Процессы сорбции и десорбции, согласно нашим предположениям, не зависят друг от друга, поэтому справедливо рассматривать их независимым образом.

**Процесс 1: деполимеризация.** Рассмотрим первый частный случай, когда полимеризационного роста не происходит, а изменения плотности вероятности и концентраций определяются только деполимеризационными членами. Тогда для произвольного элемента номером $m \subset \{2, M-1\}$ верно:

Скорость потока в молях через правую границу:

$$J_r = k_3 p_{m+1}(t)$$

через левую:

$$J_l = -k_3 p_m(t)$$

Итого:

$$w_m(t+dt) = w_m(t) + (J_r + J_l) \cdot \delta V \cdot N_A = w_m(t) + k_3(p_{m+1}(t) - p_m(t)) \cdot \delta V \cdot N_A \quad \text{(II.8)}$$

где $\delta V = \beta S$ – та часть объема соседней ячейки, из которой за время $dt$ осуществляется переход плюс-концов в $m$-ую ячейку. Разложив левую часть (II.8) по формуле Тейлора в окрестности точки $t$ до линейного члена и поделив обе части на $hSN_A$, получим:

$$\frac{dp_m(t)}{dt} = \frac{\beta}{h} k_3 (p_{m+1}(t) - p_m(t)) \qquad \text{(II.9)}$$

Для правой границы уравнение будет иметь несколько иной вид, т.к. она непроницаема для плюс-концов:

$$\frac{dp_M(t)}{dt} = -\frac{\beta}{h} k_3 p_M(t) \qquad \text{(II.10)}$$

Обратим внимание на корректный вывод граничного условия для левой границы, соответствующей «затравке». Будем полагать, что константы скоростей сорбции и десорбции на нуклеационном центре отличаются от аналогичных в остальном объеме и обозначим их соответственно $k_2^0, k_3^0$. Аналогично (8), для первой ячейки:

$$w_1(t+dt) = w_1(t) + \delta V \cdot N_A = w_1(t) + (k_3 p_2(t) - k_3^0 p_1(t)) \cdot \delta V \cdot N_A,$$

откуда:

$$\frac{dp_1(t)}{dt} = \frac{\beta}{h}(k_3 p_2(t) - k_3^0 p_1(t)) \qquad \text{(II.11)}$$

Согласно выдвинутому выше предположению, «затравка» не способна к деполимеризации, поэтому для изменения вероятности нахождения плюс-конца в нулевой ячейки справедливо:

$$w_0(t+dt) = w_0(t) + k_3^0 p_1(t) \cdot \delta V \cdot N_A$$

откуда:

$$\frac{dp_0(t)}{dt} = k_3^0 p_1(t) \qquad (II.12)$$

Заметим, что коэффициента $\frac{\beta}{h}$ в последнем уравнении нет, что обуславливается длиной «затравки», равной $\beta$ и, как отмечалось выше, $p_0(t) = \frac{w_0(t)}{\beta S N_A}$.

Деполимеризация вносит вклад также и в концентрационную переменную, а именно, в концентрацию Tu-ГДФ. Из условий разбиения понятно, что увеличение числа молекул Tu-ГДФ в $m$-ой ячейке за время $dt$ определяется исключительно за счет членов, зависящих от $w_m$, а соседние ячейки не успевают внести свой вклад. Согласно (II.8), уменьшение <u>вероятности нахождения плюс-концов</u> в $m$-ой ячейке, равно:

$$dp_m = k_3 p_m(t) \cdot \delta V \cdot N_A.$$

Нам же необходимо узнать количество молекул, образующихся в растворе за счет деполимеризации. Для вычисления воспользуемся уравнением материального баланса (II.7), откуда видно, что:

$$d\tilde{V}_m = k_3 p_m(t) \cdot \delta V \cdot N_A \cdot \frac{h}{\beta} dt$$

где $d\tilde{V}_m$ равно приращению количества молекул Tu-ГДФ в $m$-ой ячейке за время $dt$. Поделив обе части на $hSN_A$ и $dt$, получим:

$$\frac{dv_m(t)}{dt} = k_3 p_m(t), \quad m \in \{2, M\} \qquad (II.13)$$

и, соответственно, для первой ячейки:

$$\frac{dv_1(t)}{dt} = k_3^0 p_1(t) \qquad (\text{II}.14)$$

**Процесс 2: полимеризация.** Рассмотрим второй частный случай, когда деполимеризация отсутствует, а плотность вероятности и концентрации изменяются только за счет процессов сорбции. Для произвольного элемента номером $m \subset \{2, M-1\}$ верно следующее.

Поток в молях через правую границу:

$$J_r = -k_2 u_m(t) p_m(t)$$

через левую:

$$J_l = k_2 u_{m-1}(t) p_{m-1}(t)$$

Итого:

$$w_m(t+dt) = w_m(t) + (J_r + J_l) \cdot \delta V \cdot N_A = w_m(t) + k_2(u_{m-1}(t)p_{m-1}(t) - u_m(t)p_m(t)) \cdot \delta V \cdot N_A \qquad (\text{II}.15)$$

Аналогично переходу от (II.8) к (II.9):

$$\frac{dp_m(t)}{dt} = \frac{\beta}{h} k_2 (u_{m-1}(t)p_{m-1}(t) - u_m(t)p_m(t)) \qquad (\text{II}.16)$$

Правая граница, согласно условию непроницаемости для плюс-концов:

$$\frac{dp_M(t)}{dt} = \frac{\beta}{h} k_2 u_{M-1}(t) p_{M-1}(t) \qquad (\text{II}.17)$$

Рассмотрим происходящее на левой границе. $w_0$ фактически представляет собой отношение свободных для посадки мест при $x=\beta$ к общему количеству мест, или, что то же самое – площади свободных мест к площади занятых. Поток падающих частиц Tu-ГТФ на левую границу первого интервала приблизительно равен:

$$\frac{dZ}{dt} = \frac{1}{6} \vartheta u_1(t) \frac{S_Z}{S} \qquad (\text{II}.18)$$

где $Z$ – количество падающих частиц, $S_Z$ – площадь затравки, $\vartheta$ – скорость частиц. Прореагируют же только те, которые попадут в свободные места, следовательно, уменьшение количества свободных мест будет определяться уравнением:

$$\frac{dw_0}{dt} = -\frac{1}{6}\vartheta u_1(t)\frac{S_Z}{S}w_0$$

обозначая $k_2^0 = \frac{1}{6}\vartheta\frac{S_Z}{S}$, получим уравнение на $p_0$:

$$\frac{dp_0}{dt} = -k_2^0 u_1(t)p_0(t) \tag{II.19}$$

Согласно нормировке вероятности (сохранению общего количества плюс концов), насколько уменьшилась вероятность $p_0$ за счет полимеризации на затравке, настолько должна увеличиться вероятность в интервале $p_1$, следовательно:

$$w_1(t+dt) = w_1(t) + k_2^0 u_1(t)w_0(t), \tag{II.20}$$

разделив на $hSN_A$:

$$\frac{dp_1}{dt} = \frac{\beta}{h}k_2^0 u_1(t)p_0(t)$$

А с учетом полимеризации самого $p_1$, получим:

$$\frac{dp_1}{dt} = \frac{\beta}{h}(k_2^0 u_1(t)p_0(t) - k_2 u_1(t)p_1(t)) \tag{II.21}$$

Теперь рассмотрим, насколько уменьшилась концентрация $u_1$, только за счет сорбции на затравке. Согласно (II.20), за промежуток времени $dt$ приращение вероятности плюс-концов на первом интервале:

$$k_2^0 u_1(t)w_0(t)$$

и этому же равно уменьшение на нулевом. Нам же необходимы затраченные на это молекулы Tu-ГТФ, в штуках за период времени $dt$. Поэтому, согласно уравнению материального баланса (II.7), получим:

$$-dU_1 = dp_1\frac{\beta+h}{\beta} - dp_0\frac{\beta}{\beta} = k_2^0 u_1(t)w_0(t)\frac{\beta+h}{\beta} - k_2^0 u_1(t)w_0(t)\frac{\beta}{\beta} = k_2^0 u_1(t)w_0(t)\frac{h}{\beta}$$

Откуда:

$$\frac{du_1}{dt} = -k_2^0 u_1(t) p_0(t)$$

с учётом полимеризации самого $p_1$, в конечном виде получим:

$$\frac{du_1}{dt} = -k_2^0 u_1(t) p_0(t) - k_2 u_1(t) p_1(t) \tag{II.22}$$

**Общие уравнения.** Объединим полимеризационные и деполимеризационные члены, соответствующие каждой переменной в уравнениях (II.9)-(II.22). Также добавим реакцию между Tu-ГТФ и ГДФ и между Tu-ГДФ и ГТФ в растворе и диффузию концентрационных переменных по пространству. Результирующая дискретная аппроксимация исходной системы (II.1)-(II.5) по пространству примет вид:

$$\frac{dp_0}{dt} = k_3^0 p_1 - k_2^0 u_1 p_0 \tag{II.23}$$

$$\frac{dp_1}{dt} = \frac{\beta}{h}\left(k_3 p_2 - k_3^0 p_1 - k_2 u_1 p_1 + k_2^0 u_1 p_0\right) \tag{II.24}$$

$$\frac{dp_m}{dt} = \frac{\beta}{h}\left(k_3(p_{m+1} - p_m) + k_2(u_{m-1} p_{m-1} - p_m u_m)\right) \qquad m \in \{2, M-1\} \tag{II.25}$$

$$\frac{dp_M}{dt} = \frac{\beta}{h}\left(k_2 u_{M-1} p_{M-1} - k_3 p_M\right) \tag{II.26}$$

$$\frac{du_1}{dt} = D\frac{u_2 - u_1}{h^2} - k_2^0 u_1 p_0 - k_2 p_1 u_1 + k_1[\text{ГТФ}]v_1 - k_{-1}[\text{ГДФ}]u_1 \tag{II.27}$$

$$\frac{du_m}{dt} = D\frac{u_{m-1} - 2u_m + u_{m+1}}{h^2} - k_2 p_m u_m + k_1[\text{ГТФ}]v_m - k_{-1}[\text{ГДФ}]u_m, \quad m \in \{2, M-1\}$$
$$\tag{II.28}$$

$$\frac{du_M}{dt} = D\frac{u_{M-1} - u_M}{h^2} + k_1[\text{ГТФ}]v_M - k_{-1}[\text{ГДФ}]u_M \tag{II.29}$$

$$\frac{dv_1}{dt} = D\frac{v_2 - v_1}{h^2} + k_3^0 p_1 - k_1[\text{ГТФ}]v_1 + k_{-1}[\text{ГДФ}]u_1 \tag{II.30}$$

$$\frac{dv_m}{dt} = D\frac{v_{m-1} - 2v_m + v_{m+1}}{h^2} + k_3 p_m - k_1[\text{ГТФ}]v_m + k_{-1}[\text{ГДФ}]u_m, \quad m \in \{2, M-1\}$$
$$\tag{II.31}$$

$$\frac{dv_M}{dt} = D\frac{v_{M-1} - v_M}{h^2} + k_3 p_M - k_1[\text{ГТФ}]v_M + k_{-1}[\text{ГДФ}]u_M \tag{II.32}$$

$$\sum_{m=0}^{M} w_m = w_0 + hSN_A \sum_{m=1}^{M} p_m = 1 \tag{II.33}$$

$$w_0 + hSN_A \sum_{m=1}^{M}\left(u_m + v_m + \frac{\beta + mh}{\beta} p_m\right) = N_0 \tag{II.34}$$

**Альтернативный подсчет материального баланса.**

В настоящей схеме для подсчета материального баланса использовалась приближенная формула (II.7) (она же (II.34)). Согласно ей, вклад в общее количество молекул от вероятности нахождения плюс-конца в m-ом интервале составляет:

$$\frac{\beta + mh}{\beta} w_m \tag{II.35}$$

что соответствует такому рассмотрению, при котором все плюс-концы в *m*-ом интервале имеют длину $\beta + mh$. Более точное приближение, в случае, когда плюс-концы распределены по ячейке равномерно, дает вклад, равный:

$$\frac{\beta + (m-1)h}{\beta} w_m + \int_o^h w_m \frac{x}{\beta} dx = \left(\frac{\beta + (m-\frac{1}{2})h}{\beta}\right) w_m$$

Опуская подробный вывод (аналогичный уже проделанному выше), для этого приближения уравнения (II.23)-(II.34) будут выглядеть следующим образом:

$$\frac{dp_0}{dt} = k_3^0 p_1 - k_2^0 u_1 p_0 \tag{II.23**}$$

$$\frac{dp_1}{dt} = \frac{\beta}{h}\left(k_3 p_2 - k_3^0 p_1 - k_2^0 u_1 p_1 + k_2^0 u_1 p_0\right) \tag{II.24**}$$

$$\frac{dp_m}{dt} = \frac{\beta}{h}\left(k_3(p_{m+1} - p_m) + k_2(u_{m-1}p_{m-1} - p_m u_m)\right) \qquad m \in \{2, M-1\} \tag{II.25**}$$

$$\frac{dp_M}{dt} = \frac{\beta}{h}\left(k_2 u_{M-1} p_{M-1} - k_3 p_M\right) \tag{II.26**}$$

$$\frac{du_1}{dt} = D\frac{u_2 - u_1}{h^2} - \frac{1}{2}k_2^0 u_1 p_0 - k_2 p_1 u_1 + k_1[\text{ГТФ}]v_1 - k_{-1}[\text{ГДФ}]u_1 \qquad (\text{II.27**})$$

$$\frac{du_m}{dt} = D\frac{u_{m-1} - 2u_m + u_{m+1}}{h^2} - k_2 p_m u_m + k_1[\text{ГТФ}]v_m - k_{-1}[\text{ГДФ}]u_m, \quad m \in \{2, M-1\}$$
$$(\text{II.28**})$$

$$\frac{du_M}{dt} = D\frac{u_{M-1} - u_M}{h^2} + k_1[\text{ГТФ}]v_M - k_{-1}[\text{ГДФ}]u_M \qquad (\text{II.29**})$$

$$\frac{dv_1}{dt} = D\frac{v_2 - v_1}{h^2} + \frac{1}{2}k_3^0 p_1 - k_1[\text{ГТФ}]v_1 + k_{-1}[\text{ГДФ}]u_1 \qquad (\text{II.30**})$$

$$\frac{dv_m}{dt} = D\frac{v_{m-1} - 2v_m + v_{m+1}}{h^2} + k_3 p_m - k_1[\text{ГТФ}]v_m + k_{-1}[\text{ГДФ}]u_m, \quad m \in \{2, M-1\}$$
$$(\text{II.31**})$$

$$\frac{dv_M}{dt} = D\frac{v_{M-1} - v_M}{h^2} + k_3 p_M - k_1[\text{ГТФ}]v_M + k_{-1}[\text{ГДФ}]u_M \qquad (\text{II.32**})$$

$$\sum_{m=0}^{M} w_m = w_0 + hSN_A \sum_{m=1}^{M} p_m = 1 \qquad (\text{II.33**})$$

$$w_0 + hSN_A \sum_{m=1}^{M}\left(u_m + v_m + \frac{\beta + \left(m - \frac{1}{2}\right)h}{\beta} p_m\right) = N_0 \qquad (\text{II.34**})$$

т.е. фактически вид уравнений не изменяется, кроме добавления малой аддитивной постоянной к общему количеству молекул тубулина.